\documentclass[letterpaper,oneside,onecolumn,12pt]{article}

\usepackage{osajnl2}

\usepackage{theorem} \usepackage{IEEEtrantools} \usepackage{times}
\usepackage{dsfont} \usepackage{subfigure} \usepackage{algpseudocode}

\usepackage{float}

\usepackage[cmex10]{amsmath} 

\usepackage{graphicx}
\graphicspath{{./fig/}} 

\RequirePackage{amsmath} \RequirePackage{xspace} \RequirePackage{bbm}
\RequirePackage{oldgerm} 

\def\XS{\xspace}
\DeclareMathAlphabet{\mathb}{OML}{cmm}{b}{it}
\def\sbm#1{\ensuremath{\mathb{#1}}}                
\def\sbmm#1{\ensuremath{\boldsymbol{#1}}}          
\def\sdm#1{\ensuremath{\mathrm{#1}}}               
\def\scu#1{\ensuremath{\mathcal{#1\XS}}}           

  \def\ab{{\sbm{a}}\XS}
  \def\bb{{\sbm{b}}\XS}
  
\def\Db{{\sbm{D}}\XS}  
  
\def\Fb{{\sbm{F}}\XS}  
  
\def\Hb{{\sbm{H}}\XS}  
\def\Ib{{\sbm{I}}\XS}

\def\Pb{{\sbm{P}}\XS}  
  
\def\Rb{{\sbm{R}}\XS}

  \def\wb{{\sbm{w}}\XS}
  \def\xb{{\sbm{x}}\XS}
  \def\yb{{\sbm{y}}\XS}

\def\Uc{{\scu{U}}\XS}

  \def\dD{{\sdm{d}}\XS}

\def\LD{{\sdm{L}}\XS}

  \def\pD{{\sdm{p}}\XS}

  \def\tD{{\sdm{t}}\XS}

\def\gammab      {{\sbmm{\gamma}}\XS}      \def\Gammab    {{\sbmm{\Gamma}}\XS}
\def\deltab      {{\sbmm{\delta}}\XS}      
\def\epsilonb    {{\sbmm{\epsilon}}\XS}

     \def\Lambdab   {{\sbmm{\Lambda}}\XS}
\def\mub         {{\sbmm{\mu}}\XS}

      \def\Sigmab    {{\sbmm{\Sigma}}\XS}

\def\eR{\Rbb}

  \def\eg{{\textgoth{e}}\XS}
  
\def\XS{\xspace}

\RequirePackage{ifthen}
\RequirePackage{calc}

\newcommand{\taille}[1][\scad]{%
\ifthenelse{#1 = -5}{}{}%
\ifthenelse{#1 = -4}{\tiny}{}%
\ifthenelse{#1 = -3}{\scriptsize}{}%
\ifthenelse{#1 = -2}{\footnotesize}{}%
\ifthenelse{#1 = -1}{\small}{}%
\ifthenelse{#1 = 0}{\normalsize}{}%
\ifthenelse{#1 = 1}{\large}{}%
\ifthenelse{#1 = 2}{\Large}{}%
\ifthenelse{#1 = 3}{\LARGE}{}%
\ifthenelse{#1 = 4}{\huge}{}%
\ifthenelse{#1 = 5}{\Huge}{}}

\newboolean{@serif}
\setboolean{@serif}{true}
\def\scad{-5} 

\newcounter{taille}
\newcommand{\sca}[2][\scad]{\setcounter{taille}{#1}%
  \ifthenelse{\boolean{@serif}}
  {{\taille[\thetaille]\textsc{#2}}}
  {\setcounter{taille}{\value{taille}-1}{\uppercase{\taille[\thetaille]#2}}}}

\def\rem#1{}                    

\def\apost{\textit{a posteriori}\XS}
\def\Apost{\textit{A posteriori}\XS}

\def\aprio{\textit{a priori}\XS}

\def\eg{\textit{e.g.,}\XS}
\def\etal{\textit{et al.}\XS}

\def\ie{\textit{i.e.,}\XS}

\def\via{\textit{via}\XS}

\def\wrt{w.r.t.\XS}

\newcommand{\UPS}[1][\scad]{{Université Paris-Sud 11}\XS} 

\newcommand{\ESE}[1][\scad]{{Supélec}\XS}

\newcommand{\adresscea}[1][\scad]{4 Place du Général Leclerc, 91406 Orsay Cedex, France\XS}

\def\pth#1{\left(#1\right)}                
              
\def\cro#1{\left[#1\right]}

\def\diag{{\mathrm{diag}}}

\def\Exp#1{\exp\cro{#1}}

\newsavebox{\fminibox}
\newlength{\fminilength}
\newenvironment{fminipage}[1][\linewidth]
  {\setlength{\fminilength}{#1}
   \begin{lrbox}{\fminibox}\begin{minipage}{\fminilength}}
  {\end{minipage}\end{lrbox}\noindent\fbox{\usebox{\fminibox}}}

 \def\T{^\tD} \def\+{^\dagger}

\def\nequiv{\not\kern-.05em\equiv}
\def\egal{\kern-.5em=\kern-.5em}        
\def\propt{\kern-.2em\propto\kern-.2em} 

\def\intdouble{\int\kern-0.3em\int}
\def\inttriple{\int\kern-0.3em\int\kern-0.3em\int}

\def\rond#1{\overset{\kern-0.33em~_\circ}{#1}}
\def\rondit[#1]#2{\overset{\kern#1~_\circ}{#2}}

\def\babs{\begin{abstract}}             \def\eabs{\end{abstract}}
\def\barr{\begin{array}}                \def\earr{\end{array}}
\def\bcc{\begin{center}}                \def\ecc{\end{center}}

\def\bdes{\begin{description}}          \def\edes{\end{description}}
\def\bdoc{\begin{document}}             \def\edoc{\end{document}}
\def\ben{\begin{enumerate}}             \def\een{\end{enumerate}}
\def\beqn{\begin{eqnarray}}             \def\eeqn{\end{eqnarray}}
\def\beqnl#1{\beqn\label{#1}}           \def\eeqnl#1{\label{#1}\eeqn}
\def\beqnx{\begin{eqnarray*}}           \def\eeqnx{\end{eqnarray*}}
\def\bseqn{\begin{subeqnarray}}         \def\eseqn{\end{subeqnarray}}
\def\beq#1\eeq{\begin{equation}#1\end{equation}}
\def\bal#1\eal{\begin{align}#1\end{align}}
\def\balx#1\ealx{\begin{align*}#1\end{align*}}
\def\beqx{$$}                           \def\eeqx{$$}
\def\bfig{\protect\begin{figure}}       \def\efig{\protect\end{figure}}
\def\bfigx{\protect\begin{figure*}}     \def\efigx{\protect\end{figure*}}
\def\bfigt{\protect\begin{figurette}}   \def\efigt{\protect\end{figurette}}
\def\bfl{\begin{flushleft}}             \def\efl{\end{flushleft}}
\def\bfr{\begin{flushright}}            \def\efr{\end{flushright}}
\def\bit{\begin{itemize}}               \def\eit{\end{itemize}}
\def\bmi{\begin{minipage}}              \def\emi{\end{minipage}}
\def\bfmi{\begin{fminipage}}            \def\efmi{\end{fminipage}}
\def\bpic{\begin{picture}}              \def\epic{\end{picture}}
\def\bqu{\begin{quote}}                 \def\equ{\end{quote}}
\def\bqun{\begin{quotation}}            \def\equn{\end{quotation}}
\def\bsl{\begin{slide}}                 \def\esl{\end{slide}}
\def\btabb{\begin{tabbing}}             \def\etabb{\end{tabbing}}
\def\btabl{\begin{table}}               \def\etabl{\end{table}}
\def\btablx{\begin{table*}}             \def\etablx{\end{table*}}
\def\btab{\begin{tabular}} 
\def\btabu{\begin{tabular}}             \def\etabu{\end{tabular}}
\def\btabx{\begin{tabular*}}            \def\etabx{\end{tabular*}}
\def\bbib{\begin{thebibliography}{999}} \def\ebib{\end{thebibliography}}
\def\bver{\begin{verbatim}}             \def\ever{\end{verbatim}}
\def\bca{\begin{cases}}                  \def\eca{\end{cases}}

\def\phi{\varphi}
\def\eps{\varepsilon}
\def\fftdd{\textsc{fft-2d}\XS}
\def\dftdd{\textsc{dft-2d}\XS}
\def\ifft{\textsc{ifft}\XS}
\def\fft{\textsc{fft}\XS}

\def\Imzero{\rond x_0}
\def\Im{\rond \xb}
\def\Ime{\rond \xb_{*}}

\def\Data{\rond \yb}
\def\Datazero{\rond y_0}

\def\GO{\gamma_0} \def\G1{\gamma_1}	\def\GN{\gamma_{\epsilon}}
\def\hGN{\widehat{\GN}}
\def\hG1{\widehat{\G1}}

\def\wA{w_\alpha} \def\wB{w_\beta}
\def\hWA{\widehat{\wA}}	\def\hWB{\widehat{\wB}}	

\def\hPhi{\widehat{\varphi}}

\def\Xeap{\hat \xb}

\def\H{\Hb}
\def\LH{\Lambdab_{\Hb}}
\def\Hzero{\rond h_0}

\def\D{\Db}
\def\LD{\Lambdab_{\Db}}
\def\LDe{\Lambdab_{\Db*}}

\def\LR{\Lambdab_{\Rb}}

\def\One{\mathds{1}}
\def\LO{\Lambdab_{\One}}

\def\F{\Fb}
\def\ensbR{\mathds{R}}
\def\wrt{w.r.t.\XS}

\newtheorem{remark}{Remark}
\newtheorem{property}{Property}
\newtheorem{example}{Example}

\def\figWidth{3.7}
\def\figWidthMed{5}
\def\figWidthLarge{5.5}

\definecolor{Rouge}{rgb}{1,0,0} 	
\definecolor{Bleu}{rgb}{0,0,1} 	
\definecolor{Vert}{rgb}{0,1,0} 	
\definecolor{Truc}{rgb}{1,0,1} 	

\def\eR{\mathds{R}}

\begin{document}

\title{Bayesian estimation of regularization and PSF parameters for
  Wiener-Hunt deconvolution}

\author{Fran\c cois Orieux,$^{1,*}$ Jean-Fran\c{c}ois Giovannelli,$^2$
  Thomas Rodet,$^1$}

\address{$^1$ Laboratoire des Signaux et Syst\`emes (\textsc{cnrs} --
  \textsc{supelec} -- Univ.  Paris-Sud 11), \textsc{supelec}, Plateau
  de Moulon, 3 rue Joliot-Curie, 91\,192 Gif-sur-Yvette, France}
\address{$^2$ Laboratoire d'Int\'egration du Mat\'eriau au Syst\`eme
  (\textsc{cnrs} -- \textsc{enseirb} -- Univ. Bordeaux 1 --
  \textsc{enscpb}), 351 cours de la Libération, 33405 Talence, France}

\address{$^*$Corresponding author: orieux@lss.supelec.fr}

\begin{abstract}
    This paper tackles the problem of image deconvolution with joint estimation of PSF parameters and hyperparameters. Within a Bayesian framework, the solution is inferred \via a global \apost law for unknown parameters and object. The estimate is chosen as the posterior mean, numerically calculated by means of a Monte-Carlo Markov chain algorithm. The estimates are efficiently computed in the Fourier domain and the effectiveness of the method is shown on simulated examples.  Results show precise estimates for PSF parameters and hyperparameters as well as precise image estimates including restoration of high-frequencies and spatial details, within a global and coherent approach.
\end{abstract}


\maketitle

\section{Introduction}
\label{sec:introduction}

Image deconvolution has been an active research field for several
decades and recent contributions can be found in papers such
as~\cite{Idier08,Molina06,Campisi07}. Examples of application are
medical imaging, astronomy, nondestructive testing and more generally
imagery problems. In these applications, degradations induced by the
observation instrument limit the data resolution while the need of
precise interpretation can be of major importance. For example, this
is particularly critical for long-wavelength astronomy (see
\eg~\cite{Rodet08}). In addition, the development of a high quality
instrumentation system must rationally be completed by an equivalent
level of quality in the development of data processing
methods. Moreover, even for poor performance systems, the restoration
method can be used to bypass instrument limitations.

When the deconvolution problem is ill-posed a possible solution relies
on regularization, \ie introduction of information in addition to the
data and the acquisition model~\cite{Tikhonov77,Twomey62}. As a
consequence of regularization, deconvolution methods are specific to
the class of image in accordance with the introduced information. From
this standpoint, the present paper is dedicated to relatively smooth
images encountered for numerous applications in
imagery~\cite{Rodet08,Jalobeanu02,OSullivan95}. The second order
consequence of ill-posedness and regularization is the need to balance
the compromise between different sources of information.

In the Bayesian approach~\cite{Idier08,Demoment89}, information about
unknowns is introduced by means of probabilistic models. Once these
models are designed, the next step is to build the \apost law, given
the measured data. The solution is then defined as a representative
point of this law and the two most classical are (1) the maximizer,
and (2) the mean. From a computational standpoint, the first leads to
a numerical optimization problem and the latter leads to a numerical
integration problem. However, the resulting estimate depends on two
sets of variables in addition to the data.
\begin{enumerate}

\item Firstly, the estimate naturally depends on the response of the
    instrument at work, namely the point spread function (PSF). The
    literature is predominantly devoted to deconvolution in the case
    of known PSF. On the contrary, the present paper is devoted to the
    case of unknown or poorly known PSF and there are two main
    strategies to tackle its estimation from the available data set
    (without extra measurements).

    \begin{itemize}
    \item[(i)] In most practical cases, the instrument can be modeled
        using physical operating description. It is thus possible to
        find the equation for the PSF, at least in a first
        approximation. This equation is usually driven by a relatively
        small number of parameters. It is a common case in optical
        imaging where a Gaussian-shaped PSF is often
        used~\cite{pankajakshan09}. It is also the case in other
        fields: interferometry \cite{Thiebaut95}, magnetic resonance
        force microscopy~\cite{dobigeon09}, fluorescence
        microscopy~\cite{zhang06},\dots Nevertheless, in real
        experiments, the parameter values are unknown or imperfectly
        known and need to be estimated or adjusted in addition to the
        image of interest: the question is namely \textit{myopic}
        deconvolution.

    \item[(ii)] The second strategy forbears the use of the parametric
        PSF deduced from the physical analysis and the PSF then
        naturally appears in a non-parametric form. Practically, the
        non-parametric PSF is unknown or imperfectly known and needs
        to be estimated in addition to the image of interest: the
        question is referred to as \textit{blind} deconvolution for
        example in interferometry
        \cite{Mugnier04,Thiebaut08,Fusco99,Conan98a}.
    \end{itemize}

    From an inference point of view, the difficulty of both myopic and
    blind problems lies in the possible lack of information resulting
    in ambiguity between image and PSF, even in the noiseless case. In
    order to resolve the ambiguity, information must be
    added~\cite{Campisi07,Likas04} and it is crucial to make inquiries
    based on any available source of information. To this end, the
    knowledge of the parametric PSF represents a precious means to
    structure the problem and possibly resolve the
    degeneracies. Moreover, due to instrument design process, a
    nominal value as well as an uncertainty are usually available for
    the PSF parameters.

    In addition, from a practical and algorithmic standpoint, the
    myopic case, \ie the case of parametric PSF, is often more
    difficult due to the non-linear dependence of the observation
    model with respect to the PSF parameters.  On the contrary, the
    blind case, \ie the case of non-parametric PSF, yields a simpler
    practical and algorithmic problem since the observation model
    remains linear \wrt the unknown elements given the object.

    Despite the superior technical difficulty, the present paper is
    devoted to the myopic format since it is expected to be more
    efficient than the blind format from an information standpoint.
    Moreover, the blind case has been extensively studied and a large
    amount of paper is available \cite{bishop08,Lam00,Xu09}, while the
    myopic case has been less investigated, though it is of major
    importance.

\item Secondly, the solution depends on the probability law parameters
    named hyperparameters (means, variances, parameters of correlation
    matrix,\dots). These parameters adjust the shape of the laws and
    in the same time they tune the compromise between the information
    provided by the \aprio and the information provided by the
    data. In real experiments, their values are unknown and need to be
    estimated: the question is namely \textit{unsupervised}
    deconvolution.

\end{enumerate}

For both families of parameters (PSF parameters and hyperparameters),
two approaches are available. In the first one, the parameter values
are empirically tuned or estimated in a preliminary step (with Maximum
Likelihood~\cite{Jalobeanu02} or calibration~\cite{Cannon76} for
example), then the values are used in a second step devoted to image
restoration given the parameters. In the second one, the parameters
and the object are jointly estimated~\cite{bishop08,Molina06}.

For the myopic problem, Jalobeanu \etal~\cite{Jalobeanu02a} address
the case of a symmetric Gaussian PSF. The width parameter and the
noise variance are estimated in a preliminary step by
Maximum-Likelihood. A recent paper~\cite{chen09} addresses the
estimation of a Gaussian blur parameter, as in our experiment, with an
empirical method. They found the Gaussian blur parameter by minimizing
the absolute derivatives of the restored images Laplacian.

The present paper addresses the myopic and unsupervised deconvolution
problem. We propose a new method that jointly estimates the PSF
parameters, the hyperparameters, and the image of interest. It is
built in a coherent and global framework based on an extended \apost
law for all the unknown variables. The posterior law is obtained via
the Bayes rule, founded on \aprio laws: Gaussian for image and noise,
uniform for PSF parameters and gamma or Jeffreys for hyperparameters.

Regarding the image prior law, we have paid special attention to the
parametrization of the covariance matrix in order to facilitate law
manipulations such as integration, conditioning or hyperparameter
estimation. The possible degeneracy of the \apost law in some limit
cases is also studied.

The estimate is chosen as the mean of the posterior law and is
computed using Monte-Carlo simulations. To this end, Monte-Carlo
Markov chain (MCMC) algorithms~\cite{Robert04} enable to draw samples
from the posterior distribution despite its complexity and especially
the non-linear dependence \wrt the PSF parameters.

The paper is structured in the following
manner. Sec.~\ref{sec:direct-model} presents the notations and states
the problem. The three following sections describe our methodology:
firstly the Bayesian probabilistic models are detailed in
Sec.~\ref{sec:bayesian-model}; then a proper posterior law is
established in Sec.~\ref{sec:proper-posterior-law}; an MCMC algorithm
to compute the estimate is described in
Sec.~\ref{sec:mcmc-computation}. Numerical results are shown in
Sec.~\ref{sec:deconv-results}. Finally, Sec.~\ref{sec:conclusion} is
devoted to conclusion and perspectives.

\section{Notations and convolution model}
\label{sec:direct-model}

Consider $N$ pixels real square images represented in lexicographic
order by vector $\xb \in \mathds{R}^N$, with generic elements
$x_{n}$. The forward model writes
\begin{equation}
    \label{eq:1}
    \yb = \H_\wb \,\xb + \epsilonb
\end{equation}
where $\yb \in \ensbR^N$ is the vector of data, $\H_\wb$ a convolution
matrix, $\xb$ the image of interest and $\epsilonb$ the modelization
errors or the noise. Vector $\wb \in \ensbR^P$ stands for the PSF
parameters, such as width or orientation of a Gaussian PSF.

The matrix $\H_\wb$ is block-circulant with circulant-block (BCCB) for
computational efficiency of the convolution in the Fourier space. The
diagonalization~\cite{Hunt71} of $\H_\wb$ writes $\LH = \F \H_\wb
\F^\dag$ where $\F$ is the unitary Fourier matrix and $\dag$ is the
transpose conjugate symbol. The convolution, in the Fourier space, is
then
\begin{equation}
    \label{eq:2}
    \Data = \LH \, \Im + \rond \epsilonb
\end{equation}
where $\Im = \F \xb$, $\Data = \F \yb$ and $\rond \epsilonb = \F \epsilonb$ are
the 2D discrete Fourier transform (\dftdd) of image, data and noise,
respectively.

Since $\LH$ is diagonal, the convolution is computed with a term-wise
product in the Fourier space. There is a strict equivalence between a
description in spatial domain~(Eq.~(\ref{eq:1})) and in Fourier
domain~(Eq.~(\ref{eq:2})). Consequently, for coherent description and
computational efficiency, all the developments are equally done in the
spatial space or in the Fourier space.

For notational convenience, let us introduce the component at
null-frequency $\Imzero\in\eR$ and the vector of component at non-null
frequencies $\Im_* \in \mathds{C}^{N-1}$ so that the whole set of
components writes $\Im = [\Imzero , \Im_* ]$.

Let us note $\One$ the vector of $N$ components equal to $1/N$, so
that $\One\T \xb$ is the empirical mean level of the image. The
Fourier components are the $\rond \One_n$ and we have: $\rond \One_0 =
1$ and $\rond \One_n = 0$ for $n\neq0$. Moreover, $\LO = \F \One
\One\T \F^\dag$ is a diagonal matrix with only one non-null
coefficient at null frequency.

\section{Bayesian probabilistic model}
\label{sec:bayesian-model}

This section presents the prior law for each set of
parameters. Regarding the image of interest, in order to account for
smoothness, the law introduces high-frequency penalization through a
differential operator on the pixel. A conjugate law is proposed for
the hyperparameters and a uniform law is considered for the PSF
parameters.

Moreover, we have paid a special attention to the image prior law
parametrization. In the next section we present several
parametrization in order to facilitate law manipulations such as
integration, conditioning or hyperparameter estimation. Moreover, the
correlation matrix of the image law may become singular in some limit
cases resulting in a degenerated prior law (when $p(\xb)=0$ for all
$\xb\in\eR^N$). Based on this parametrization,
Sec.~\ref{sec:proper-posterior-law} studies the degeneracy of the
posterior in relation with the parameters of the prior law.

\subsection{Image prior law}
\label{sec:image-aprio-law}

The probability law for the image is a Gaussian field with a given
precision matrix $\Pb$ parametrized by a vector $\gammab$. The pdf
reads
\begin{equation} 
     \label{eq:GaussPDF}
     p(\xb | \gammab) = (2\pi)^{-N/2} \det[\Pb]^{1/2} \,
     \Exp{-\frac{1}{2} \, \xb\T\Pb\xb} \,.
\end{equation}
For computational efficiency, the precision matrix is designed (or
approximated) in a toroidal manner, and it is diagonal in the Fourier
domain $\Lambdab_\Pb = \F \Pb \F^\dag$. Thus, the law for $\xb$ also
writes
\begin{align}
    \label{eq:3}
    p(\xb | \gammab) & =
    (2\pi)^{-N/2}\det[\F]\det[\Lambdab_\Pb]^{1/2}\det[\F^{\dag}] \,
    \Exp{-\frac{1}{2} \xb^t \F^{\dag} \Lambdab_\Pb\F \xb} \\
    & = (2\pi)^{-N/2}\det[\Lambdab_\Pb]^{1/2} \, \Exp{-\frac{1}{2}
      \Im^\dag\Lambdab_\Pb\Im}
\end{align}
and it is sometimes referred to~\cite{Calder97} as a Whittle
approximation (see also~\cite[p.133]{Brockwell91}) for the Gaussian
law. The filter obtained for fixed hyperparameters is also the
Wiener-Hunt filter~\cite{Hunt72b}, as described in
Sec.~\ref{sec:sampling-object}.

This paper focuses on smooth images, thus on positive correlation
between pixels. It is introduced by high-frequencies penalty using
any circulant differential operator: $p$-th differences between
pixels, Laplacian, Sobel\dots The differential operator is denoted by
$\D$ and its diagonalized form by $\LD = \F \D \F^\dag$. Then, the
precision matrix writes $\Pb =\G1\Db\T \Db$ and its Fourier
counterpart writes
\begin{equation}
    \label{eq:4}
    \Lambdab_\Pb = \G1\LD^\dag\LD = \diag \left(0, \G1 |\rond d_1|^2,
        \hdots, \G1 |\rond d_{N-1}|^2\right)
\end{equation}
where $\G1$ is a positive scale factor, $\diag$ builds a diagonal
matrix from elementary components and $\rond d_n$ is the $n$-th \dftdd
coefficient of $\D$.

Under this parametrization of $\Pb$, the first eigenvalue is equal to
zero corresponding to the absence of penalty for the null frequency
$\Imzero$, \ie ~no information accounted for about the empirical mean
level of the image. As a consequence, the determinant vanishes
$\det[\Pb] = 0$ resulting in a degenerated prior. To manage this
difficulty, several approaches have been proposed.

Some authors~\cite{Molina06,Mardia92} still use this prior despite its
degeneracy and this approach can be analyzed in two ways.

\begin{enumerate}
\item On the one hand, it can be seen as a non-degenerated law for
    $\Ime$, the set of non-null frequency components only. In this
    format, the prior does not affect any probability to the null
    frequency component and the Bayes rule does not apply to this
    component. Thus, this strategy yields an incomplete posterior law,
    since the null frequency is not embedded in the methodology.

\item On the other hand, it can be seen as a degenerated prior for the
    whole set of frequencies. The application of the Bayes rule is
    then somewhat confusing due to degeneracy. In this format, the
    posterior law cannot be guaranteed to remain non-degenerated.
\end{enumerate}

Anyway, none of the two standpoints yields a posterior law that is
both non-degenerated and addressing the whole set of frequencies.

An alternative parametrization relies on the energy of $\xb$. An extra
term ${\GO} \Ib$, tuned by $\GO>0$, in the precision
matrix~\cite{Bouman93}, introduces information for all the frequencies
including $\Imzero$. The precision matrix writes
\begin{align}
    \Lambda_\Pb
    & = \GO \Ib + \G1 \LD^\dag\LD \nonumber \\
    & = \diag \left(\GO, \GO + \G1 |\rond d_1|^2, \hdots, \GO + \G1
        |\rond d_{N-1}|^2\right) \label{eq:7}
\end{align}
with a determinant
\begin{equation}
    \det[\Lambda_\Pb] = \prod_{n=0}^{N-1} \left (\GO + \G1 |\rond
        d_n|^2 \right). \label{eq:8}
\end{equation}
The obtained Gaussian prior is not degenerated and undoubtedly leads
to a proper posterior. Nevertheless, the determinant Eq.~(\ref{eq:8})
is not separable in $\GO$ and $\G1$. Consequently, the conditional
posterior for these parameters is not a classical law and future
development will be more difficult. Moreover, the non-null frequencies
$\Ime$ are controlled by two parameters $\GO$ and $\G1$
\begin{equation}
    \label{eq:9}
    p(\Im | \GO, \G1 ) = p(\Imzero | \GO ) p(\Ime | \GO, \G1 ).
\end{equation}

The proposed approach to manage the degeneracy relies on the addition
of a term for the null frequency only $\LO=\diag\left(1,0, \hdots, 0
\right)$
\begin{align}
    \Lambda_\Pb
    & = \GO \LO^\dag\LO + \G1 \LD^\dag\LD. \label{eq:11}\\
    & = \diag \left(\GO, \G1 |\rond d_1|^2, \hdots, \G1 |\rond
        d_{N-1}|^2\right) \,.\nonumber
\end{align}
The determinant has a separable expression
\begin{align}
    \det[\Lambda_\Pb] & = \GO \G1^{N-1} \prod_{n=1}^{N-1} |\rond
    d_n|^2 \,, \label{eq:12}
\end{align}
\ie the precision parameters have been factorized. In addition, each
parameter controls a different set of frequencies:
\begin{equation}
    p(\Im | \GO, \G1 ) = p(\Imzero | \GO ) p(\Ime |  \G1 ) \,, \nonumber 
\end{equation}
$\GO$ drives the empirical mean level of the image $\Imzero$ and $\G1$
drives the smoothness $\Ime$ of the image. With the Fourier precision
structure of Eq.~(\ref{eq:11}), we have the non-degenerated prior law
for the image that addresses separately all the frequencies with a
factorized partition function \wrt $(\GO,\G1)$
\begin{equation}
    p(\xb | \GO, \G1 ) = (2 \pi)^{-N/2} \prod_{n=1}^{N-1} |\rond
    d_n| \,\GO^{1/2} \G1^{(N-1)/2} \Exp{ - \frac{\GO}{2} \|\rond
      x_0\|^2 - \frac{\G1}{2} \|\LDe \Ime\|^2}.
    \label{eq:13}
\end{equation}
where $\LDe$ is obtained from $\LD$ without the first line and
column. The next step is to write the \aprio law for the noise in an
explicit form and the other parameters, including the law parameters
$\gammab$ and the instrument parameters $\wb$.

\subsection{Noise and data laws}
\label{sec:noise-data-laws}

From a methodological standpoint, any statistic can be included for
errors (measurement and model errors). It is possible to account for
correlations in the error process or to account for a non-Gaussian
law, \eg Laplacian law, generalized Gaussian law, or other laws based
on robust norm,\dots In the present paper, the noise is modeled as
zero-mean white Gaussian vector with unknown precision parameter $\GN$
\begin{equation}
    \label{eq:14}
    p(\epsilonb | \GN) = \left( 2 \pi \right)^{-N/2} \GN^{N/2}
    \Exp{-\frac{\GN}{2} \| \epsilonb\|^2}.
\end{equation}
Consequently, the likelihood for the parameters given the observed
data writes
\begin{equation}
    \label{eq:16}
    p(\yb | \xb, \GN, \wb ) = (2 \pi)^{-N/2} \GN^{N/2} \Exp{ - \frac{
        \GN}{2} \| \yb - \Hb_{\wb} \xb \|^2 }.
\end{equation}
It naturally depends on the image $\xb$, on the noise parameter $\GN$
and on the PSF parameters $\wb$ embedded in $\Hb_{\wb}$. It clearly
involves a least squares discrepancy that can be rewritten in the
Fourier domain: $\|\yb - \Hb_{\wb} \xb\|^2 = \|\Data - \Lambdab_{\Hb}
\Im\|^2$.

\subsection{Hyperparameters law}
\label{sec:hyperparam-aprio}

A classical choice for hyperparameter law relies on conjugate
prior~\cite{MacKay03}: the conditional posterior for the
hyperparameters is in the same family as its prior. It results in
practical and algorithmic facilities: update of the laws amounts to
update of a small number of parameters.

The three parameters $\GO$, $\G1$ and $\GN$ are precision parameters
of Gaussian laws Eq.~(\ref{eq:13}) and (\ref{eq:16}) and a conjugate
law for these parameters is the Gamma law (see
Appendix~\ref{sec:gamma-law}). Given parameters $(\alpha_i, \beta_i)$,
for $i=0$, 1 or $\epsilon$, the pdf reads
\begin{equation}
    \label{eq:17}
    p(\gamma_i) = \frac{1}{\beta_i^{\alpha_i} \Gamma(\alpha_i)}
    \gamma_i^{\alpha_i - 1}\exp \left( - \gamma_i / \beta_i \right),
    \forall \gamma_i \in [0, +\infty[
\end{equation}

In addition to computational efficiency, the law allows for
non-informative priors. With specific parameter values, one obtains
two improper non-informative prior : the Jeffreys' law $p(\gamma) =
1/\gamma$ and the uniform law $p(\gamma) = \Uc_{[0, +\infty[}(\gamma)$
with $(\alpha_i, \beta_i)$ set to $(0, +\infty)$ and $(1,+\infty)$,
respectively. Jeffreys' law is a classical law for the precisions and
is considered as non-informative~\cite{Kass96}.  This law is also
invariant to power transformations: the law of
$\gamma^n$~\cite{Kass96,Jaynes03} is also a Jeffreys' law.  For these
reasons development is done using the Jeffreys' law.

\subsection{PSF parameters law}
\label{sec:instr-param-law}

Regarding the PSF parameters $\wb$, we consider that the instrument
design process or a physical study provides a nominal value
$\overline{\wb}$ with uncertainty $\deltab$, that is to say
$\wb\in[\overline{\wb}-\deltab ~,~\overline{\wb}+\deltab]$. The
"Principle of Insufficient Reason"~\cite{Kass96} leads to a uniform
prior on this interval
\begin{equation}\label{eq:18}
    p(\wb)  =  \Uc_{\overline{\wb},\deltab}(\wb)
\end{equation}
where $\Uc_{\overline{\wb},\deltab}$ is a uniform pdf on
$[\overline{\wb}-\deltab ~,~\overline{\wb}+\deltab]$. Nevertheless,
within the proposed framework, the choice is not limited and other
laws, such as Gaussian, are possible. Anyway other choices do not
allow easier computation because of the non-linear dependency of the
observation model \wrt PSF parameters.

\section{Proper posterior law}
\label{sec:proper-posterior-law}

At this point, the prior law of each parameter is available: the PSF
parameters, the hyperparameters and the image. Thus, the joint law for
all the parameters is built by multiplying the likelihood
Eq.~(\ref{eq:16}) and the \aprio laws Eq.~(\ref{eq:13}), (\ref{eq:17})
and (\ref{eq:18})
\begin{equation}
    \label{eq:19}
    p(\Im, \GN, \GO, \G1, \wb, \Data) = p(\Data | \Im, \GN,
    \wb)p(\Im|\GO, \G1)p(\GN)p(\GO)p(\G1)p(\wb)
\end{equation}
and explicitly 
\begin{multline}
p(\Im, \GN, \GO, \G1, \wb, \Data) = 
\frac{ (2 \pi)^{-N} \prod_{n=1}^{N-1} |\rond d_n|}{\beta_\epsilon^{\alpha_\epsilon} \Gamma (\alpha_\epsilon)\,
      \beta_0^{\alpha_0}\Gamma (\alpha_0)\, \beta_1^{\alpha_1} \Gamma (\alpha_1)} \\
\GN^{\alpha_\epsilon + N/2 - 1 } \GO^{\alpha_0 - 1/2} \G1^{\alpha_1 + (N-1)/2 - 1} 
\Exp{-\frac{\GN}{\beta_\epsilon} - \frac{\GO}{\beta_0}- \frac{\G1}{\beta_1}} \Uc_{\overline{\wb},\deltab}(\wb)\\
\Exp{- \frac{\GN}{2} \|\Data - \LH \Im\|^2 - \frac{\GO}{2}\|\Imzero\|^2 - \frac{\G1}{2}\|\LD \Im\|^2}. \label{eq:20}
\end{multline}
According to the Bayes rule, the \apost law reads
\begin{equation}
    p(\Im, \GN, \GO, \G1, \wb | \Data) =  
    \frac{p(\Im, \GN, \GO, \G1, \wb, \Data)}{p(\Data)} 
    \label{eq:21}
\end{equation}
where $p(\Data)$ is a normalization constant
\begin{equation}
    \label{eq:22}
    p(\Data) = \int p(\Data, \Im, \gammab, \wb)
    \,\dD \Im \, \dD \gammab \, \dD \wb.
\end{equation}

As described before, setting $\GO = 0$ leads to degenerated prior and
joint laws. However, when the observation system preserves the null
frequency $\GO$ can be considered as a nuisance parameter. In
addition, only prior information on the smoothness is available.

In Bayesian framework, a solution to eliminate the nuisance parameters
is to integrate them out in the \apost law.  According to our
parametrization Sec.~\ref{sec:image-aprio-law}, the integration of
$\GO$ is the integration of a Gamma law.  Application of
Appendix~\ref{sec:marginalisation} on $\GO$ in the \apost law
Eq.~(\ref{eq:21}) provides
\begin{equation}
    p(\Im, \GN, \G1, \wb | \Data) =  \frac{ p(\Imzero) 
      p(\Data, \Ime, \GN, \G1, \wb | \Imzero )}{ \displaystyle \int
      p(\Imzero) p(\Data, \Ime, \GN, \G1, \wb | \Imzero ) \,\dD \GN
      \,\dD \G1 \,\dD \wb \,\dD
      \Ime \,\dD \Imzero }\label{eq:23}
\end{equation}
with
\begin{align}
    p(\Imzero) & = \displaystyle \int p(\Imzero | \GO) p(\GO)
    \,\dD \GO \nonumber \\
    & = \left( 1 +
        \frac{\beta_0 \Imzero^2}{2} \right)^{-\alpha_0 -1/2}\label{eq:24}.
\end{align}
Now the parameter is integrated, the parameters $\alpha_0$ and
$\beta_0$ are set to remove the null frequency penalization. Since we
have $\alpha_0 > 0$ and $\beta_0 > 0$ we get $( 1 + \beta_0
\Imzero^2/2)^{-\alpha_0 -1/2} \leq 1$ and the joint law is majored
\begin{equation}
    \left( 1 + \frac{\beta_0 \Imzero^2}{2} \right)^{-\alpha_0 -1/2}
    p(\Data, \Ime, \GN, \G1, \wb |
    \Imzero ) \leq p(\Data, \Ime, \GN, \G1, \wb |
    \Imzero ).  \label{eq:25}
\end{equation}
Consequently, by the dominated convergence theorem~\cite{Lang93}, the
limit of the law with $\alpha_0 \rightarrow 1$ and $\beta_0
\rightarrow 0$ can be placed under the integral sign at the
denominator.  Then the null-frequency penalization $p(\Imzero)$ from
the numerator and denominator are removed. It is equivalent with the
integration of the $\GO$ parameter under a Dirac (see
appendix~\ref{sec:gamma-law}).  The equation is simplified and the
integration with respect to $\Imzero$ in the denominator
Eq.~(\ref{eq:22})
\begin{align}
    \label{eq:26}
    \int_{\ensbR} p(\Data | \Im, \GN, \wb)p(\Ime | \G1)p(\G1, \GN,
    \wb)\, \dD \Imzero & \propto \int_{\ensbR} p(\rond y_0 | \Imzero,
    \GN, \wb)\,\dD \Imzero \\
    \label{eq:27}
    & \propto \int_{\ensbR} \Exp{-\frac{\GN}{2} \left(  \Datazero  -
          \Hzero \Imzero  \right)^2} \,\dD \Imzero
\end{align}
converges if and only if $\Hzero \neq 0$: the null frequency is
observed. If this condition is met, Eq.~(\ref{eq:23}) with $\beta_0 =
0$ and $\alpha_0$ = 1 is a proper posterior law for the image, the
precision parameters and the PSF parameters. In other words, if the
average is observed, the degeneracy of the \aprio law is not
transmitted to the \apost law.

Then, the obtained \apost law writes
\begin{equation}
    \begin{split}
        p(\Im, \GN, \G1, \wb | \Data) & = \frac{p(\Im, \GN, \G1, \wb, \Data)}{p(\Data)}\\
        & \propto \GN^{\alpha_\epsilon + N/2 - 1 } \G1^{\alpha_1 + (N-1)/2 - 1}  \Uc_{\overline{\wb},\deltab}(\wb) \\
        & \qquad \Exp{- \frac{\GN}{2} \|\Data - \LH \Im\|^2 - \frac{\G1}{2}\|\LDe \Ime\|^2} 
              \Exp{-\frac{\GN}{\beta_\epsilon} - \frac{\G1}{\beta_1}}.
    \end{split} \label{eq:28}
\end{equation}
Finally, inference is done on this law Eq.~(\ref{eq:28}). If the null
frequency is not observed, or information must be added, the previous
Eq.~(\ref{eq:21}) can be used.

\section{Posterior mean estimator and law exploration}
\label{sec:mcmc-computation}

This section presents the algorithm to explore the posterior law
Eq.~(\ref{eq:21}) or (\ref{eq:28}) and to compute an estimate of the
parameters. For this purpose, Monte Carlo Markov chain is used to
provide samples. Firstly, the obtained samples are used to compute
different moments of the law. Afterwards, they are also used to
approximate marginal laws as histograms. These two representations are
helpful to analyse the \apost law, the structure of the available
information and the uncertainty. They are used in
Sec.~\ref{sec:instr-param-char} to illustrate the mark of the
ambiguity in the myopic problem.

Here, the samples of the \apost law are obtained by a Gibbs
sampler~\cite{Robert04,Bremaud99,Geman84}: it consists in iteratively
sampling the conditional posterior law for a set of parameters given
the other parameters (obtained at previous iteration). Typically, the
sampled laws are the law of $\Im$, $\gamma_i$ and $\wb$. After a
burn-in time, the complete set of samples are under the joint \apost
law. The three next sections present each sampling step.

\subsection{Sampling the image}
\label{sec:sampling-object}
The conditional posterior law of the image is a Gaussian law
\begin{align}
    \Im^{(k+1)} & \sim p \left(\Im | \Data, \GN^{(k)}, \GO^{(k)},
        \G1^{(k)}, \wb^{(k)}\right)\label{eq:29}\\
    \label{eq:30}
    & \sim \mathcal{N}\left(\mub^{(k+1)},
        \Sigmab^{(k+1)}\right).
\end{align}
The covariance matrix is diagonal and writes
\begin{equation}
    \label{eq:31}
    \Sigmab^{(k+1)} = \left( \GN^{(k)} |\LH^{(k)}|^2 + \GO^{(k)} |\LO|^2 +
        \G1^{(k)} |\LD|^2 \right)^{-1}
\end{equation}
and the mean
\begin{equation}
    \label{eq:32}
    \mub^{(k+1)} = \GN^{(k)} \Sigmab^{(k+1)} {\LH^\dag}^{(k)} \Data.
\end{equation}
where $\dag$ is the transpose conjugate symbol. The vector
$\mub^{(k+1)}$ is the regularized least square solution at the current
iteration (or the Wiener-Hunt filter). Clearly, if the null-frequency
is not observed $\rond h_0 = 0$ and if $\GO = 0$, the covariance
matrix $\Sigmab$ is not invertible and the estimate is not defined as
described Sec.~\ref{sec:proper-posterior-law}.

Finally, since the matrix is diagonal, the sample $\Im^{(k+1)}$ is
obtained by a term-wise product of $\Fb \epsilonb$ (where $\epsilonb$
is white Gaussian) with the standard deviation matrix
$\left(\Sigmab^{(k+1)}\right)^{1/2}$ followed by the addition of the
mean $\mub^{(k+1)}$ also computed with term-wise products
Eq.~(\ref{eq:32}). Consequently, the sampling of the image is
effective even with high-dimensional object.

\subsection{Sampling precision parameters}
\label{sec:sampl-inverse-vari}
The conditional posterior laws of the precisions are Gamma
corresponding to their prior law with parameters updated by the
likelihood
\begin{align}
    \gamma_i^{(k+1)} & \sim p \left(\gamma_i | \Data, \Im^{(k+1)},
        \wb^{(k)}\right) \label{eq:33}\\
    & \sim \mathcal{G}\left(\gamma_i | \alpha_i^{(k+1)},
        \beta_i^{(k+1)} \right). \label{eq:34}
\end{align}
For $\GN, \GO$ and $\G1$ the parameters law are, respectively,
\begin{IEEEeqnarray}{r;c;l"t"l;c;l}
    \label{eq:35}
    \alpha_\epsilon^{(k+1)} & = & \alpha_\epsilon + N/2 & and & \beta_\epsilon^{(k+1)} & =
    & \left(\beta_\epsilon^{-1} + \frac{1}{2}\|\Data - \LH^{(k)}
        \Im^{(k+1)}\|^2\right)^{-1},  \\
    \alpha_0^{(k+1)} & = & \alpha_0 + 1/2 & and & \beta_0^{(k+1)} & =
    & \left(\beta_0^{-1} + \frac{1}{2}\left(\Imzero^{(k+1)}\right)^2
    \right)^{-1},
    \label{eq:36}\\
    \alpha_1^{(k+1)} & = & \alpha_1 + (N-1)/2 & and & \beta_1^{(k+1)}
    & = & \left(\beta_1^{-1} + \frac{1}{2}\|\LD \Im^{(k+1)}\|^2
    \right)^{-1}.\label{eq:37}
\end{IEEEeqnarray}
In the case of Jeffreys' prior, the parameters are
\begin{IEEEeqnarray}{r;c;l"t"l;c;l}
    \alpha_\epsilon^{(k+1)} & = & N/2 & and & \beta_\epsilon^{(k+1)} & = & 2/\|\Data
    - \LH^{(k)} \Im^{(k+1)}\|^2,
    \label{eq:38} \\
    \alpha_0^{(k+1)} & = & 1/2 & and &
    \beta_0^{(k+1)} & = & 2/\left(\Imzero^{(k+1)}\right)^2, \label{eq:39} \\
    \alpha_1^{(k+1)} & = & (N-1)/2 & and & \beta_1^{(k+1)} & = &
    2/\|\LD \Im^{(k+1)}\|^2. \label{eq:40}
\end{IEEEeqnarray}

\begin{remark}~---~ If the \apost law Eq.~(\ref{eq:28}) without $\GO$
    is considered, there is no need to sample this parameter
    (Eq.~(\ref{eq:36}) and~(\ref{eq:39}) are not useful) and
    $\GO^{(k)} = 0$ in Eq.~(\ref{eq:31}).
\end{remark}

\subsection{Sample PSF parameters}
\label{sec:sample-instr-paramt}
The conditional law for PSF parameters writes
\begin{align}
    \wb^{(k+1)} & \sim p \left(\wb | \Data, \Im^{(k+1)}, \GN^{(k+1)}
    \right)\label{eq:41}\\
    \label{eq:42}
    & \propto \Exp{-\frac{\GN^{(k+1)}}{2} \|\Data -  \Lambdab_{\Hb,
        \wb}~ \Im^{(k+1)}\|^2}
\end{align}
where parameters $\wb$ are embedded in the PSF $\LH$.  This law is not
standard and intricate: no algorithm exists for direct sampling and we
use the Metropolis-Hastings (M.-H.) method to bypass this
difficulty. In M.-H. algorithm, a sample $\wb_{\pD}$ is proposed and
accepted with a certain probability.  This probability depends on the
ratio between the likelihood of the proposed value and the likelihood
of the current value $\wb^{(k)}$.  In practice, in the independent
form described in appendix \ref{sec:metr-hast-algor}, with prior law
as proposition law, it is divided in several steps.
\begin{enumerate}
\item \textsc{Proposition:} Sample a proposition
    \begin{equation}
        \label{eq:43}
        \wb_{\pD} \sim p(\wb)=\Uc_{[\ab~\bb]}(\wb).       
    \end{equation}
\item \textsc{Probability of acceptation:} Calculate the criterion
    \begin{equation}
        \label{eq:44}
        J\left(\wb^{(k)},\wb_{\pD}\right) = \frac{\GN^{(k+1)}}{2}
        \left( \| \Data - \Lambdab_{\Hb, \wb^{(k)}}~\Im^{(k+1)} \|^2 - \|\Data -
            \Lambdab_{\Hb, \wb_{\pD}}~\Im^{(k+1)}\|^2 \right).
    \end{equation}
\item \textsc{Update:} Sample $t \sim \Uc_{[0~1]}$
    and takes
    \begin{equation}
        \label{eq:45}
        \wb^{(k+1)} = \left\{
            \begin{array}{ll}
                \wb_{\pD} & \text{if~} \log t  < J \\
                \wb^{(k)} & \text{otherwise}.
            \end{array}\right.
    \end{equation}
\end{enumerate}

\subsection{Empirical mean}
\label{sec:empirical-mean}

The sampling of $\Im$, $\gammab$ and $\wb$ are repeated iteratively
until the law has been sufficiently explored. These samples $\left[
    \Im^{(k)}, \gammab^{(k)}, \wb^{(k)} \right]$ follow the global
\apost law of Eq.~(\ref{eq:21}). By the large numbers law, the
estimate, defined as the posterior mean, is approximated by
\begin{equation}
    \hat \xb = \F^\dag \mathds{E}[\Im] \approx \F^\dag \left[
        \frac{1}{K} \sum_{k = 0}^{K-1} \Im^{(k)} \right]. \label{eq:46}
\end{equation}
As described by Eq.~(\ref{eq:46}), to obtain an estimate of the image
in the spatial space, all the computation are achieved recursively in
the Fourier space with a single \ifft at the end.  An implementation
example in pseudo code is described~Fig.~\ref{fig:algo}.

\section{Deconvolution results}
\label{sec:deconv-results}

This section presents numerical results obtained by the proposed
method.  In order to completely evaluate the method, true value of all
parameters $\xb$, $\wb$, $\GN$ but also $\G1, \GO$ is needed. In order
to achieve this, an entirely simulated case is studied: image and
noise are simulated under their respective prior laws
Eq.~(\ref{eq:13}) and~(\ref{eq:14}) with given values of $\GO$, $\G1$
and $\GN$. Thanks to this protocol, all experimental conditions are
controlled and the estimation method is entirely evaluated.

The method has also been applied in different conditions (lower signal
to noise ratio, broader PSF, different and realistic (non-simulated)
images, \dots) and showed similar behaviour. However, in the case of
realistic images, since the true value of the hyperparameters $\GO$
and $\G1$ is unknown, the evaluation cannot be complete.

\subsection{Practical experimental conditions}
\label{sec:exper-cond}

Concretely, a $128\times 128$ image is generated in the Fourier space
as the product of a complex white Gaussian noise and the \aprio
standard deviation matrix $\Sigmab = (\GO \LO^\dag\LO + \G1
\LD^\dag\LD)^{-1/2}$, given by Eq.~(\ref{eq:11}). The chosen matrix
$\LD$ results from the \fftdd of the Laplacian operator $\left[0 ~
    1~0;1 -4~1;0~1~0\right]/8$ and the parameter values are $\GO = 1$
and $\G1 = 2$.

These parameters provide the image shown in Fig.~\ref{fig:True} : it
is an image with smooth features similar to a cloud.  Pixels have
numerical values between $-100$ and $150$, and the profile line 68
shows fluctuations around a value of $-40$.

The \aprio law for the hyperparameters are set to the non-informative
Jeffreys' law by fixing the $(\alpha_i, \beta_i)$ to $(0,+\infty)$, as
explained in Sec.~\ref{sec:hyperparam-aprio}. In addition, the PSF is
obtained in the Fourier space by discretization of a normalized
Gaussian shape
\begin{multline}
    \label{eq:47}
    \rond h(\nu_\alpha, \nu_\beta) = \exp \bigg ( - 2 \pi^2 \Big(
    \nu_\alpha^2 ( \wA \cos^2 \varphi +
    \wB \sin^2 \varphi )  \\
    + \nu_\beta^2 ( \wA \sin^2 \varphi + \wB \cos^2
    \varphi ) \\
    + 2 \nu_\alpha \nu_\beta \sin \varphi \cos \varphi \left(\wA - \wB
    \right) \Big) \bigg )
\end{multline}
with frequencies $(\nu_\alpha, \nu_\beta) \in \left[ -0.5 ; 0.5
\right]^2$. This low-pass filter, illustrated in Fig.~\ref{fig:RI}, is
controlled by three parameters:
\begin{itemize}

\item two width parameters $\wA$ and $\wB$ set to 20 and 7,
    respectively. Their \aprio laws are uniform: $p(\wA) =
    \Uc_{[19~21]}(\wA)$ and $p(\wB) = \Uc_{[6~8]}(\wA)$ corresponding
    to an uncertainty of about~5\% and 15\% around the nominal value
    (see Sec~\ref{sec:instr-param-law}).

\item a rotation parameter $\varphi$ set to $\pi/3$. The \aprio law is
    also uniform $p(\varphi) = \Uc_{[\pi/4~\pi/2]}(\varphi)$
    corresponding to~50\% uncertainty.

\end{itemize}

Then, the convolution is computed in the Fourier space and the data
are obtained by adding a white Gaussian noise with precision $\GN =
0.5$. Data are shown Fig.~\ref{fig:Data}: they are naturally smoother
than the true image and the small fluctuations are less visible and
corrupted by the noise. The empirical mean level of the image is
correctly observed (the null frequency coefficient of $\H_\wb$ is
$\rond h_0 = 1$) so the parameter $\GO$ is considered as a nuisance
parameter. Consequently it is integrated out under a Dirac (see
Sec.~\ref{sec:proper-posterior-law}). This is equivalent to fix its
value to 0 in the algorithm Fig.~\ref{fig:algo}, line 4.

Finally, the method is evaluated on two different situations.
\begin{enumerate}

\item The unsupervised and non-myopic case: the parameters $\wb$ are
    known. Consequently, there is no Metropolis-Hastings step
    (Sec.~\ref{sec:sample-instr-paramt}): lines 9 to 16 are ignored in
    the algorithm of Fig.~\ref{fig:algo} and $\wb$ is set to its true
    value. To obtain sufficient law exploration, the algorithm is run
    until the difference between two successive empirical means is
    less than $10^{-3}$. In this case, 921 samples are necessary and
    they are computed in approximately 12 seconds on a processor at
    2.66 GHz with Matlab,

\item The unsupervised and myopic case: all the parameters are
    estimated. To obtain sufficient law exploration, the algorithm is
    run until the difference between two successive empirical means is
    less than $5 \times 10^{-5}$. In this case, 18~715 samples are
    needed and they are computed in approximately 7 minutes.

\end{enumerate}

\begin{remark}~---~ The algorithm has also been run for up to
    1\,000\,000 samples, in both cases, without perceptible
    qualitative changes.
\end{remark}

\subsection{Estimation results}
\label{sec:estimation-results}

\subsubsection{Images}

The two results for the image are given Figs.~\ref{fig:Xeap} and
\ref{fig:XeapMyope} for the non-myopic and the myopic cases,
respectively.

The effect of deconvolution is notable on the image, as well as on the
shown profile. The object is correctly positioned, the orders of
magnitude are respected and the mean level is correctly
reconstructed. The image is restored, more details are visible and the
profiles are closer matching to the true image than data. More
precisely, the pixels 20-25 of the 68-th line in
Fig.~\ref{fig:Results} show the restoration of the original dynamic
whereas it is not visible in the data. Between pixels 70 and 110,
fluctuations not visible in data are also correctly restored.

In order to visualize and study the spectral contents of the images,
circular average of empirical power spectral density is considered and
called ``spectrum'' hereafter. The subjacent spectral variable is a
radial frequency $f$ such as $f^2=\nu_\alpha^2+\nu_\beta^2$. The
spectrum of the true object, data and restored object are shown
Figs.~\ref{fig:psdNonMyopic} and~\ref{fig:psdMyopic} in non-myopic and
myopic cases, respectively. It is clear that the spectrum of the true
image is correctly retrieved, in both cases, up to the radial
frequency $f \approx 0.075$. Above this frequency, noise is clearly
dominant and information about the image is almost lost. In other
words, the method produces correct spectral equalization in the
properly observed frequency band. The result is expected from a
Wiener-Hunt method but the achievement is the joint estimation of
hyperparameter and instrument parameters in addition to the correct
spectral equalization.

Concerning a comparison between non-myopic and myopic cases, there is
no visual differences. The spectrum Figs.~\ref{fig:psdNonMyopic}
and~\ref{fig:psdMyopic} in non-myopic and myopic cases respectively
are visually indistinguishable. This is also the case when comparing
Figs.~\ref{fig:Xeap} and \ref{fig:XeapMyope} and especially 68-th
line. From a more precise quantitative evaluation, a slight difference
is observed and detailed below.

In order to quantify performances, a normalized euclidean distance
\begin{equation}\label{eq:48}
    e = \| \xb  - \xb^{*}\|/\|\xb^{*}\|
\end{equation}
between an image $\xb$ and the true image $\xb^{*}$ is considered. It
is computed between true image and estimate images as well as between
true image and data. Results are reported in Tab.~\ref{tab:VarIm} and
confirm that the deconvolution is effective with an error of
approximately 6 \% in myopic case compared to 11 \% with data. Both
non-myopic and myopic deconvolution reduce error by a factor 1.7 with
respect to the observed data.

Regarding a comparison between non-myopic and myopic case, the errors
are almost the same, with a slightly lower value for the non-myopic
case, as expected. This difference is coherent with the intuition:
more information are injected in the non-myopic case through the true
PSF parameters values.

\subsubsection{Hyperparameters and instrument parameters}

Concerning the other parameters, their estimates are close to the true
values and are reported in Tab.~\ref{tab:ResultPrior}. The $\GN$
estimate is very close to the true value with $\hGN = 0.49$ instead of
0.5 in the two cases.  The error for the PSF parameters are 0.35\%,
2.7\% and 1.9\% for $\wA$, $\wB$ and $\phi$, respectively.  The value
of $\G1$ is underestimated in the two cases with approximately 1.7
instead of 2.  All the true values fall in the $\hat \mu \pm 3 \hat
\sigma$ interval.

In order to deepen the numerical study, the paper evaluates the
capability of the method to accurately select the best values for
hyperparameters and instrument parameters. To this end, we compute the
estimation error Eq. (\ref{eq:48}) for a set of ``exhaustive'' values
of the parameters $[\GN, \G1, \wA, \wB, \phi]$. The protocol is the
following: 1) choose a new value for a parameter ($\GN$ for example)
and fix the other parameters to the value provided by our algorithm,
2) compute the Wiener-Hunt solution (Sec.~\ref{sec:sampling-object})
and 3) compute the error index.

Results are reported in Fig.~\ref{fig:bestparam}. In each case, smooth
variation of error is observed when varying hyperparameters and
instrument parameters and an unique optimum is visible. By this way,
one can find the value of the parameters that provide the best
Wiener-Hunt solution when the true image $\xb^\star$ is known. It is
reported on Tab.~\ref{tab:VarIm} and shows almost imperceptible
improvement: optimization of the parameters (based on the true image
$\xb^\star$) allow negligible improvement (smaller than 0.02 \% as
reported in Tab. \ref{tab:VarIm}).

So, the main conclusion is that, the unsupervised and myopic proposed
approach is a relevant tool in order to tune parameters: it works
(without the knowledge of the true image), as well as an optimal
approach (based on the knowledge of the true image).

\subsection{\Apost law characteristics}
\label{sec:apost-law-estimator}

This section describes the \apost law using histograms, means and
variances of the parameters. The sample histograms,
Figs.~\ref{fig:hyperparam} and \ref{fig:intrumentParam}, provide an
approximation of the marginal posterior law for each parameter.
Tabs.~\ref{tab:VarIm} and \ref{tab:ResultPrior} report the variance
for the image and law parameters respectively and thus allow to
quantify the uncertainty.

\subsubsection{Hyperparameter characteristics}
\label{sec:hyperp-char}

The histograms for $\GN$ and $\G1$, Fig.~\ref{fig:hyperparam}, are
concentrated around a mean value in both non-myopic and myopic
cases. The variance for $\GN$ is lower than the one for $\G1$ and it
can be explained as follows.

The observed data are directly impacted by noise (present at the
system output) whereas they are indirectly impacted by the object
(present at the system input). The convolution system damages the
object and not the noise: as a consequence, the parameter $\GN$ (that
drives noise law) is more reliably estimated than $\G1$ (that drives
object law).

A second observation is the smaller variance for $\G1$ in the
non-myopic case Fig.~\ref{fig:G1Gibbs} than in the myopic case
Fig.~\ref{fig:G1Myope}. It is the consequence of the addition of
information in the non-myopic case w.r.t. the myopic one, through the
value of the PSF parameters. In the myopic case, the estimates are
founded on the knowledge of an interval for the values of the
instrument parameters, whereas in the non-myopic case, the estimates
are founded on the true values for the instrument parameters.

\subsubsection{PSF parameter characteristics}
\label{sec:instr-param-char}

Fig.~\ref{fig:intrumentParam} gives histograms for the three PSF
parameters and their appearances are quite different from the one for
hyperparameters. The histograms for $\wA$ and $\wB$,
Figs.~\ref{fig:Wa} and~\ref{fig:Wb} are not as concentrated as the one
of Fig.~\ref{fig:hyperparam} for hyperparameters. Their variances are
quite large with regards to the interval of the prior law. On the
contrary, the histogram for the parameter $\varphi$,
Fig.~\ref{fig:Phi}, has the smallest variance. It is analyzed as a
consequence of a larger sensitivity of the data w.r.t. the parameter
$\varphi$ than w.r.t. the parameters $\wA$ and $\wB$. In an equivalent
manner, the observed data are more informative about the parameter
$\varphi$ than about the parameters $\wA$ and $\wB$.

\subsubsection{Mark of the myopic ambiguity}
\label{sec:myopic-ambiguity}

Finally, a correlation between parameters $(\G1,\wA)$ and $(\G1, \wB)$
is visible on their joint histograms Fig.~\ref{fig:JointHist}. It can
be interpreted as a consequence of the ambiguity in the primitive
myopic deconvolution problem, in the following manner: the parameters
$\G1$ and $\wb$ both participate in the interpretation of the spectral
content of data, $\G1$ as a scale factor and $\wb$ as a shape factor.
An increase of $\wA$ or $\wB$ results in a decrease of the cutoff
frequency of the observation system. In order to explain the spectral
content of a given data set, the spectrum of the original image must
contain more high frequencies, \ie~a smaller $\G1$.  This is also
observed on the histogram illustrated Fig.~\ref{fig:JointG1Wa}.

\subsection{MCMC algorithm characteristics}
\label{sec:algorithm-characteristic}

Globally, the chains of Figs.~\ref{fig:hyperparam} and
\ref{fig:intrumentParam}, have a Markov feature (correlated) and
explore the parameter space. They have a burn-in period followed by a
stationary state. This characteristic has always been observed
regardless the initialization. For fixed experimental conditions, the
stationary state of multiple runs was always around the same
value. Considering different initializations, the only visible change
is on the length of the burn-in period.

More precisely, the chain of $\GN$ is concentrated in a small
interval, the burn-in period is very short (less than 10 samples) and
its evolution seems independent of the other parameters.  The chain of
$\G1$ has a larger exploration, the burn-in period is longer
(approximately 200 samples) and the histogram is larger. This is in
accordance with the analysis of Section~\ref{sec:hyperp-char}.

About the PSF parameters, the behaviour is different for $(\wA, \wB)$
and $\varphi$. The chain of the two width parameters has a very good
exploration with quasi-instantaneous burn-in period. Conversely, the
chain of $\varphi$ is more concentrated and its burn-in period is
approximately 4~000 samples.  This is also in accordance with previous
analysis (Section~\ref{sec:instr-param-char}).

Acceptation rates in the Metropolis-Hastings algorithm are reported in
Tab.~\ref{tab:ResultAcceptRate}: they are quite small, especially for
the rotation parameter. This is due to the structure of the
implemented algorithm: an independant Metropolis-Hastings algorithm
with the prior law as a proposition law. The main advantage of this
choice is its simplicity but as a counterpart, a high rejection rate
is observed due to a large \aprio interval for the angle parameter. A
future work will be devoted to the design of more accurate proposition
law.

\subsection{Robustness of prior image model}
\label{sec:robustn-prior-inform}

Fig.~\ref{fig:lena} illustrates the proposed method on a more
realistic image with heterogeneous spatial structures. The original is
the Lena image and the data has been obtained with the same Gaussian
PSF and also corruption by white Gaussian noise. The
Fig.~\ref{fig:eapLena} shows that the restored image is closer to the
true one than the data. Smaller structures are visible and edges are
sharper, for example around pixel $200$. The estimated parameters are
$\widehat{\GN} =1.98$ while the true value is $\GN^\star =
2$. Concerning the PSF parameters, the results are $\widehat{\wA} =
19.3$, $\widehat{\wB} = 7.5$ and $\widehat{\phi} = 1.15$ while the
true values are respectively $\wA^\star = 20$, $\wB^\star = 7$ and
$\phi^\star = 1.05$ as in the previous section. Here again, the
estimated PSF parameters are close to the true values giving a first
assessment of the capability of the method in a more realistic
context.

\section{Conclusion and perspectives}
\label{sec:conclusion}

This paper presents a new global and coherent method for myopic and
unsupervised deconvolution of relatively smooth images. It is built
within a Bayesian framework and a proper extended \apost law for the
PSF parameters, the hyperparameters and the image. The estimate,
defined as the posterior mean, is computed by means of an MCMC
algorithm in less than a few minutes.

Numerical assessment testifies that the parameters of the PSF and the
parameters of the prior laws are precisely estimated. In addition,
results also demonstrate that the myopic and unsupervised deconvolved
image is closer to the true image than the data and show true restored
high-frequencies as well as spatial details.

The paper focuses on linear invariant model often encountered in
astronomy, medical imaging, nondestructive testing and especially in
optical problems. Non-invariant linear models can also be considered in
order to address other applications such as
spectrometry~\cite{Rodet08} or fluorescence
microscopy~\cite{zhang06}. The loss of invariance property precludes
entirely Fourier-based computations but the methodology remains valid
and practicable. In particular, it is possible to draw samples of the
image by means of an optimization algorithm \cite{Orieux10a}.

Gaussian law, related to $\mbox{L}_2$ penalization, is known for
possible excessive sharp edges penalization in the restored
object. The use of convex $\mbox{L}_2-\mbox{L}_1$
penalization~\cite{Kunsch94,Charbonnier97,Giovannelli08} or non convex
$\mbox{L}_2-\mbox{L}_0$ penalization~\cite{Geman95} can overcome this
limitation. In these cases a difficulty occurs in the development of
myopic and unsupervised deconvolution: the partition function of the
prior law for the image is in intricate or even unknown dependency
\wrt the parameters~\cite{Idier08,Descombes99a,Jalobeanu02}. However a
recent paper~\cite{Giovannelli08} overcome the difficulty resulting in
an efficient unsupervised deconvolution and we plan to extend this
work for the myopic case.

Regarding noise, Gaussian likelihood limits robustness to outliers or
aberrant data and it is possible to appeal to robust law such as Huber
penalization in order to bypass the limitation. Nevertheless, the
partition function for the noise law is again difficult or impossible
to manage and it is possible to resort to the idea proposed
in~\cite{Giovannelli08} to overcome the difficulty.

Finally, estimation of parameters of correlation matrix (cutoff
frequency, attenuation coefficients,\dots) is possible within the same
methodological framework. This could be achieved for the correlation
matrix of the object or the noise. As for the PSF parameters, the
approach could rely on an extended \apost law, including the new
parameters and a Metropolis-Hastings sampler.

\section{Acknowledgment}
\label{sec:acknowledgment}

The authors would like to thank Professor Alain Abergel in IAS
laboratory at Universit\'e Paris-Sud 11, France, for fruitful
discussions and constructive suggestions. The authors are also
grateful to Cornelia Vacar, IMS laboratory, for carefully reading the
paper.

\appendix

\section{Law in Fourier space}
\label{sec:law-fourier-space}

For a Gaussian vector $\xb \sim \mathcal{N} (\mub, \Sigmab)$, the law
for $\Im = \F \xb$ (the \fft of $\xb$) is also Gaussian whose first
two moments are the following:
\begin{itemize}
\item The mean is
    \begin{equation}
        \label{eq:49}
         \rond \mub = \mathds{E}[\rond \xb] = \Fb \mathds{E}[\rond \xb] = \Fb \mub.
    \end{equation}
\item The covariance matrix is
    \begin{equation}
        \rond \Sigmab = \mathds{E}[(\rond \xb - \rond \mub)(\rond \xb
        - \rond \mub)^\dag] = \Fb \Sigmab \Fb^\dag.
        \label{eq:50}
    \end{equation}
\end{itemize}
Moreover, if the covariance matrix $\Sigmab$ is circulant it writes
\begin{equation}
    \label{eq:51}
    \rond \Sigmab = \Fb \Sigmab \Fb^\dag = \Lambdab_\Sigmab.
\end{equation}
\ie the covariance matrix $\rond \Sigmab$ is diagonal.

\section{The Gamma probability density}
\label{sec:gamma-law}

\subsection{Definition}
\label{sec:definition}

The Gamma pdf for $\gamma > 0$, with given parameter $\alpha > 0$ and
$\beta > 0$, is written
\begin{equation}
    \label{eq:GammaPDF}
    \mathcal{G}(\gamma|\alpha, \beta) = \frac{1}{\beta^\alpha
      \Gamma(\alpha)}\gamma^{\alpha - 1}\exp \left( - \gamma / \beta
    \right).
\end{equation}
Tab.~\ref{tab:limitelaw} gives three limit cases for $(\alpha,
\beta)$. The following properties hold:
\begin{itemize}
\item The mean is $\mathds{E}_{\mathcal{G}} [\gamma] = \alpha \beta$
\item The variance is $\mathds{V}_{\mathcal{G}} [\gamma] = \alpha
    \beta^2$
\item The maximiser is $\beta (\alpha - 1)$ if and only if $\alpha >
    1$
\end{itemize}

\subsection{Marginalisation}
\label{sec:marginalisation}

First consider a $N$ dimensional zero-mean Gaussian vector with a
given precision matrix $\gamma \Gammab$ with $\gamma > 0$. The pdf
reads
\begin{equation}
    \label{eq:GaussPDF2}
    p( \xb | \gamma ) = (2 \pi)^{-N/2} \gamma^{N/2} \det[ \Gammab
    ]^{1/2} \, \Exp{ - \gamma \xb^t \Gammab \xb \,/2} \,.
\end{equation}
So consider the conjugate pdf for $\gamma$ as a Gamma law with
parameter $(\alpha,\beta)$ (see previous Annex). The joint law for
$(\xb , \gamma)$ is the product of the pdf given by
Eq.~(\ref{eq:GammaPDF}) and Eq.~(\ref{eq:GaussPDF2}): $p(\xb,\gamma) =
p(\xb | \gamma)p(\gamma)$. The marginalization of the joint law is
known~\cite{Box72}:
\begin{align}
    \begin{split}
        p(\xb) & = \int_{\ensbR_{+}} p(\xb | \gamma)p(\gamma)\,\dD
        \gamma \\
        & = \frac{\beta^{N/2}\det[\Gammab]^{1/2} \Gamma \left( \alpha +
              N/2 \right)}{(2 \pi)^{N/2} \Gamma(\alpha)} \left( 1 +
            \frac{\beta \xb^t \Gammab \xb}{2}
        \right)^{-\alpha - N/2} 
    \end{split}
    \label{eq:53}
\end{align}
which is a $N$ dimensional \textit{t}-Student law of $2\alpha$ degrees
of freedom with a $\beta \Gamma$ precision matrix. Finally, the
conditional law reads:
\begin{equation}
    p( \gamma | \xb) = \frac{(2 \pi)^{-N/2} \det[ \Gammab
      ]^{1/2}}{\beta^\alpha \Gamma(\alpha)} ~~
    \gamma^{\alpha+N/2-1} \, \Exp{ - \gamma \pth{\xb^{t}\Gammab\xb \,/2 +
        1/\beta}} \,.
\end{equation}
Thanks to conjugacy, it is also a Gamma pdf with parameters
$\bar{\alpha}\,,\,\bar{\beta}$ given by $\bar{\alpha}=\alpha+N/2$ and
$\bar{\beta}^{-1}=\beta^{-1} + 2 /(\xb^t\Gammab\xb)$.

\section{The Metropolis-Hastings algorithm}
\label{sec:metr-hast-algor}

The Metropolis-Hastings algorithm provides samples of a target law
$f(\wb)$ that cannot be directly sampled but can be evaluated, at
least up to a multiplicative constant. Using the so called
``instrument law'' $q\left(\wb_\pD | \wb^{(t)}\right)$, samples of the
target law are obtained by the following iterations.
\begin{enumerate}
\item Sample a proposition $\wb_\pD \sim q\left(\wb_\pD |
        \wb^{(t)}\right)$.
\item Compute the probability
    \begin{equation}
        \label{eq:54}
        \rho = \min \left\{ \frac{f\left(\wb_\pD\right)}{f\left(\wb^{(t)}\right)}
            \frac{q\left(\wb^{(t)}|\wb_\pD\right)}{q\left(\wb_\pD | \wb^{(t)}\right)}, 1
        \right\}.
    \end{equation}
\item Take
    \begin{equation}
        \label{eq:55}
        \wb^{(t+1)} = \left\{
            \begin{array}{cl}
                \wb_\pD & \text{with~} \rho \text{~probability~} \\
                \wb^{(t)} & \text{with~} 1 - \rho
                \text{~probability~}.
            \end{array}\right.
    \end{equation}
\end{enumerate}
At convergence, the samples follow the target law
$f(\wb)$~\cite{Robert04,Bremaud99}. When $q\left(\wb_\pD|
    \wb^{(t)}\right) = q(\wb_\pD)$ the algorithm is named independent
Metropolis-Hastings. In addition, if the instrument law is uniform,
the acceptance probability gets simpler in
\begin{equation}
    \label{eq:57}
    \rho = \min \left\{ \frac{f\left(\wb_\pD\right)}{f\left(\wb^{(t)}\right)}, 1
    \right\}.
\end{equation}


\bibliographystyle{osajnl}

\begin{thebibliography}{10}
\newcommand{\enquote}[1]{``#1''}

\bibitem{Idier08}
J.~Idier, ed., \emph{Bayesian Approach to Inverse Problems} (ISTE Ltd and John
  Wiley \& Sons Inc., London, 2008).

\bibitem{Molina06}
R.~Molina, J.~Mateos, and A.~K. Katsaggelos, \enquote{{B}lind deconvolution
  using a variational approach to parameter, image, and blur estimation,}
  \uppercase{ieee} {T}rans. {I}mage {P}rocessing \textbf{15}, 3715--3727
  (2006).

\bibitem{Campisi07}
P.~Campisi and K.~Egiazarian, eds., \emph{Blind Image Deconvolution} (CRC
  Press, 2007).

\bibitem{Rodet08}
T.~Rodet, F.~Orieux, J.-F. Giovannelli, and A.~Abergel, \enquote{Data inversion
  for over-resolved spectral imaging in astronomy,} \uppercase{ieee} {J}. of
  {S}elec. {T}opics in {S}ignal {P}roc. \textbf{2}, 802--811 (2008).

\bibitem{Tikhonov77}
A.~Tikhonov and V.~Arsenin, \emph{Solutions of Ill-Posed Problems} (Winston,
  Washington, \sca{dc}, 1977).

\bibitem{Twomey62}
S.~Twomey, \enquote{On the numerical solution of {F}redholm integral equations
  of the first kind by the inversion of the linear system produced by
  quadrature,} J. Assoc. Comp. Mach. \textbf{10}, 97--101 (1962).

\bibitem{Jalobeanu02}
A.~Jalobeanu, L.~Blanc-F\'eraud, and J.~Zerubia, \enquote{Hyperparameter
  estimation for satellite image restoration by a {MCMC} maximum likelihood
  method,} Pattern Recognition \textbf{35}, 341--352 (2002).

\bibitem{OSullivan95}
J.~A. O'Sullivan, \enquote{Roughness penalties on finite domains,}
  \uppercase{ieee} {T}rans. {I}mage {P}rocessing \textbf{4}, 1258--1268 (1995).

\bibitem{Demoment89}
G.~Demoment, \enquote{Image reconstruction and restoration: Overview of common
  estimation structure and problems,} \uppercase{ieee} {T}rans. {A}coust.
  {S}peech, {S}ignal {P}rocessing \textbf{\sca{assp}-37}, 2024--2036 (1989).

\bibitem{pankajakshan09}
P.~Pankajakshani, B.~Zhang, L.~Blanc-F\'{e}raud, Z.~Kam, J.-C. Olivo-Marin, and
  J.~Zerubia, \enquote{Blind deconvolution for thin-layered confocal imaging,}
  Appl. Opt. \textbf{48}, 4437--4448 (2009).

\bibitem{Thiebaut95}
E.~Thiébaut and J.-M. Conan, \enquote{Strict a priori constraints for maximum
  likelihood blind deconvolution,} \josaa \textbf{12}, 485--492 (1995).

\bibitem{dobigeon09}
N.~Dobigeon, A.~Hero, and J.-Y. Tourneret, \enquote{Hierarchical bayesian
  sparse image reconstruction with application to {MRFM},} \uppercase{ieee}
  {T}rans. {I}mage {P}rocessing  (2009).

\bibitem{zhang06}
B.~Zhang, J.~Zerubia, and J.-C. Olivo-Marin, \enquote{Gaussian approximations
  of fluorescence microscope point-spread function models,} Appl. Opt.
  \textbf{46}, 1819--1829 (2007).

\bibitem{Mugnier04}
L.~Mugnier, T.~Fusco, and J.-M. Conan, \enquote{{MISTRAL}: a myopic
  edge-preserving image restoration method, with application to astronomical
  adaptive-optics-corrected long-exposure images,} {J}. {O}pt. {S}oc. {A}mer.
  \textbf{21}, 1841--1854 (2004).

\bibitem{Thiebaut08}
E.~Thi\'ebaut, \enquote{{MiRA}: an effective imaging algorithm for optical
  interferometry,} in \enquote{proc. {SPIE}: Astronomical Telescopes and
  Instrumentation,} , vol. 7013 (2008), vol. 7013, pp. 70131--I.

\bibitem{Fusco99}
T.~Fusco, J.-P.~V. ran, J.-M. Conan, and L.~M. Mugnier, \enquote{Myopic
  deconvolution method for adaptive optics images of stellar fields,} Astron.
  Astrophys. Suppl. Ser. \textbf{134}, 193 (1999).

\bibitem{Conan98a}
J.-M. Conan, L.~Mugnier, T.~Fusco, V.~Michau, and R.~G., \enquote{Myopic
  deconvolution of adaptive optics images by use of object and point-spread
  function power spectra,} Applied Optics \textbf{37}, 4614--4622 (1998).

\bibitem{Likas04}
A.~C. Likas and N.~P. Galatsanos, \enquote{A variational approach for
  {B}ayesian blind image deconvolution,} \uppercase{ieee} {T}rans. {I}mage
  {P}rocessing \textbf{52}, 2222--2233 (2004).

\bibitem{bishop08}
T.~Bishop, R.~Molina, and J.~Hopgood, \enquote{Blind restoration of blurred
  photographs via {AR} modelling and {MCMC},} in \enquote{Image Processing,
  2008. ICIP 2008. 15th IEEE Int. Conference on,}  (2008).

\bibitem{Lam00}
E.~Y. Lam and J.~W. Goodman, \enquote{Iterative statistical approach to blind
  image deconvolution,} J. Opt. Soc. Am. A \textbf{17}, 1177--1184 (2000).

\bibitem{Xu09}
Z.~Xu and E.~Y. Lam, \enquote{Maximum a posteriori blind image deconvolution
  with {H}uber--{M}arkov random-field regularization,} Opt. Lett. \textbf{34},
  1453--1455 (2009).

\bibitem{Cannon76}
M.~Cannon, \enquote{Blind deconvolution of spatially invariant image blurs with
  phase,} \uppercase{ieee} {T}rans. {A}coust. {S}peech, {S}ignal {P}rocessing
  \textbf{24}, 58--63 (1976).

\bibitem{Jalobeanu02a}
A.~Jalobeanu, L.~Blanc-Feraud, and J.~Zerubia, \enquote{Estimation of blur and
  noise parameters in remote sensing,} in \enquote{{P}roc. \uppercase{ieee}
  \uppercase{icassp},} , vol.~4 (2002), vol.~4, pp. 3580--3583.

\bibitem{chen09}
F.~Chen and J.~Ma, \enquote{An empirical identification method of {G}aussian
  blur parameter for image deblurring,} Signal Processing, {IEEE} Trans. on
  (2009).

\bibitem{Robert04}
C.~P. Robert and G.~Casella, \emph{{M}onte-{C}arlo Statistical Methods},
  Springer Texts in Statistics (Springer, New York, \sca{ny}, 2000).

\bibitem{Hunt71}
B.~R. Hunt, \enquote{A matrix theory proof of the discrete convolution
  theorem,} \uppercase{ieee} {T}rans. {A}utomat. {C}ontr. \textbf{AC-19},
  285--288 (1971).

\bibitem{Calder97}
M.~Calder and R.~A. Davis, \enquote{Introduction to {W}hittle (1953) `{T}he
  analysis of multiple stationary time series',} Breakthroughs in Statistics
  \textbf{3}, 141--148 (1997).

\bibitem{Brockwell91}
P.~J. Brockwell and R.~A. Davis, \emph{Time Series: Theory and Methods}
  (Springer-Verlag, New York, 1991).

\bibitem{Hunt72b}
B.~R. Hunt, \enquote{Deconvolution of linear systems by constrained regression
  and its relationship to the {W}iener theory,} \uppercase{ieee} {T}rans.
  {A}utomat. {C}ontr. \textbf{AC-17}, 703--705 (1972).

\bibitem{Mardia92}
K.~Mardia, J.~Kent, and J.~Bibby, \emph{Multivariate Analysis} (San Diego :
  Academic Press, 1992), chap.~2, pp. 36--43.

\bibitem{Bouman93}
C.~A. Bouman and K.~D. Sauer, \enquote{A generalized {G}aussian image model for
  edge-preserving \sca{map} estimation,} \uppercase{ieee} {T}rans. {I}mage
  {P}rocessing \textbf{2}, 296--310 (1993).

\bibitem{MacKay03}
D.~MacKay, \emph{Information Theory, Inference, and Learning Algorithms}
  (Cambridge University Press, 2003).

\bibitem{Kass96}
R.~E. Kass and L.~Wasserman, \enquote{The selection of prior distributions by
  formal rules,} {J}. {A}mer. {S}tatist. {A}ssoc. \textbf{91}, 1343--1370
  (1996).

\bibitem{Jaynes03}
E.~T. Jaynes, \emph{Probability Theory: The Logic of Science} (Cambridge
  University Press, 2003).

\bibitem{Lang93}
S.~Lang, \emph{Real and functional analysis} (Springer, 1993).

\bibitem{Bremaud99}
P.~Br{\'e}maud, \emph{{M}arkov Chains. {G}ibbs fields, Monte Carlo Simulation,
  and Queues}, Texts in Applied Mathematics~31 (Spinger, New York, \sca{ny},
  1999).

\bibitem{Geman84}
S.~Geman and D.~Geman, \enquote{Stochastic relaxation, {G}ibbs distributions,
  and the {B}ayesian restoration of images,} \uppercase{ieee} {T}rans.
  {P}attern {A}nal. {M}ach. {I}ntell. \textbf{6}, 721--741 (1984).

\bibitem{Orieux10a}
F.~Orieux, O.~F\'eron, and J.-F. Giovannelli, \enquote{Stochastic sampling of
  large dimension non-stationnary gaussian field for image restoration,}
  Submitted to ICIP2010.

\bibitem{Kunsch94}
H.~R. K{\"u}nsch, \enquote{Robust priors for smoothing and image restoration,}
  {A}nn. {I}nst. {S}tat. {M}ath. \textbf{46}, 1--19 (1994).

\bibitem{Charbonnier97}
P.~Charbonnier, L.~Blanc-F\'eraud, G.~Aubert, and M.~Barlaud,
  \enquote{De{\-}ter{\-}ministic edge-preserving regularization in computed
  imaging,} \uppercase{ieee} {T}rans. {I}mage {P}rocessing \textbf{6}, 298--311
  (1997).

\bibitem{Giovannelli08}
J.-F. Giovannelli, \enquote{Unsupervised {B}ayesian convex deconvolution based
  on a field with an explicit partition function,} \uppercase{ieee} {T}rans.
  {I}mage {P}rocessing \textbf{17}, 16--26 (2008).

\bibitem{Geman95}
D.~Geman and C.~Yang, \enquote{Nonlinear image recovery with half-quadratic
  regularization,} \uppercase{ieee} {T}rans. {I}mage {P}rocessing \textbf{4},
  932--946 (1995).

\bibitem{Descombes99a}
X.~Descombes, R.~Morris, J.~Zerubia, and M.~Berthod, \enquote{Estimation of
  {M}arkov random field prior parameters using {M}arkov chain {M}onte {C}arlo
  maximum likelihood,} \uppercase{ieee} {T}rans. {I}mage {P}rocessing
  \textbf{8}, 954--963 (1999).

\bibitem{Box72}
G.~E.~P. Box and G.~C. Tiao, \emph{{B}ayesian inference in statistical
  analysis} (Addison-Wesley publishing, 1972).

\end{thebibliography}

\newpage{}

\begin{table}[H]
    \caption{Error $e$ (Eq. (\ref{eq:48})) and averaged standard deviation $\hat
      \sigma$ of the posterior image law. The ``Best'' error has been
      obtained with the knowledge of the true image.}  \label{tab:VarIm}
    \centering
    \begin{tabular}{l|c|c|c|c}
        \hline
        & Data & Non-myopic & Myopic & Best \\
        \hline
        Error ($e$) & 11.092 \% &  6.241 \% & 6.253 \% & 6.235 \%\\
        $\hat \sigma$ of $\xb$ law & - & 3.16 & 3.25 & - \\
        \hline
    \end{tabular}
\end{table}

\newpage{}

\begin{table}[H]
  \caption{Quantitative evaluation: true and estimated values of
    hyperparameters and PSF parameters.}
    \label{tab:ResultPrior}
    \centering
    \begin{tabular}{l|l|c|c|c|c|c}
    \hline
    & & $\widehat\GN \pm \hat\sigma$ & $\widehat\G1 \pm \hat\sigma$ & $\widehat\wA \pm \hat\sigma$ & $\widehat\wB \pm \hat\sigma$ & $\widehat\varphi \pm \hat\sigma$\\
    \hline
    & \textbf{True value} & \textbf{0.5} & \textbf{2} & \textbf{20} & \textbf{7} & \textbf{1.05} ($\pi/3$)  \\
    \hline
    \textbf{Non-myopic} & Estimate & 0.49 $\pm 0.0056$ & 1.78 $\pm 0.14$ & - & - & -   \\
    & Error & 2.0 \%  & 11 \% & - & - & -   \\
    \hline
    \textbf{Myopic} & Estimate & 0.49 $\pm 0.0056$ & 1.65 $\pm 0.15$ & 20.07 $\pm 0.53$ & 7.19 $\pm 0.38$ & 1.03 $\pm 0.04$  \\
    & Error & 2.0 \% &  18 \%&  0.35 \% &  2.7 \%&  1.9 \% \\
    \hline
    \end{tabular}
\end{table}

\newpage{}

\begin{table}[H]
    \caption{Acceptation rate.}
    \label{tab:ResultAcceptRate}
    \centering
    \begin{tabular}{l|c|c|c}
    \hline
    Parameter			& $\wA$		& $\wB$	& $\varphi$	\\ \hline
    Acceptation rate		& 14.50 \%	& 9.44 \%	& 2.14
    \%		\\ \hline
    \end{tabular}
\end{table}

\newpage{}

\begin{table}[H]
    \caption{Specific laws obtained as limit of the Gamma pdf.}
    \label{tab:limitelaw}
    \centering
    \begin{tabular}{c|cc}
        \hline
        & $\alpha$ & $\beta$ \\
        \hline
        Jeffreys & 0 & $+ \infty$ \\
        Uniform & 1 & $+ \infty$ \\
        Dirac & - & 0\\
        \hline
    \end{tabular}
\end{table}

\begin{figure}[H]
    \centering

    \subfigure[True image]{
      \begin{tabular}{c}
          \includegraphics[width=\figWidthMed cm]{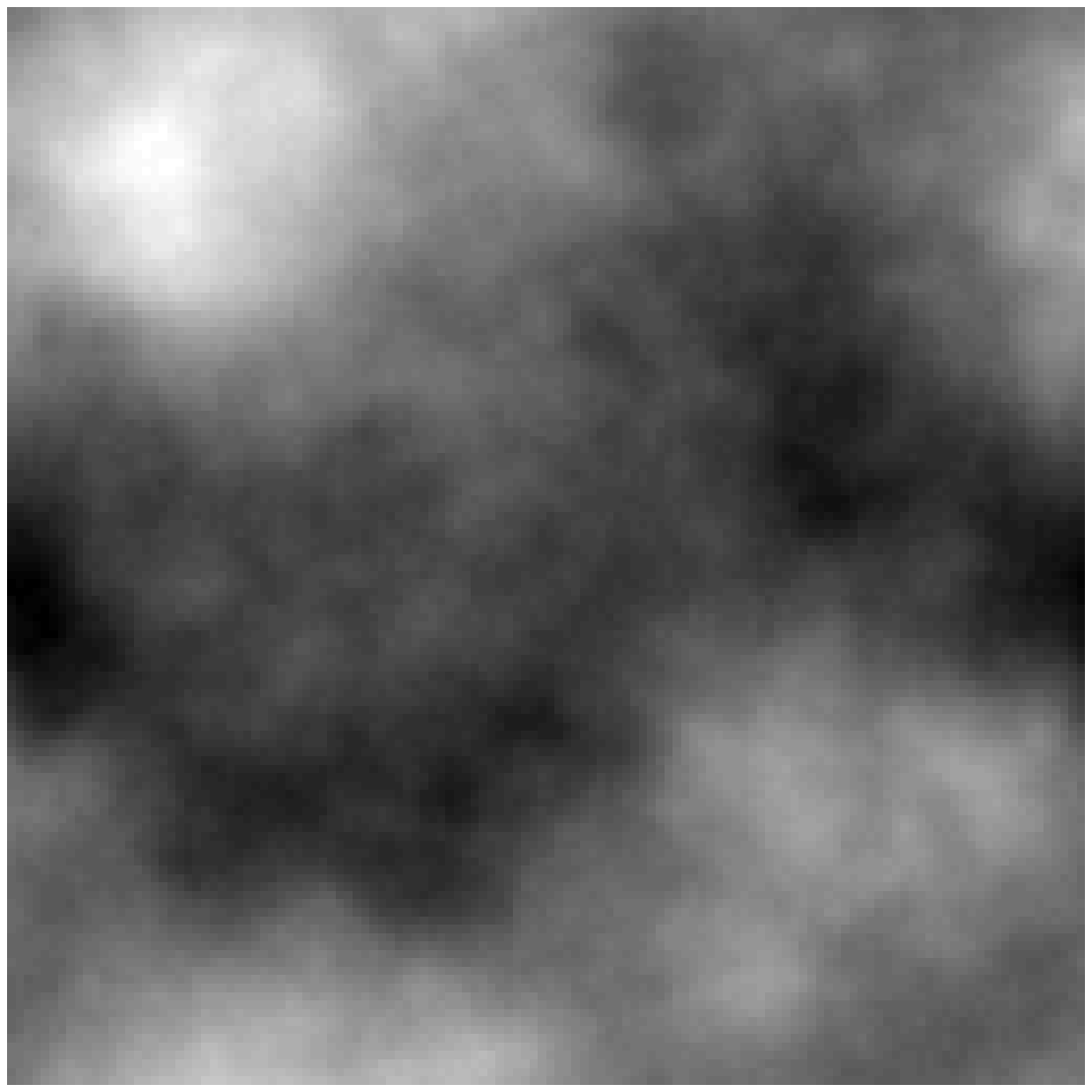}\\
          \includegraphics[width=\figWidthMed cm]{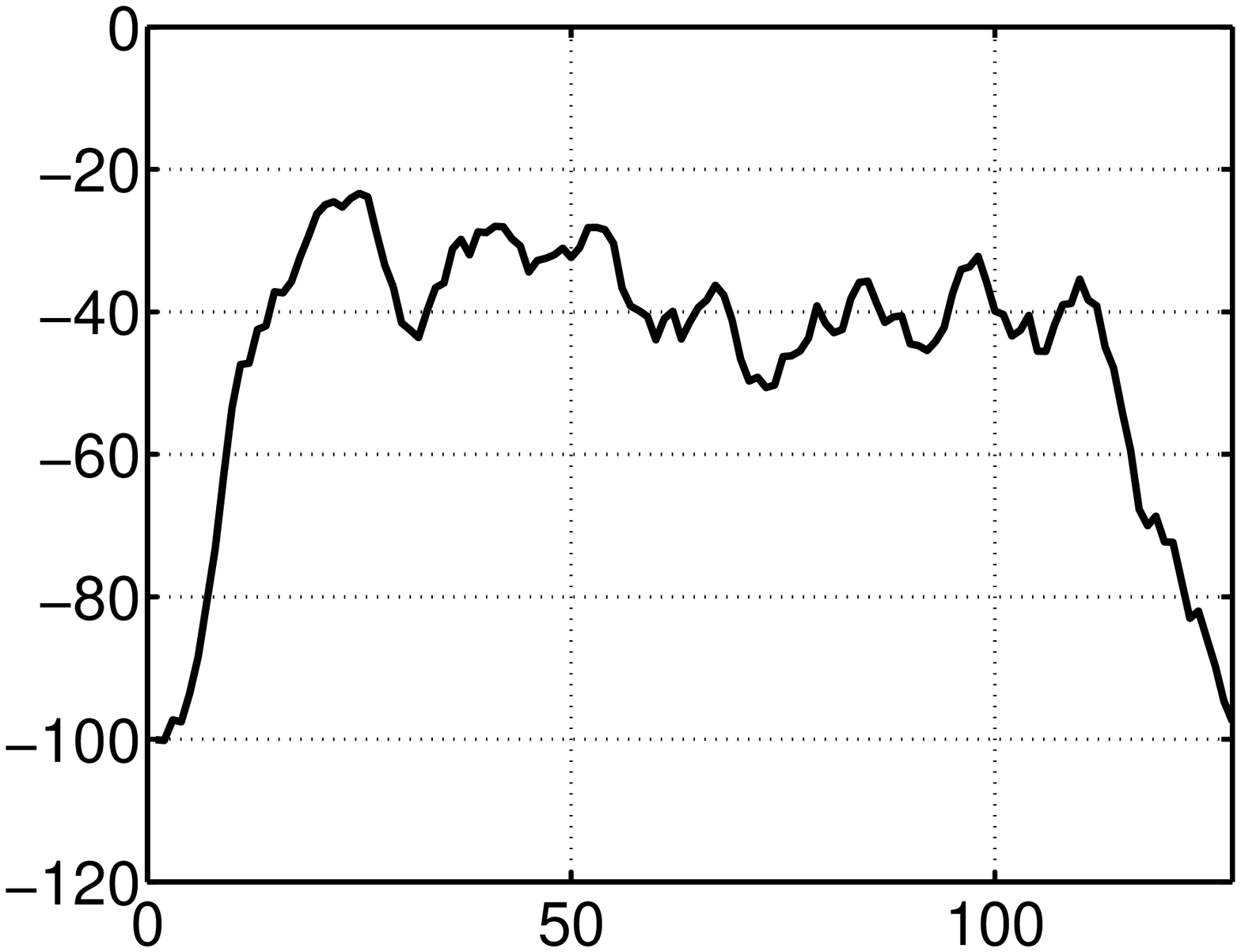}
      \end{tabular}
      \label{fig:True}
    } \hspace{2cm}
    \subfigure[Data]{
      \begin{tabular}{c}
          \includegraphics[width=\figWidthMed cm]{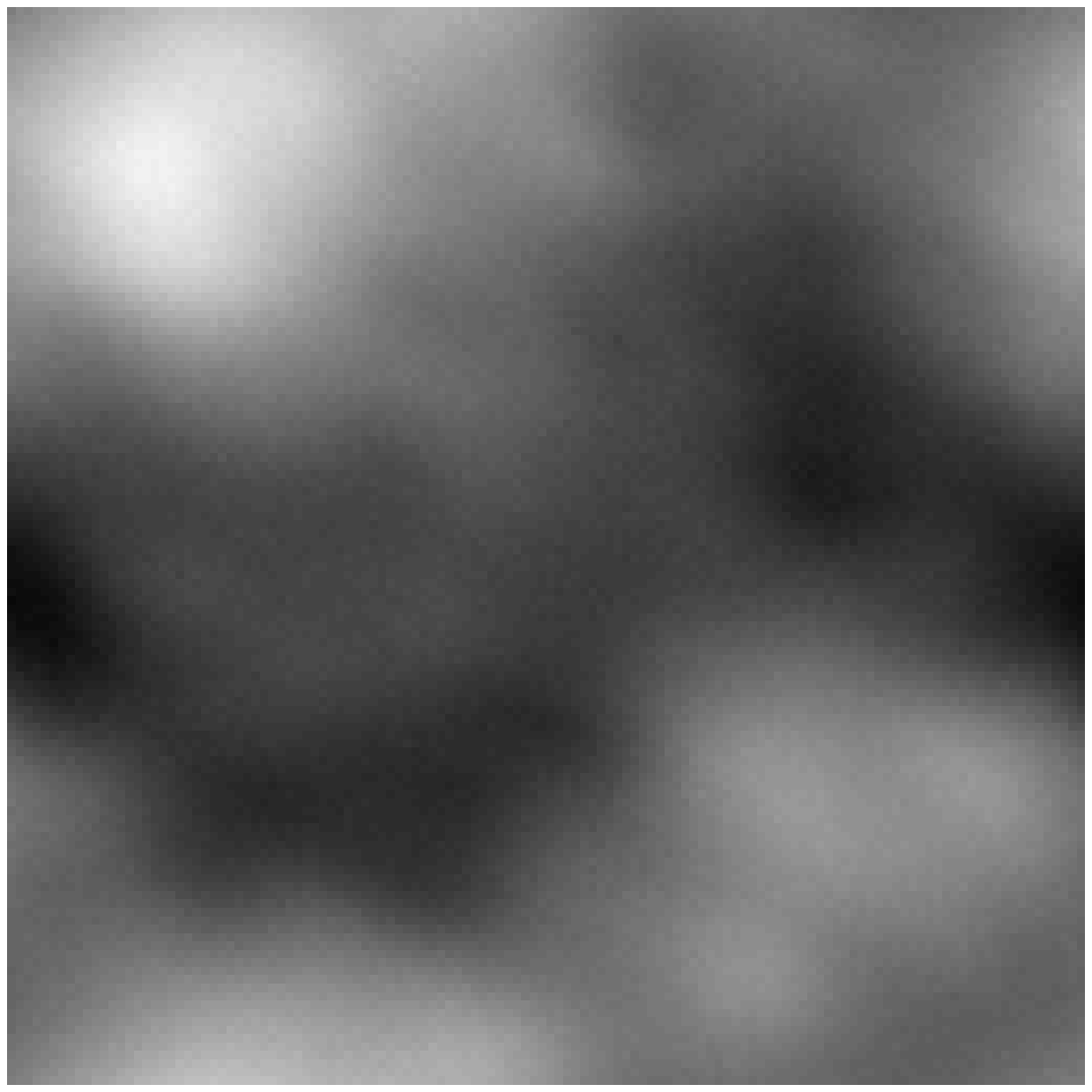} \\
          \includegraphics[width=\figWidthMed cm]{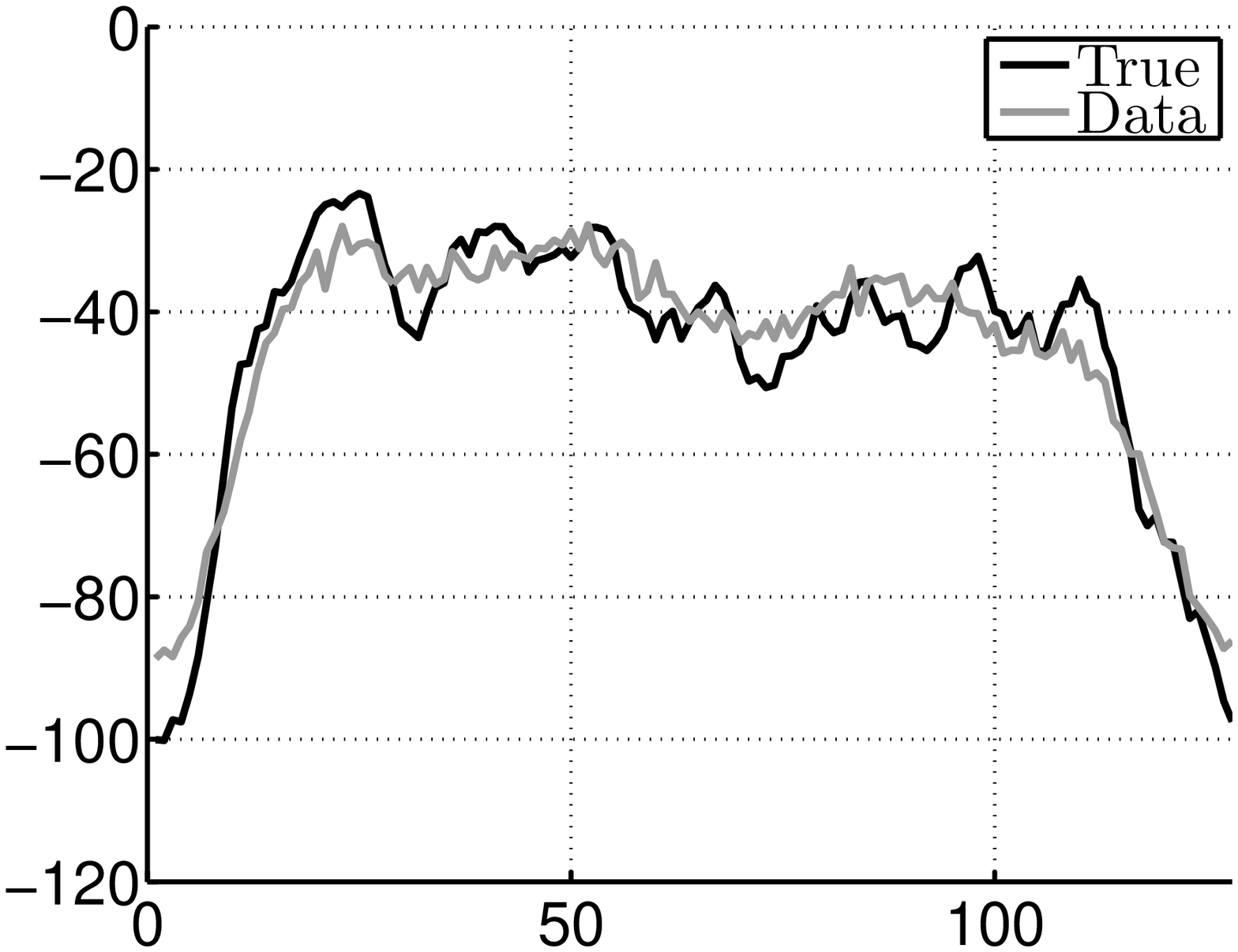}
      \end{tabular}
      \label{fig:Data}
    }

    \subfigure[Non-myopic]{
      \begin{tabular}{c}
          \includegraphics[width=\figWidthMed cm]{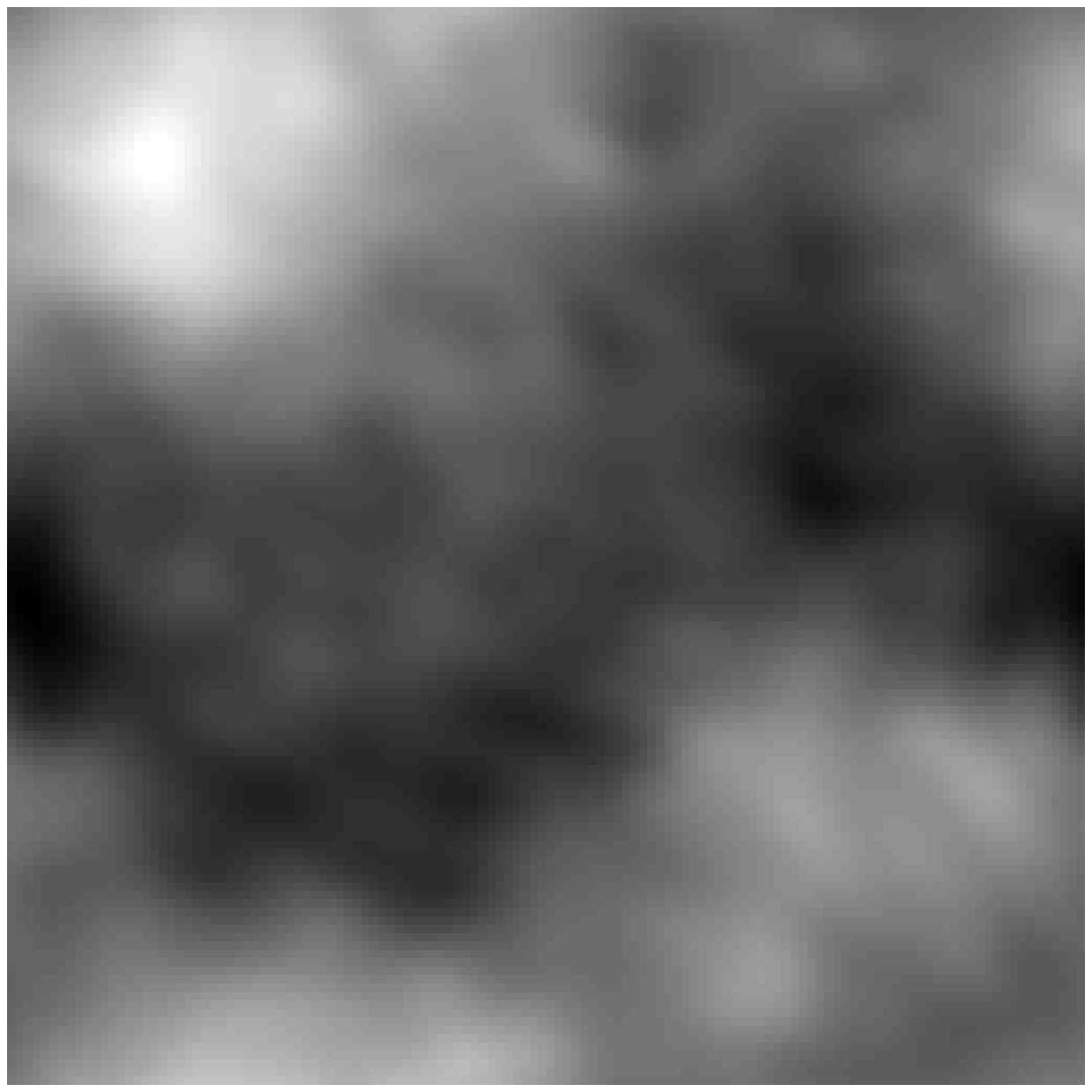} \\
          \includegraphics[width=\figWidthMed cm]{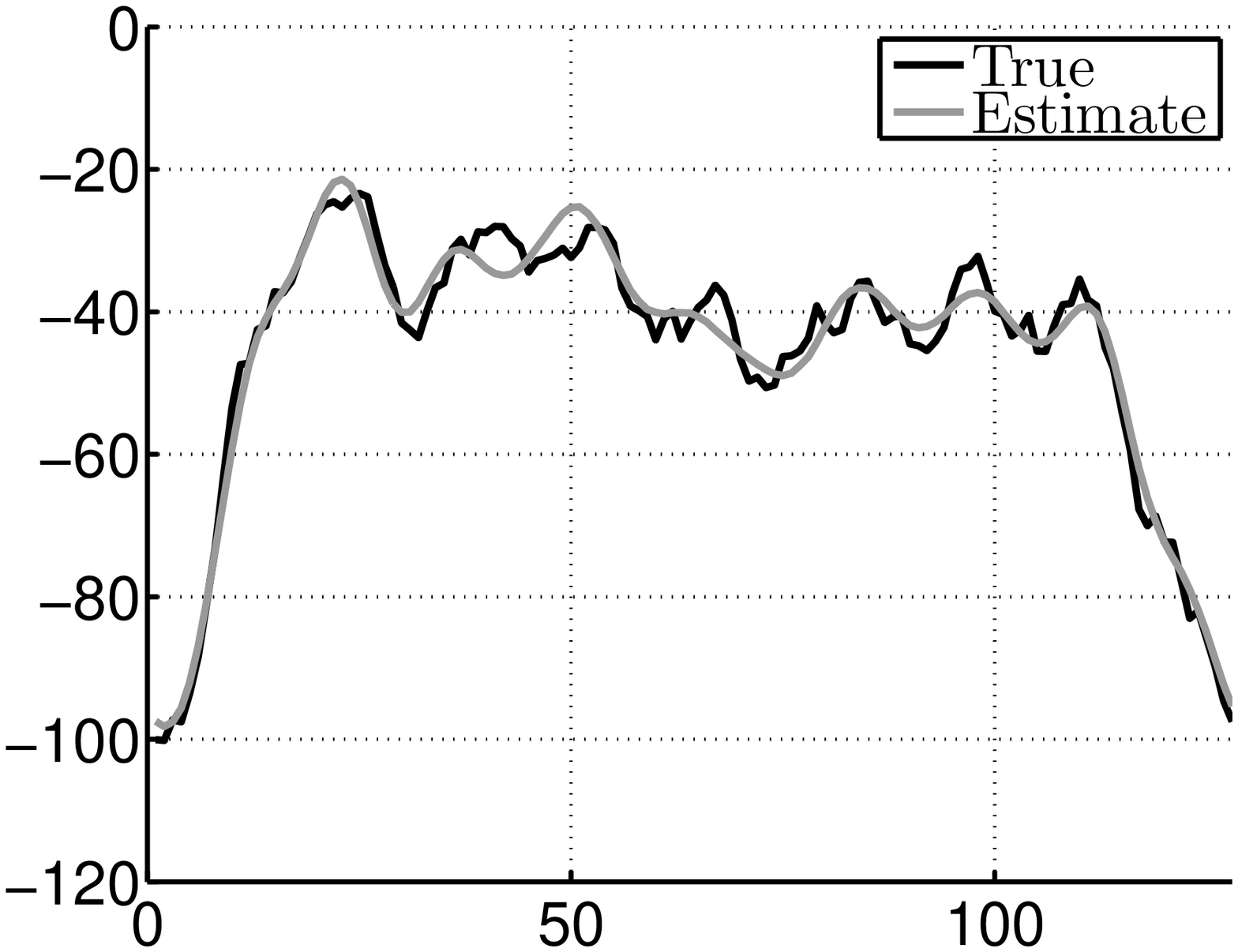}
      \end{tabular}
      \label{fig:Xeap}
    } \hspace{2cm}
    \subfigure[Myopic]{
      \begin{tabular}{c}
          \includegraphics[width=\figWidthMed cm]{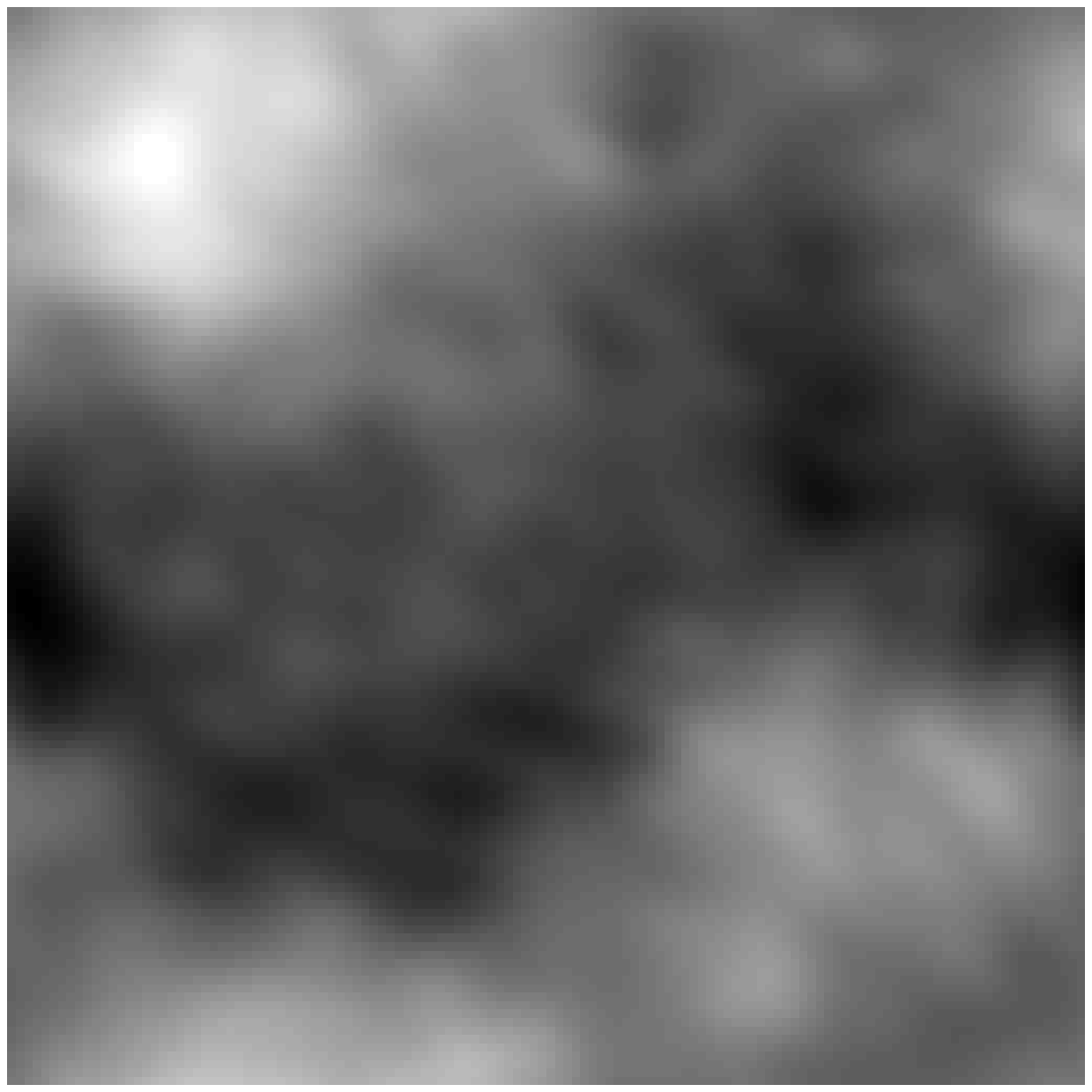} \\
          \includegraphics[width=\figWidthMed cm]{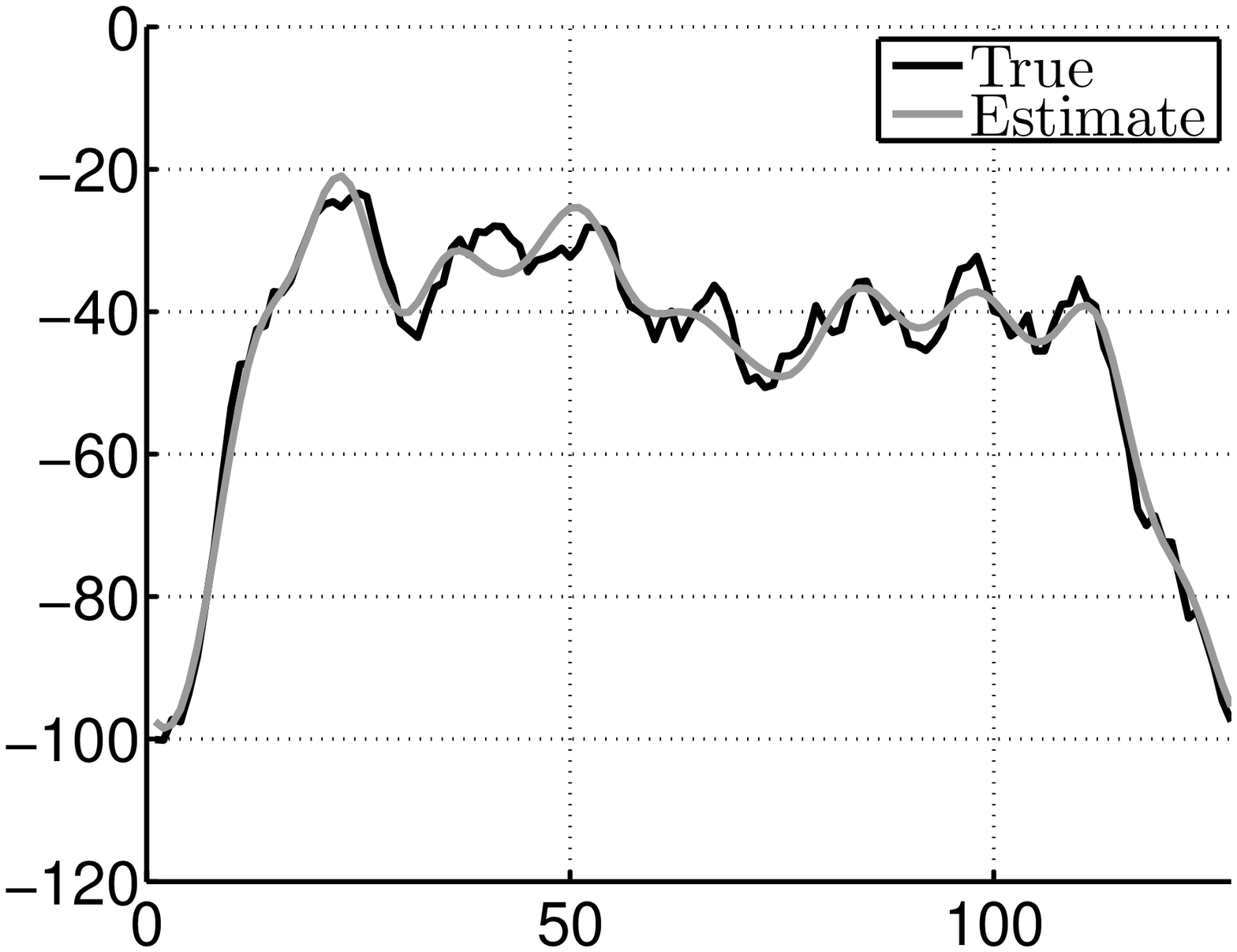}
      \end{tabular}
      \label{fig:XeapMyope}
    }
    \caption{The figure \ref{fig:True} represents a $128 \times 128$
      sample of the \aprio law for the object with $\GO = 1$ and $\G1
      = 2$. Fig.~\ref{fig:Data} is the data computed with the PSF
      shown in Fig.~\ref{fig:RI}. Figs.~\ref{fig:Xeap} and
      \ref{fig:XeapMyope} are the estimates with non-myopic and the
      myopic estimate, respectively. Profiles correspond to the 68-th
      line.}
    \label{fig:Results}
\end{figure}

\newpage{}

\begin{figure}[H]
    \centering \includegraphics[width=\figWidthLarge cm]{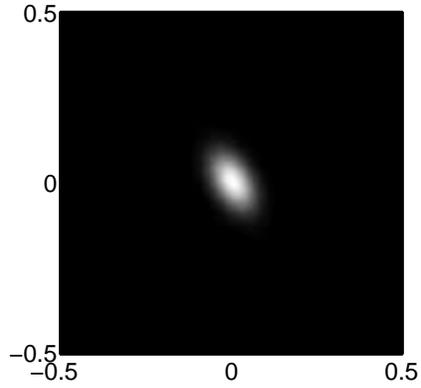}
    \caption{PSF with $\wA = 20$, $\wB = 7$ and $\varphi = \pi/3$. The
      x-axis and y-axis are reduced frequency.}
    \label{fig:RI}
\end{figure}

\begin{figure}[H]
    \centering 

    \subfigure[Non-Myopic]{\includegraphics[width=\figWidthLarge
      cm]{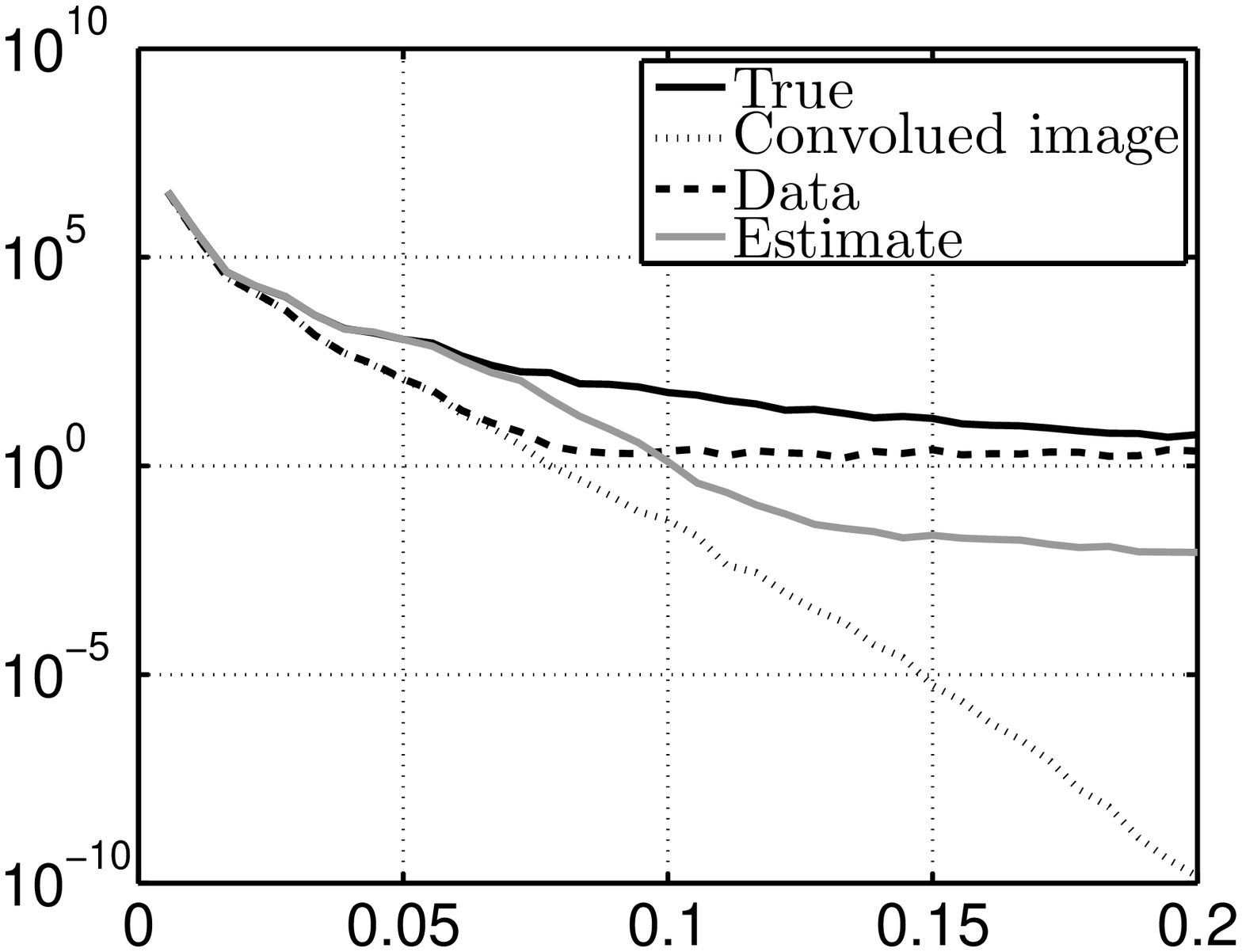}\label{fig:psdNonMyopic}}
    \subfigure[Myopic]{\includegraphics[width=\figWidthLarge
      cm]{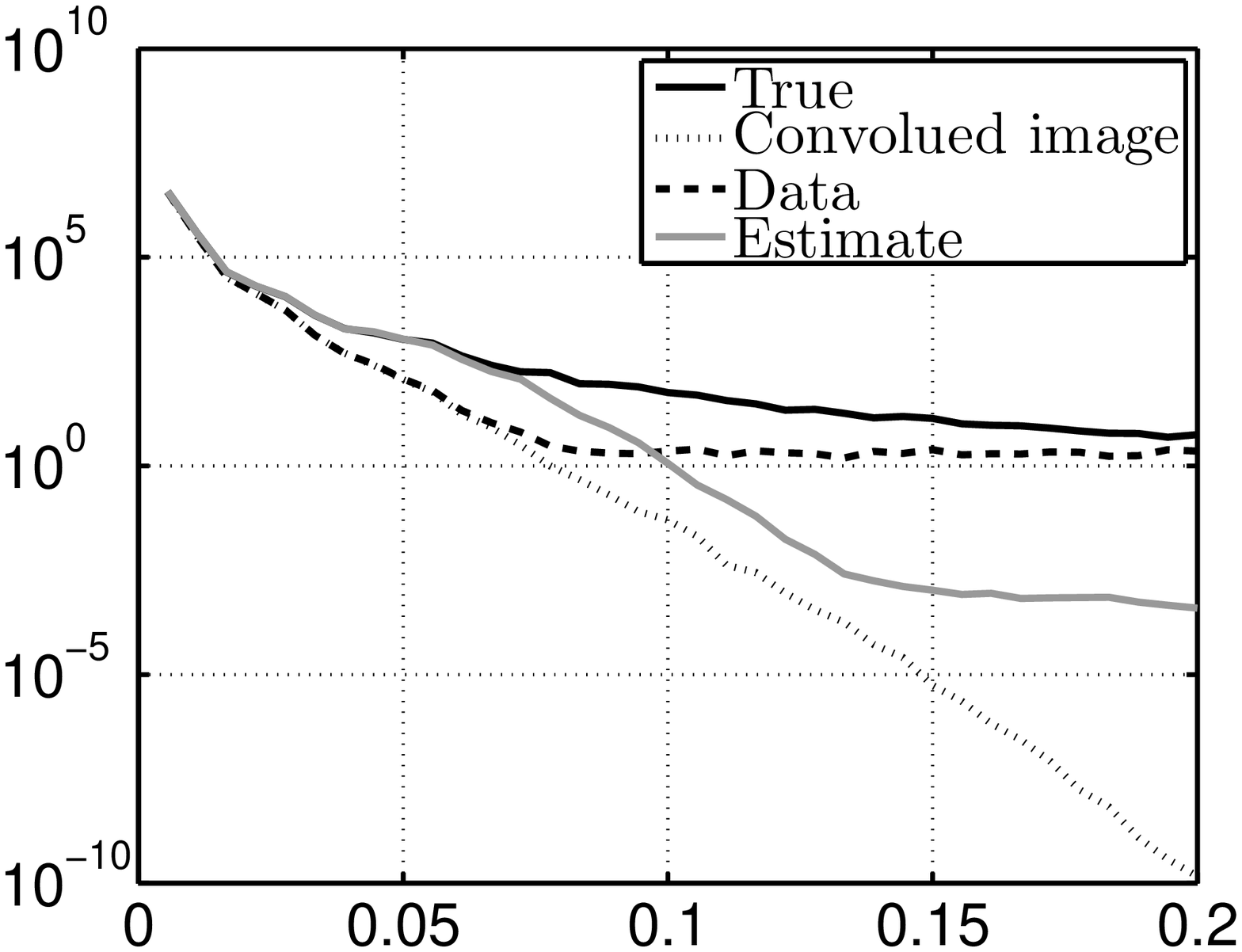}\label{fig:psdMyopic}}

    \caption{Circular average of the empirical power spectral density
      of the image, the convolued image, the data (convolued image
      corrupted by noise) and the estimates, in radial frequency with
      y-axis in logarithmic scale. The x-axis is the radial
      frequency.}
    \label{fig:meanPSD}
\end{figure}

\flushbottom{}\newpage{}

\begin{figure}[H]
    \centering

	\subfigure[$\GN$]{\includegraphics[width=\figWidth cm]{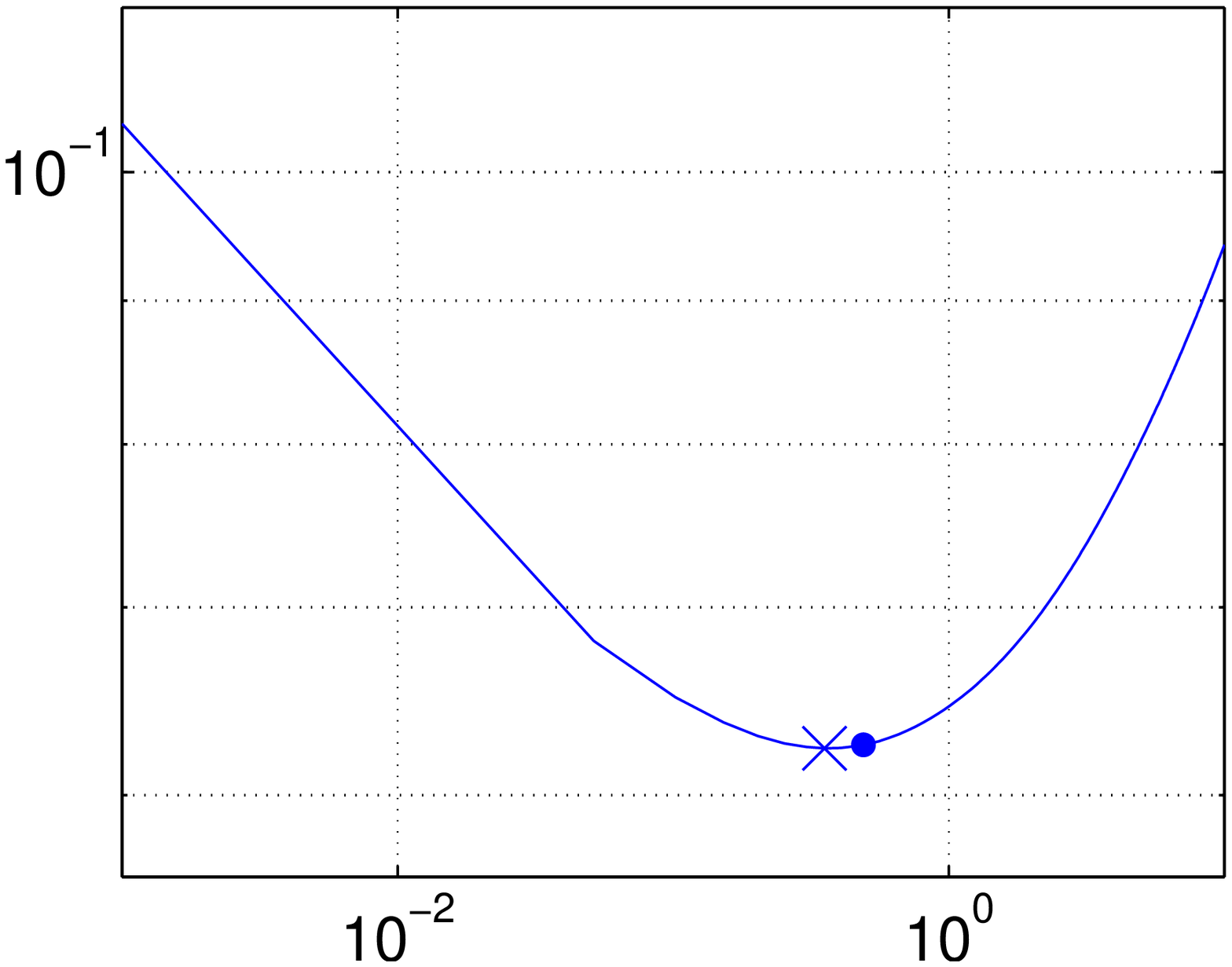}} ~~~~~
	\subfigure[$\G1$]{\includegraphics[width=\figWidth cm]{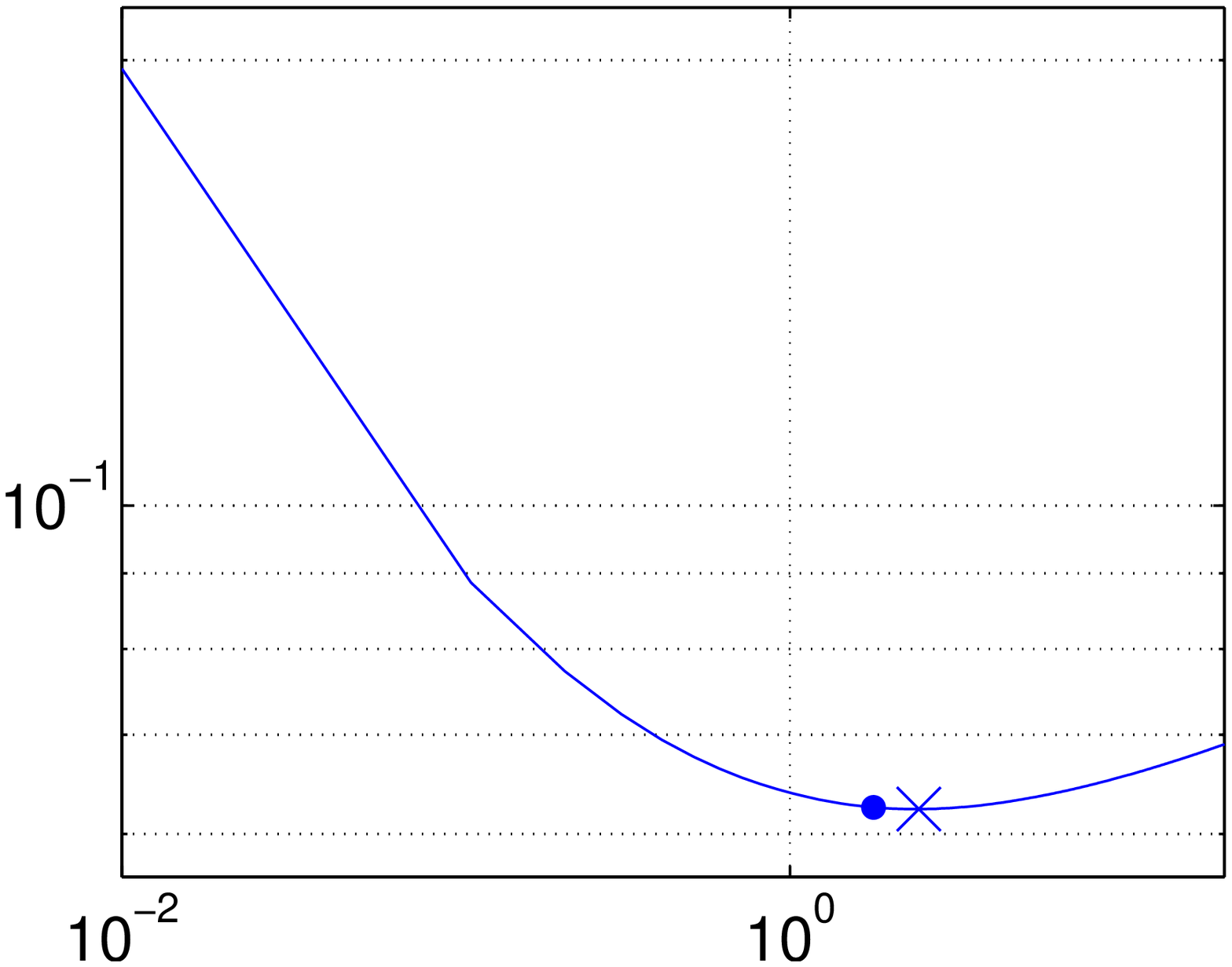}}
    
	\subfigure[$\wA$]{\includegraphics[width=\figWidth cm]{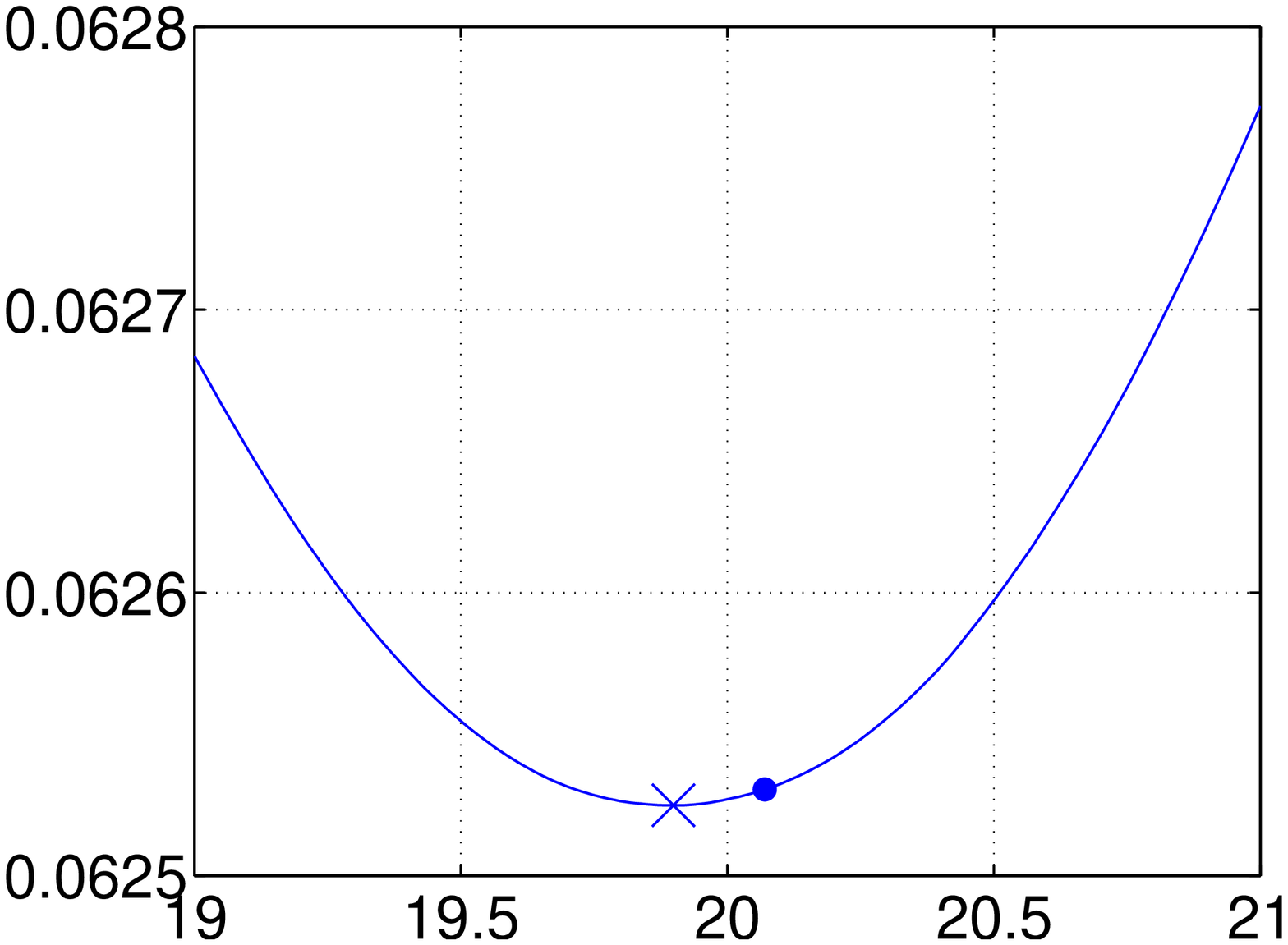}}~~~~~
	\subfigure[$\wB$]{\includegraphics[width=\figWidth cm]{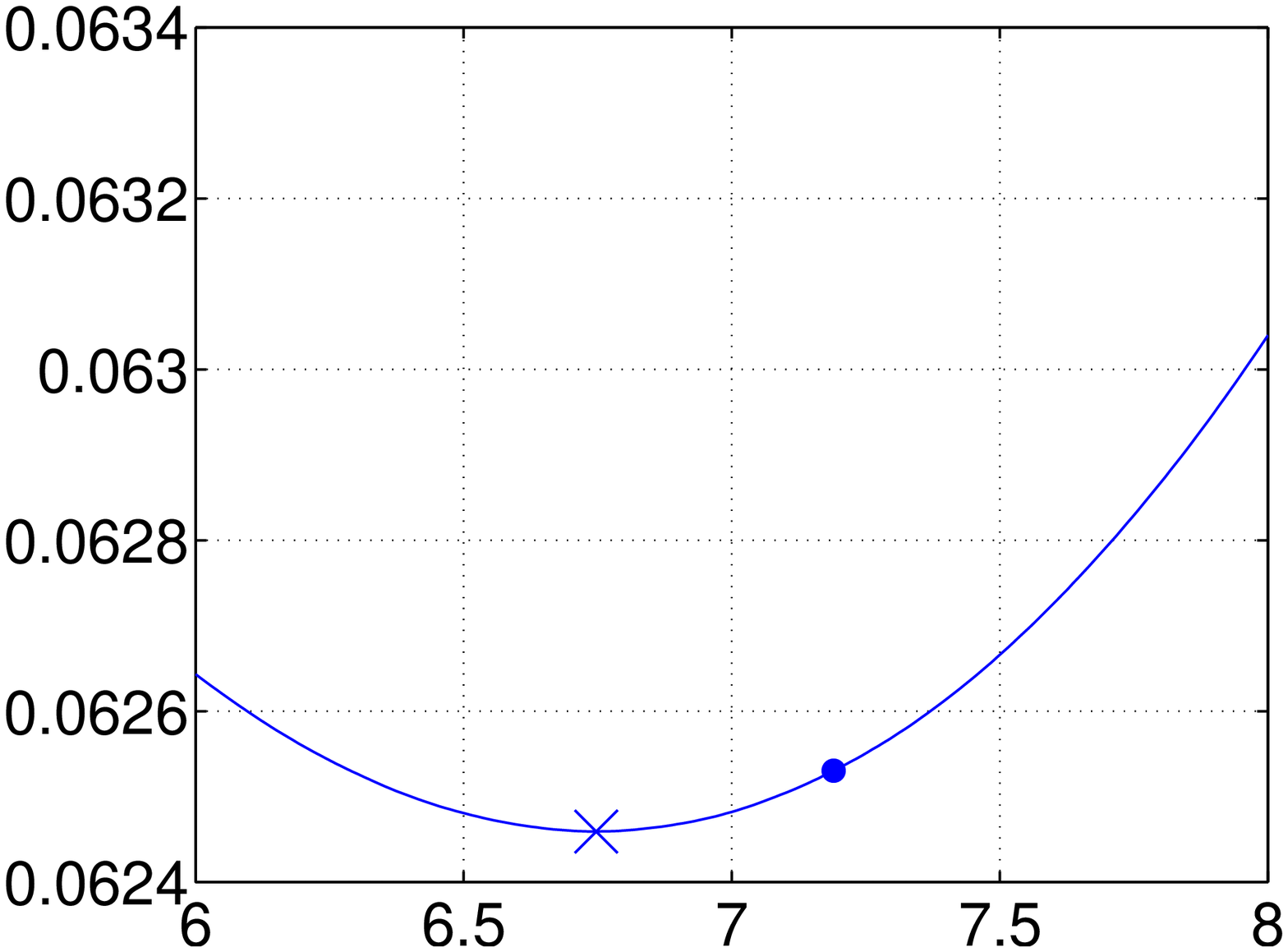}}~~~~~
	\subfigure[$\phi$]{\includegraphics[width=\figWidth cm]{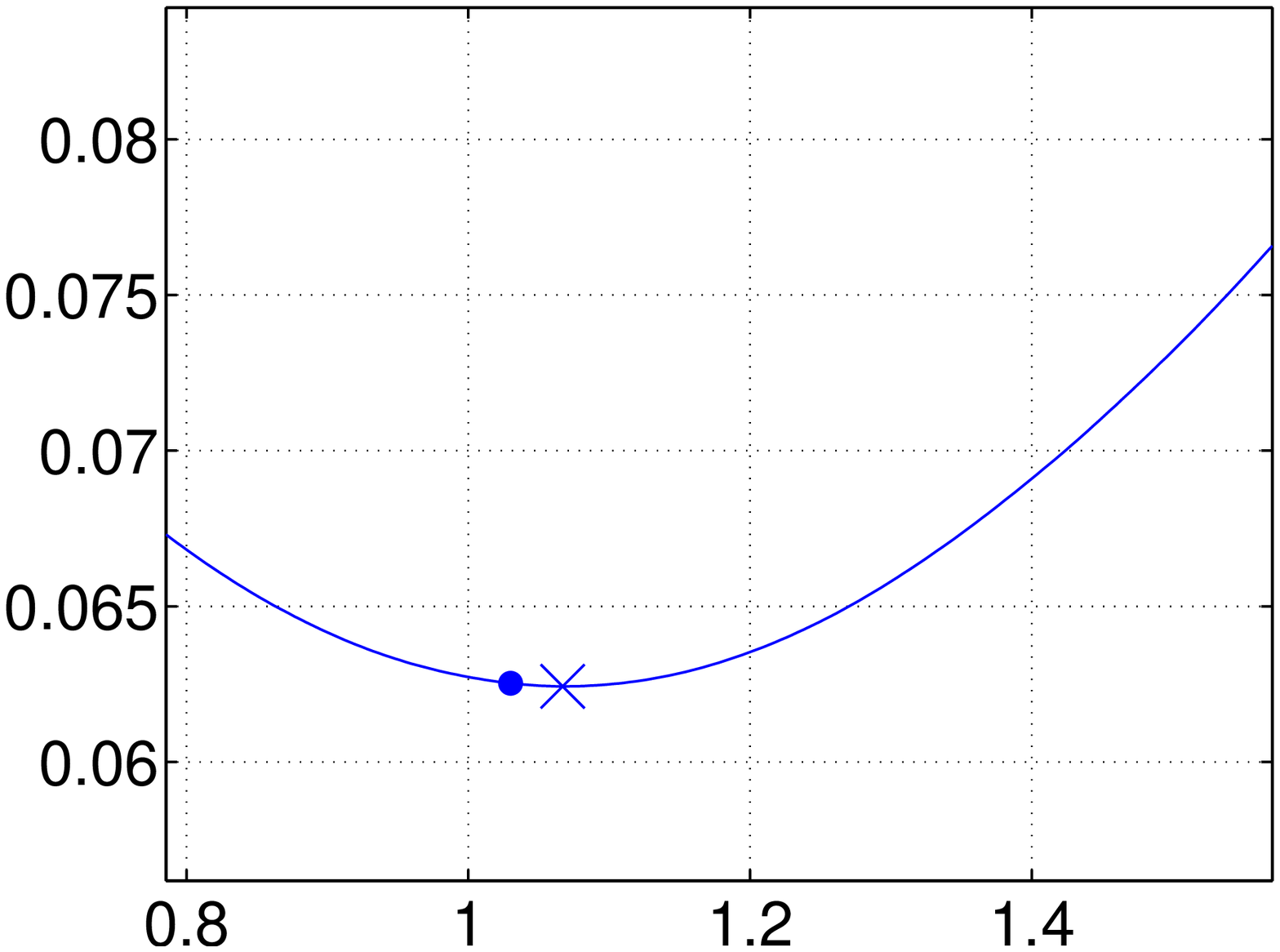}}

        \caption{Computation of the best parameters in the sense $e$
          Eq. (\ref{eq:48}). The symbol '$\times$' is the minimum and
          the symbol '.' is the estimated value by our approach. The
          y-axis of $\GN$ and $\G1$ are in logarithmic scale.}
    \label{fig:bestparam}
\end{figure}

\newpage{}

\begin{figure}[H]
    \centering

    \subfigure[$\GN$ for non-myopic case]{
      \begin{tabular}{c}
          \includegraphics[width=\figWidth cm,height=3cm]{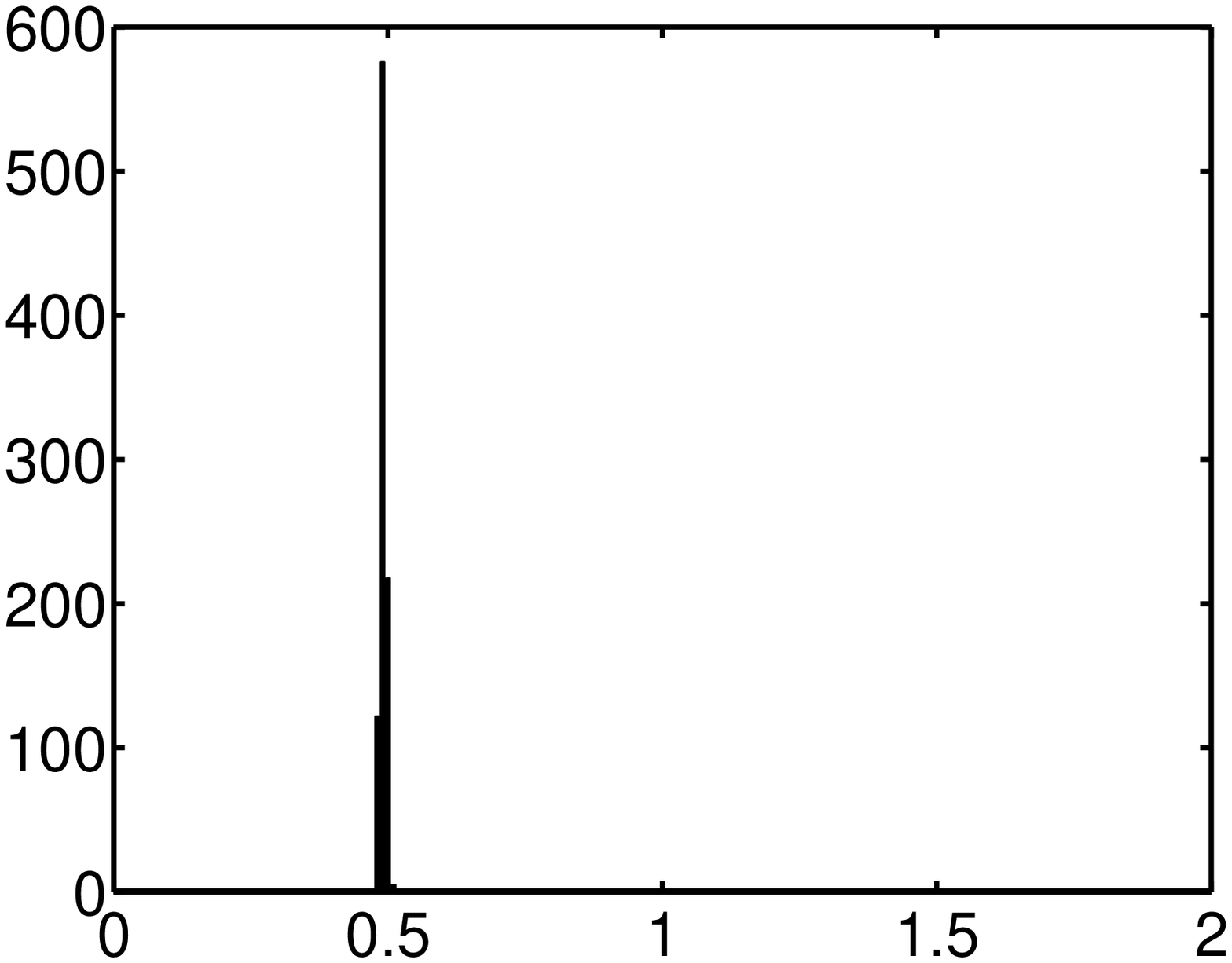} \\
          \includegraphics[width=\figWidth cm,height=3cm]{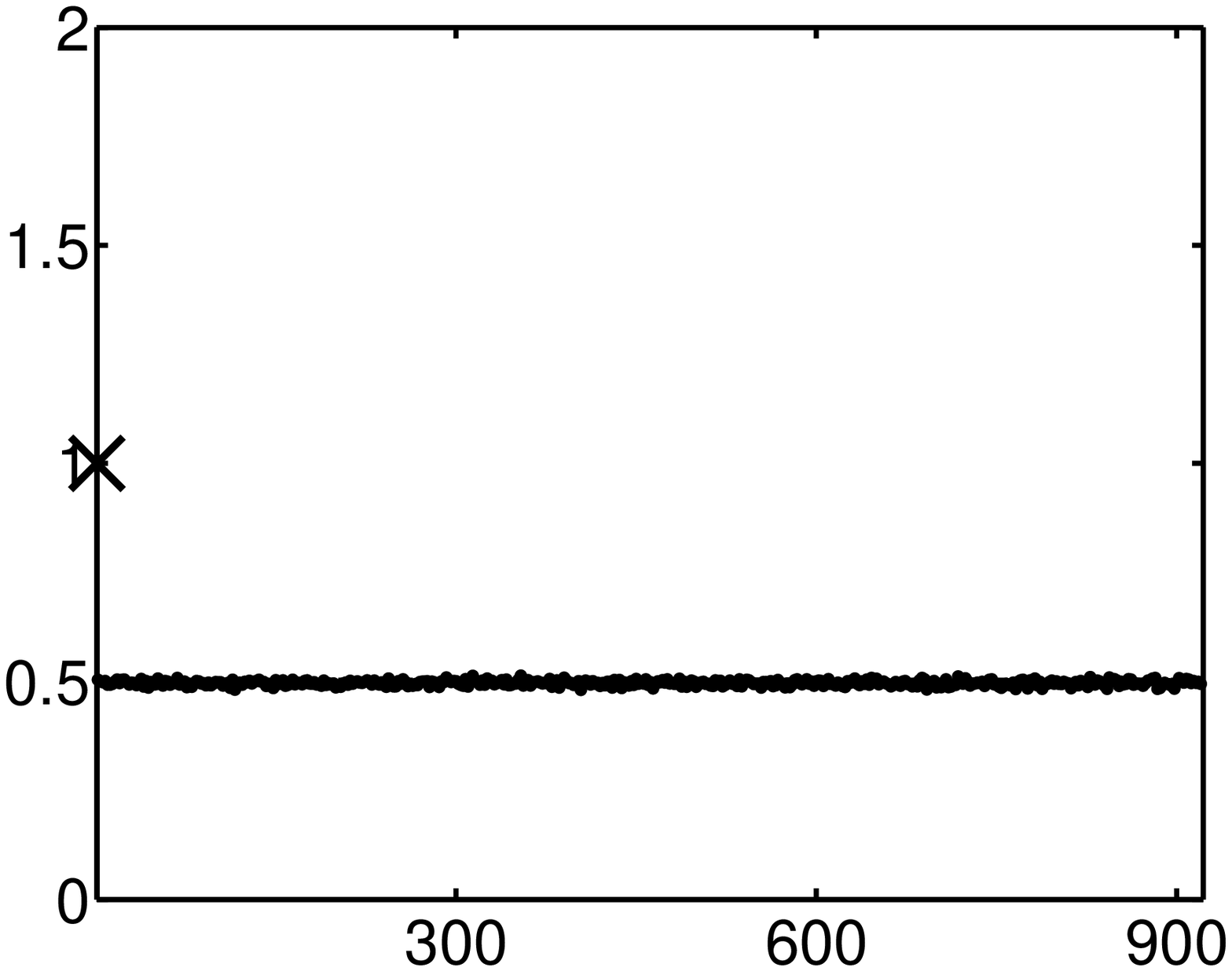}
      \end{tabular} \label{fig:GbGibbs}
    }
    \subfigure[$\GN$ for myopic
    case]{
      \begin{tabular}{c}
          \includegraphics[width=\figWidth cm,height=3cm]{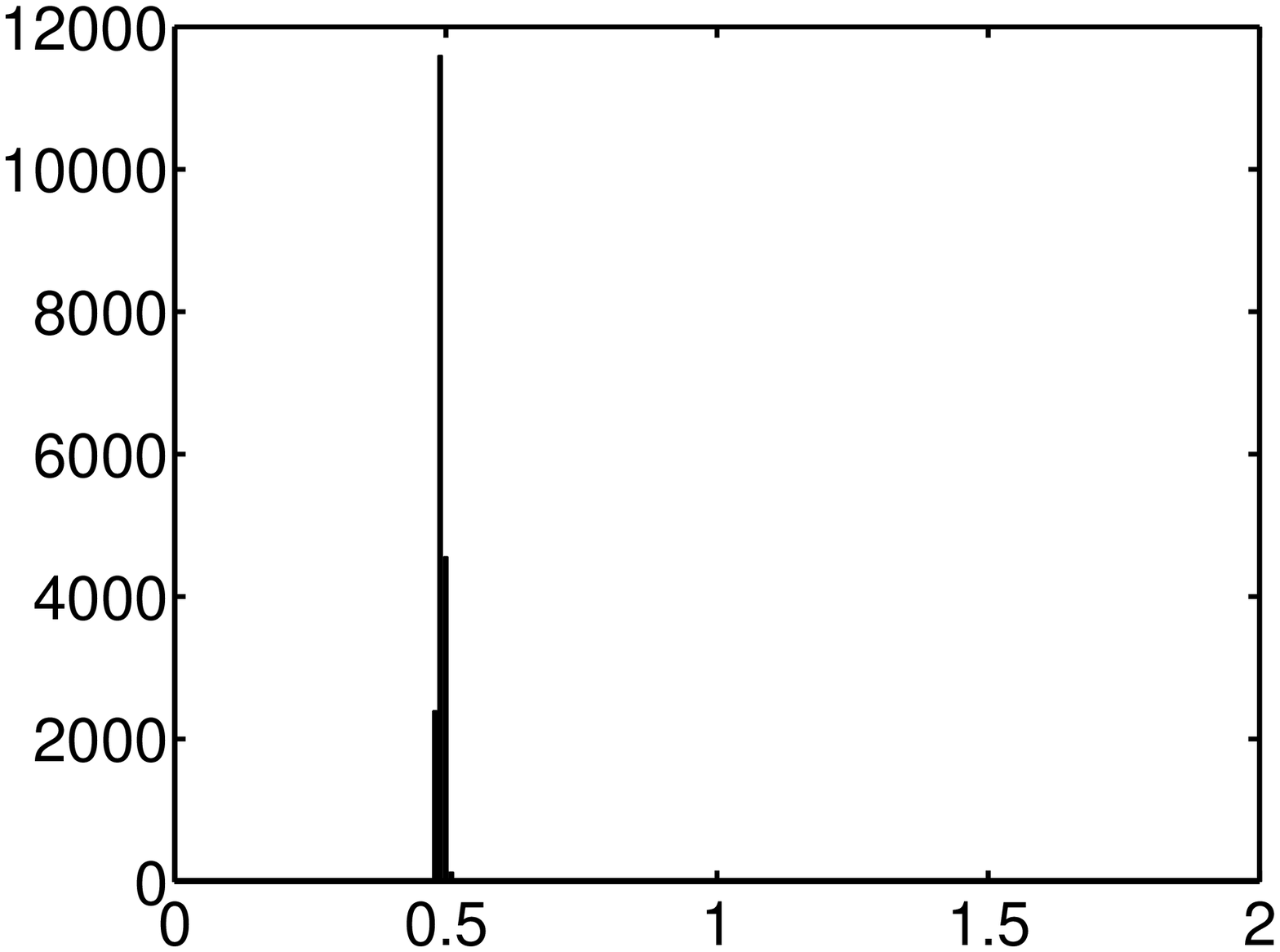} \\
          \includegraphics[width=\figWidth cm,height=3cm]{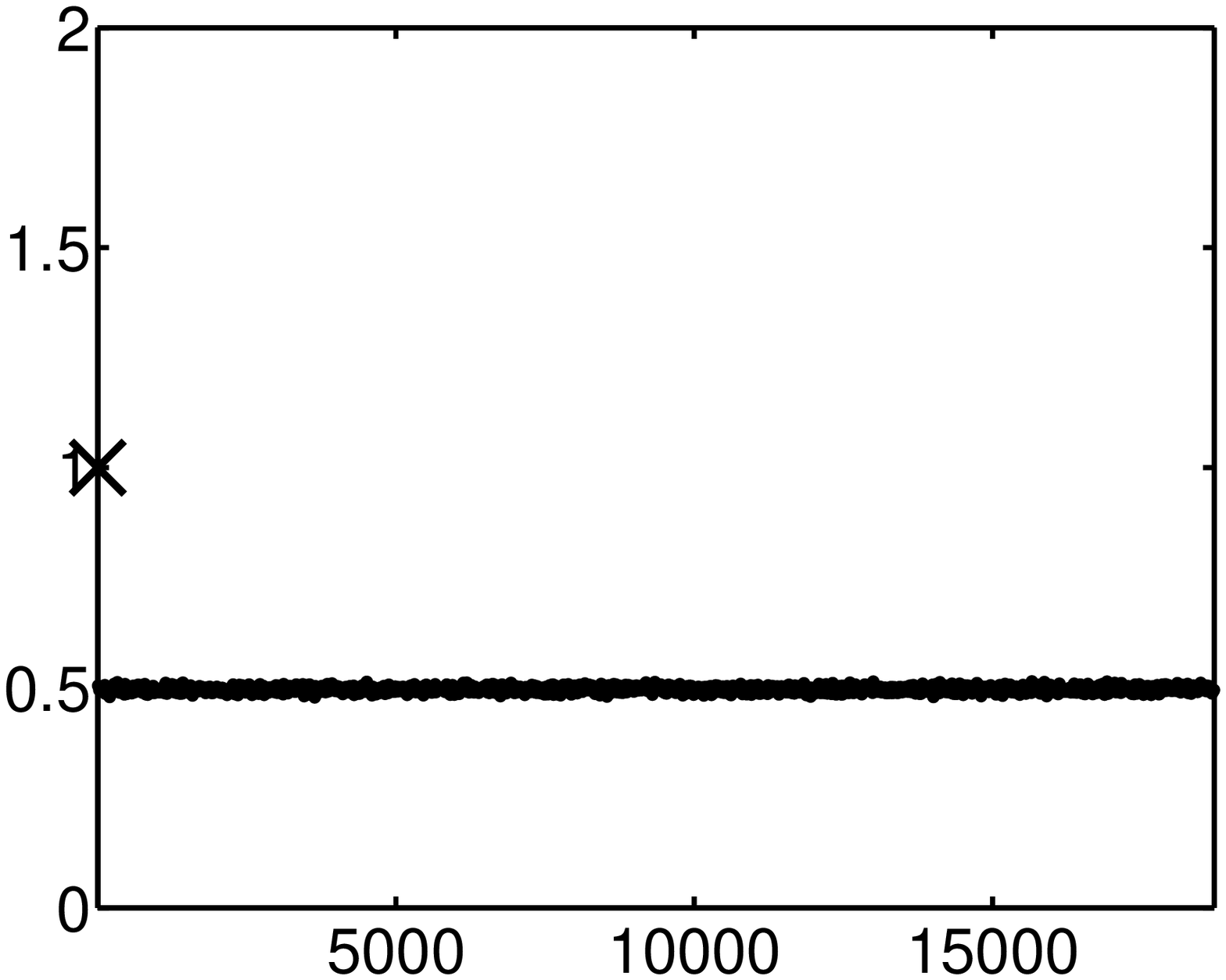}
      \end{tabular} \label{fig:GbMyope}
    }

    \subfigure[$\G1$ for non-myopic case]{
      \begin{tabular}{c}
          \includegraphics[width=\figWidth cm,height=3cm]{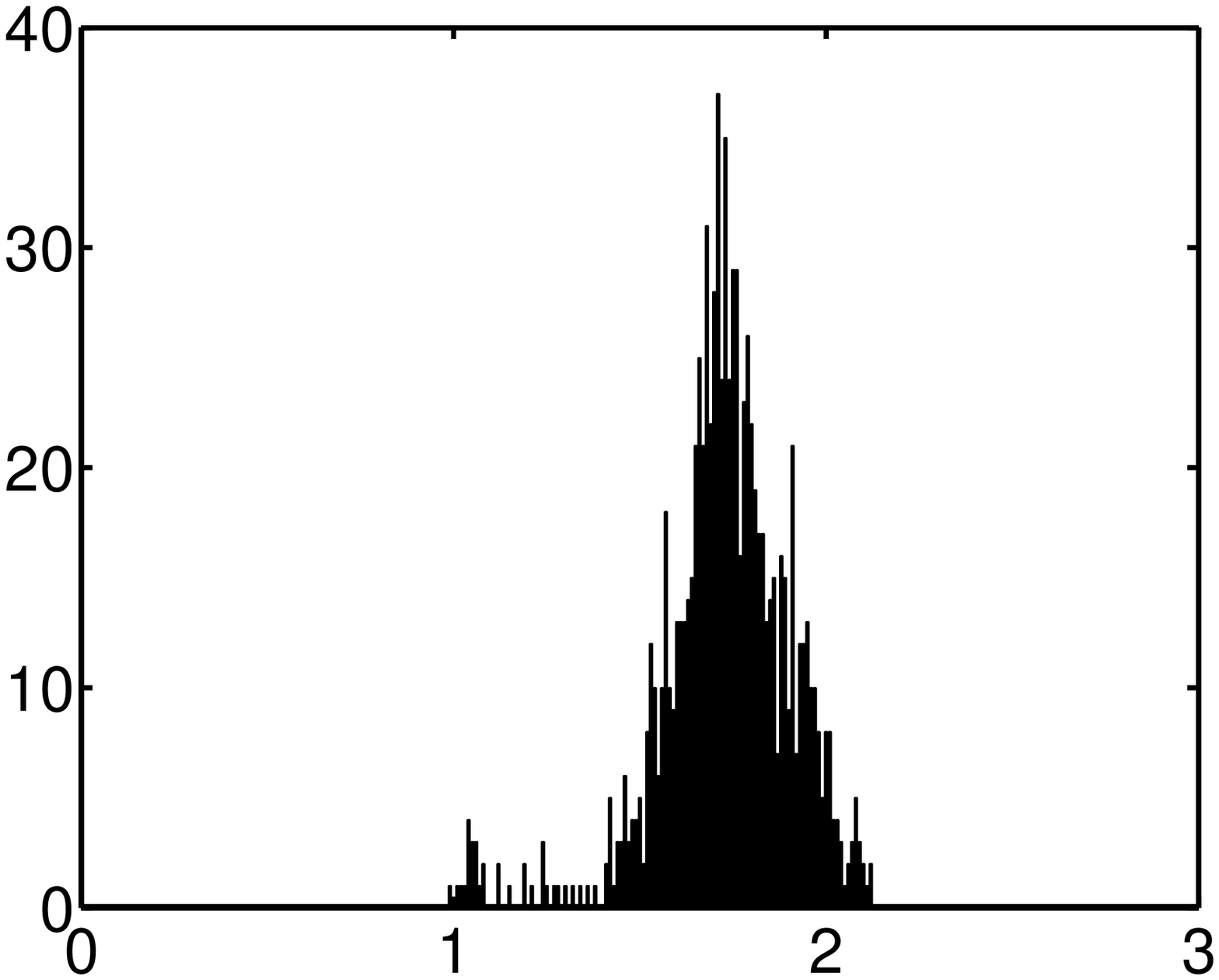} \\
          \includegraphics[width=\figWidth cm,height=3cm]{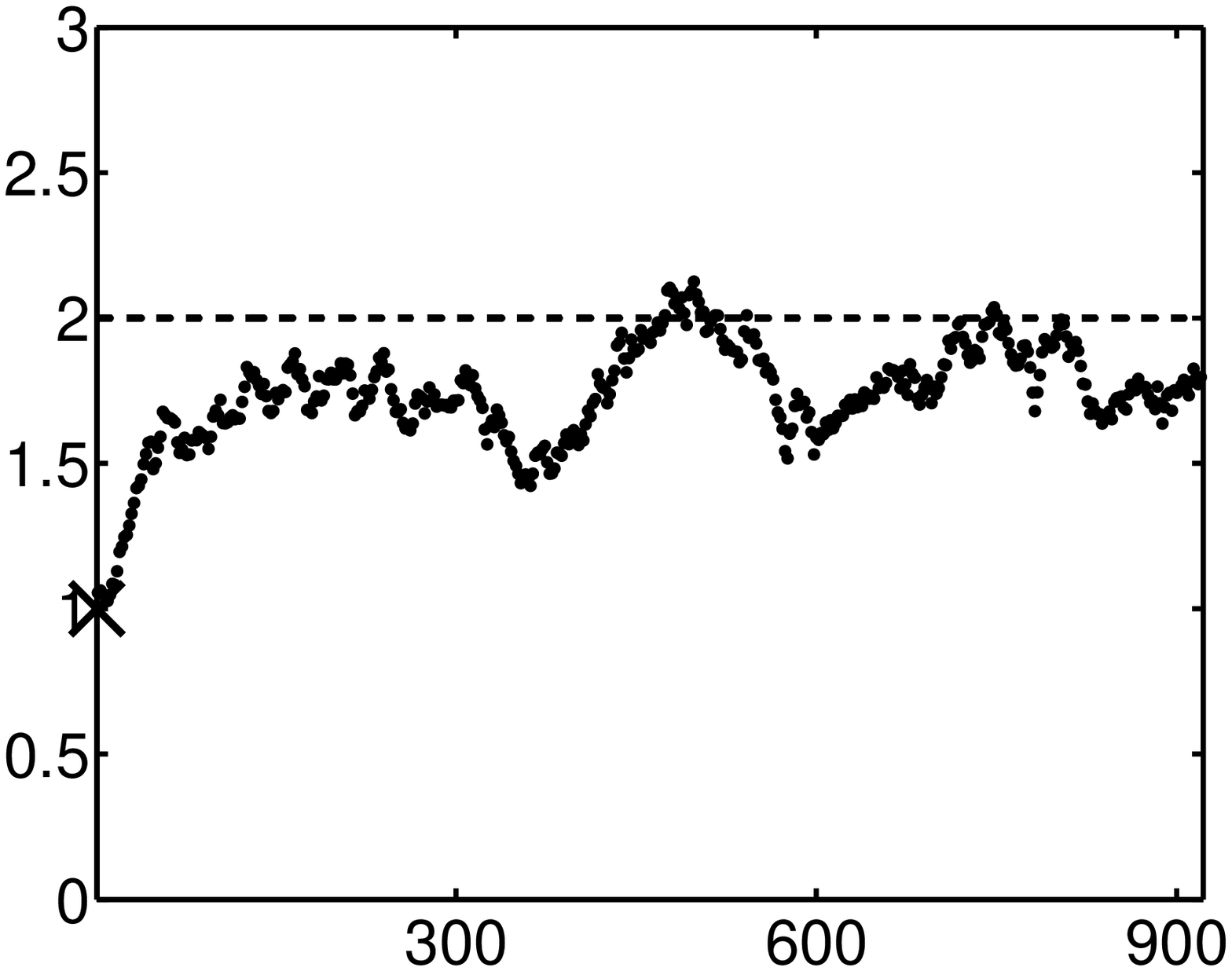}
      \end{tabular} \label{fig:G1Gibbs}
    }
    \subfigure[$\G1$ for myopic
    case]{
      \begin{tabular}{c}
          \includegraphics[width=\figWidth cm,height=3cm]{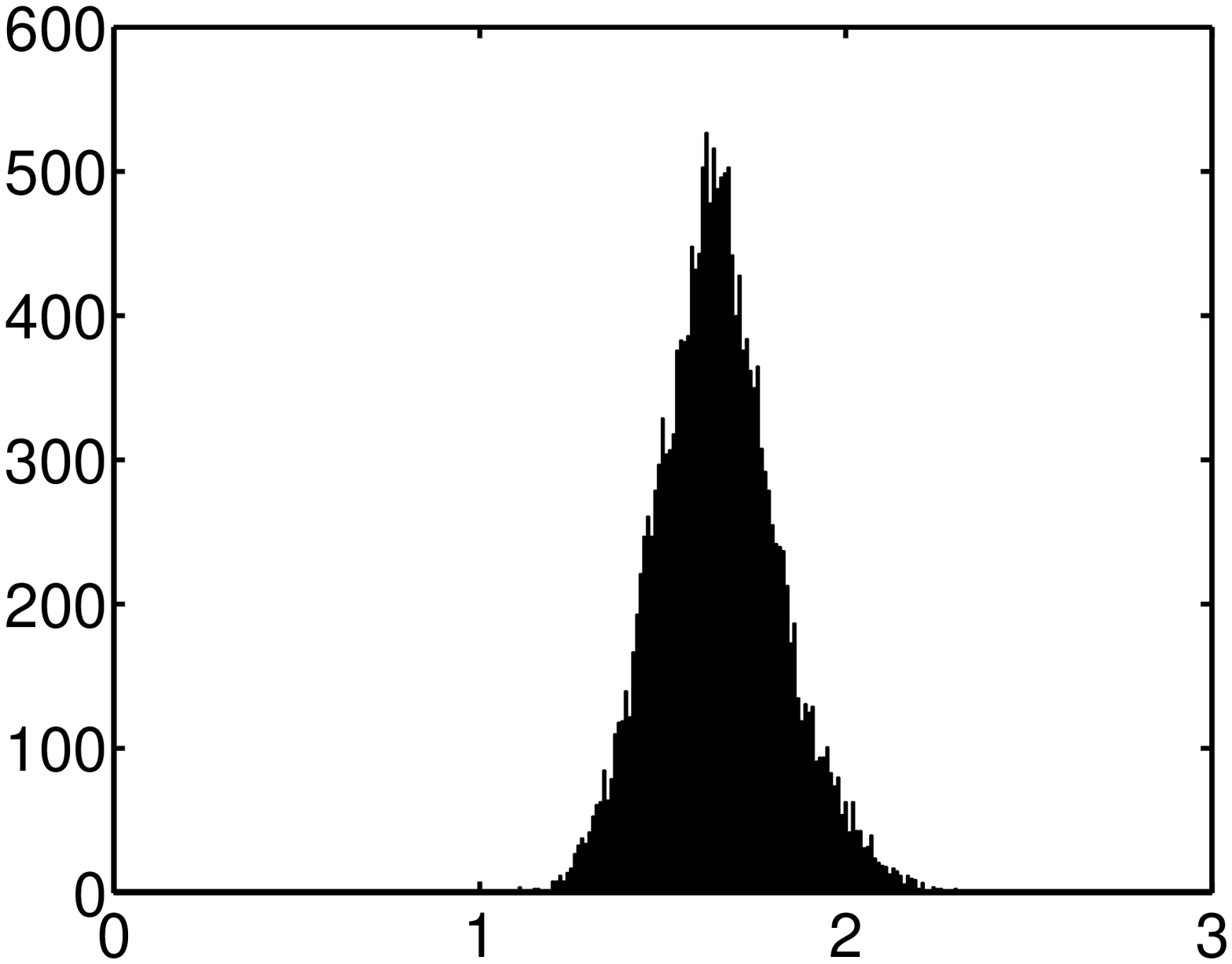} \\
          \includegraphics[width=\figWidth cm,height=3cm]{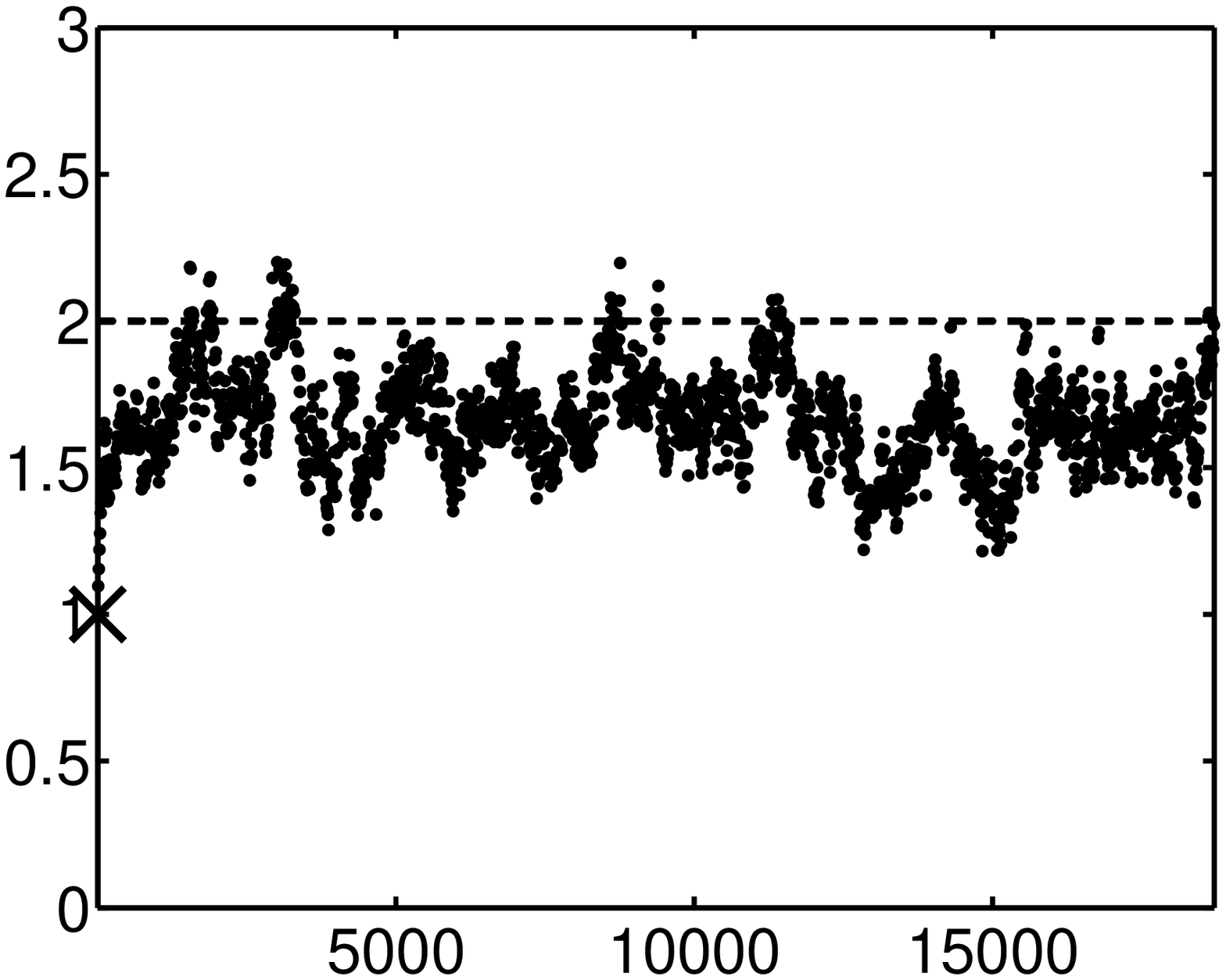}
      \end{tabular} \label{fig:G1Myope}
    }

    \caption{Histograms and chains for the non-myopic case in
      Figs.~\ref{fig:GbGibbs}-\ref{fig:G1Gibbs} and the myopic case in
      Figs.~\ref{fig:GbMyope}-\ref{fig:G1Myope} for $\GN$ and $\G1$,
      respectively. The symbol $\times$ localizes the initial value
      and the dashed line corresponds to the true value. The x-axis
      are iteration's index for the chains and parameter value for the
      histograms.}
    \label{fig:hyperparam}
\end{figure}

\newpage{}

\begin{figure}[H]
    \centering

    \subfigure[$\wA$]{
      \begin{tabular}{c}
          \includegraphics[width=\figWidth cm,height=3.5cm]{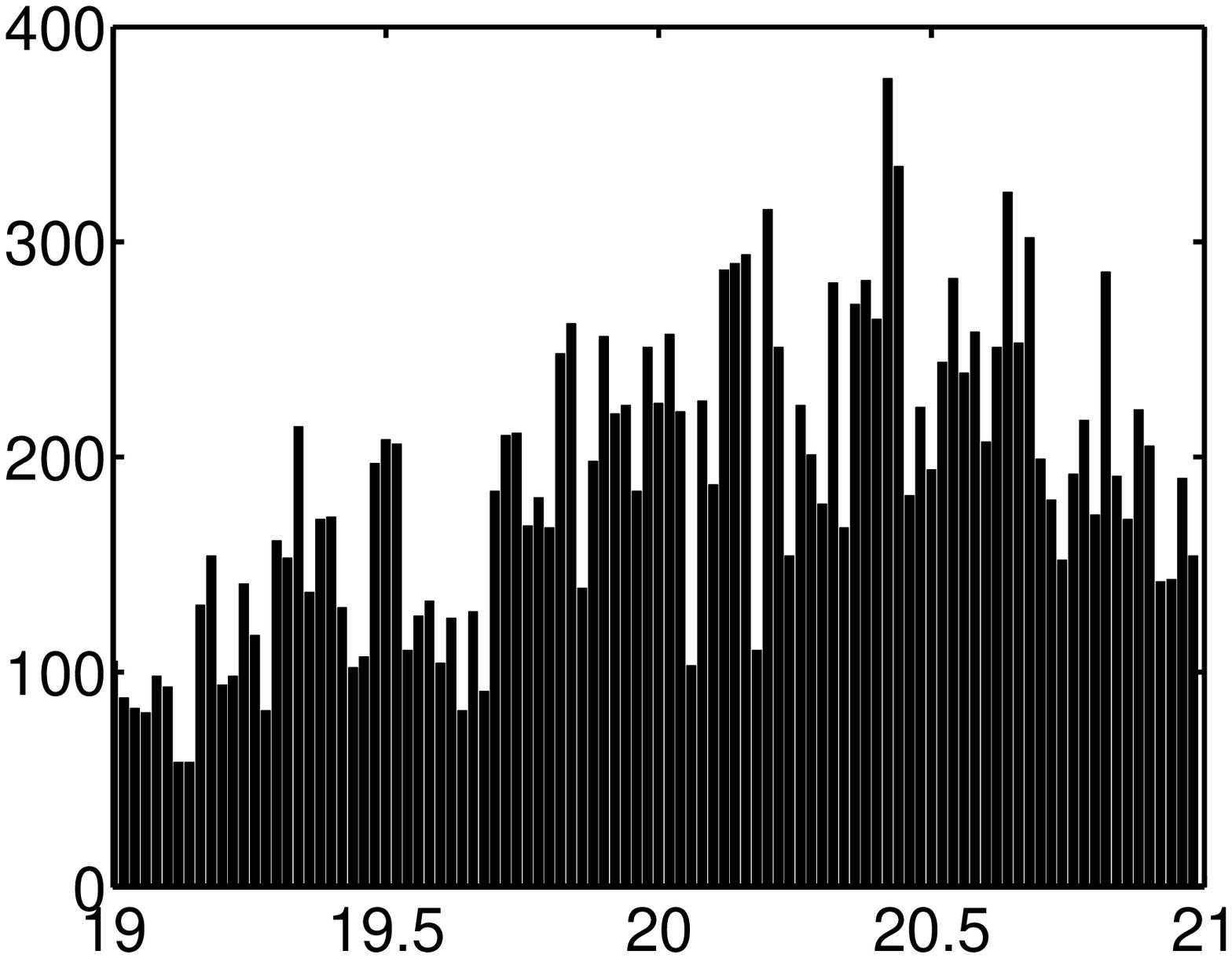} \\
          \includegraphics[width=\figWidth cm,height=3.5cm]{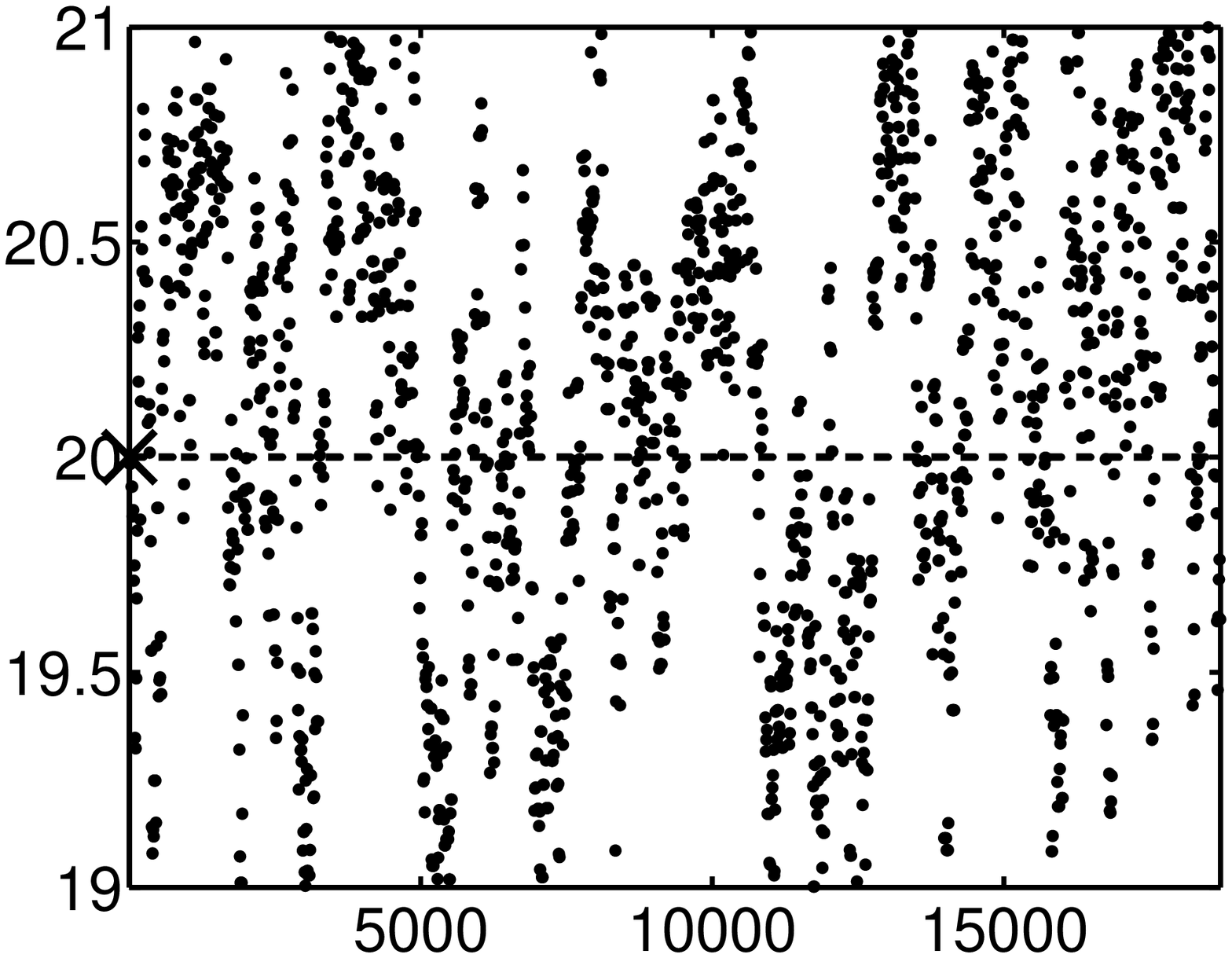}          
      \end{tabular} \label{fig:Wa}
    }
    \subfigure[$\wB$]{
      \begin{tabular}{c}
          \includegraphics[width=\figWidth cm,height=3.5cm]{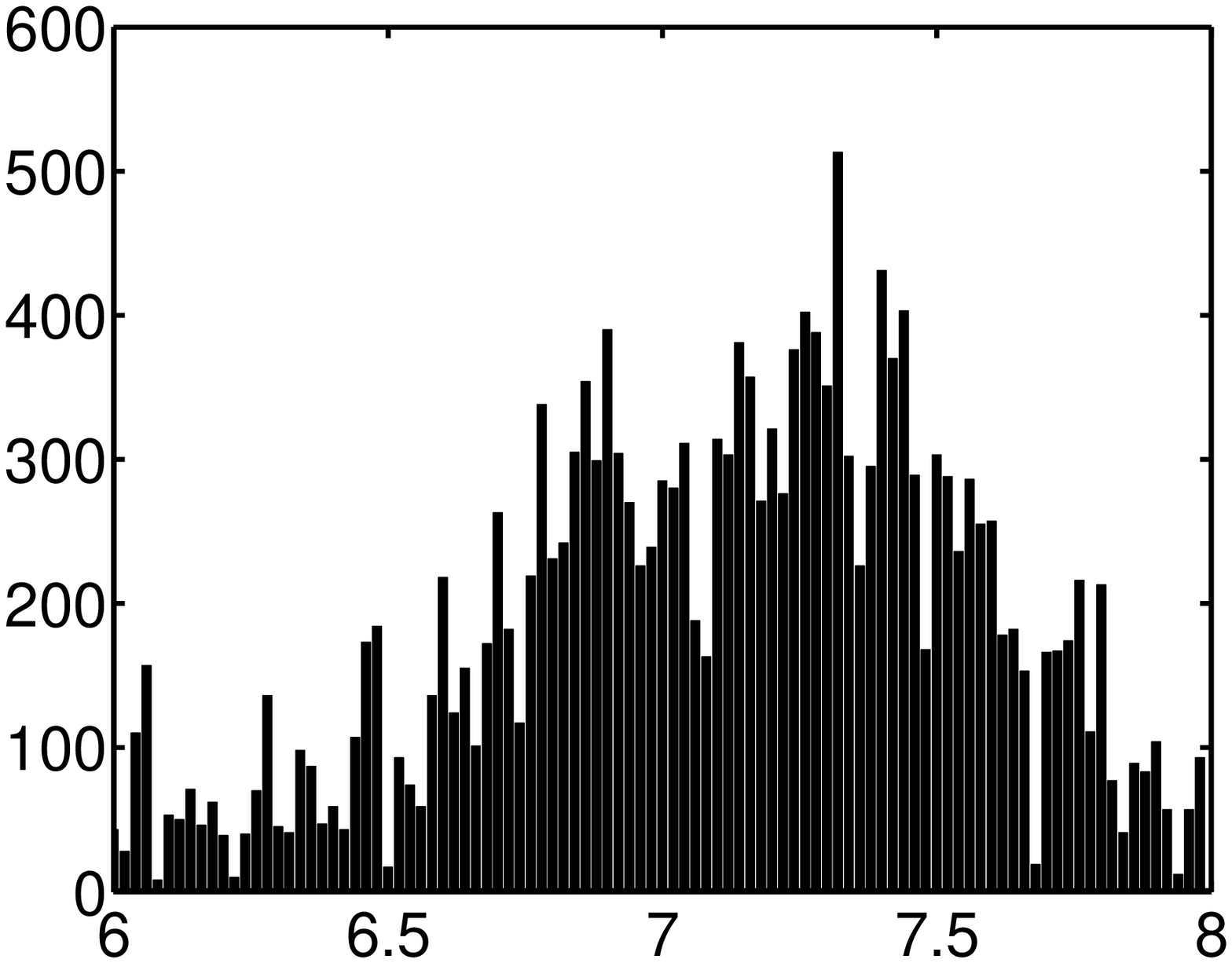} \\
          \includegraphics[width=\figWidth cm,height=3.5cm]{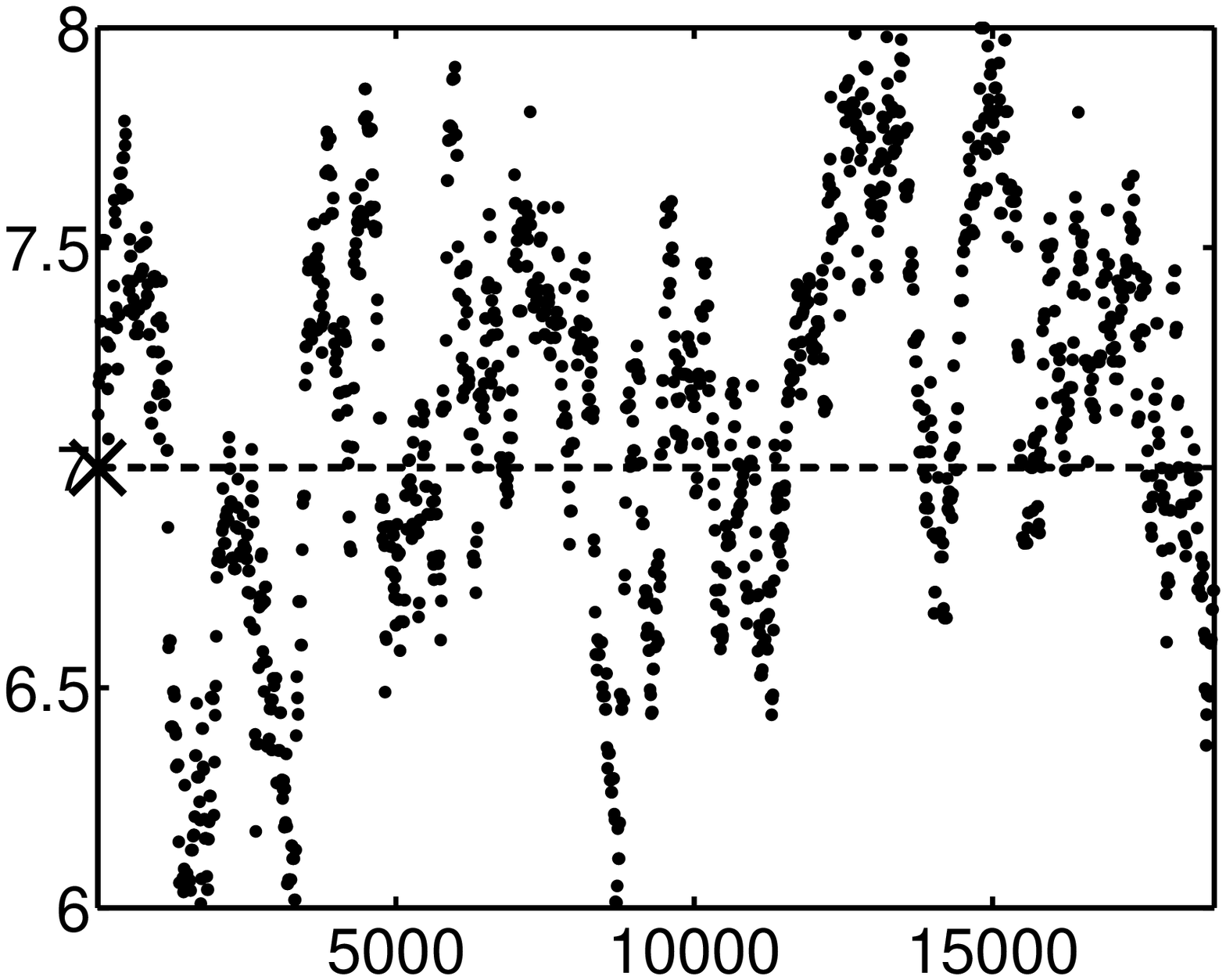}       
      \end{tabular} \label{fig:Wb}
    }
    \subfigure[$\varphi$]{
      \begin{tabular}{c}
          \includegraphics[width=\figWidth cm,height=3.5cm]{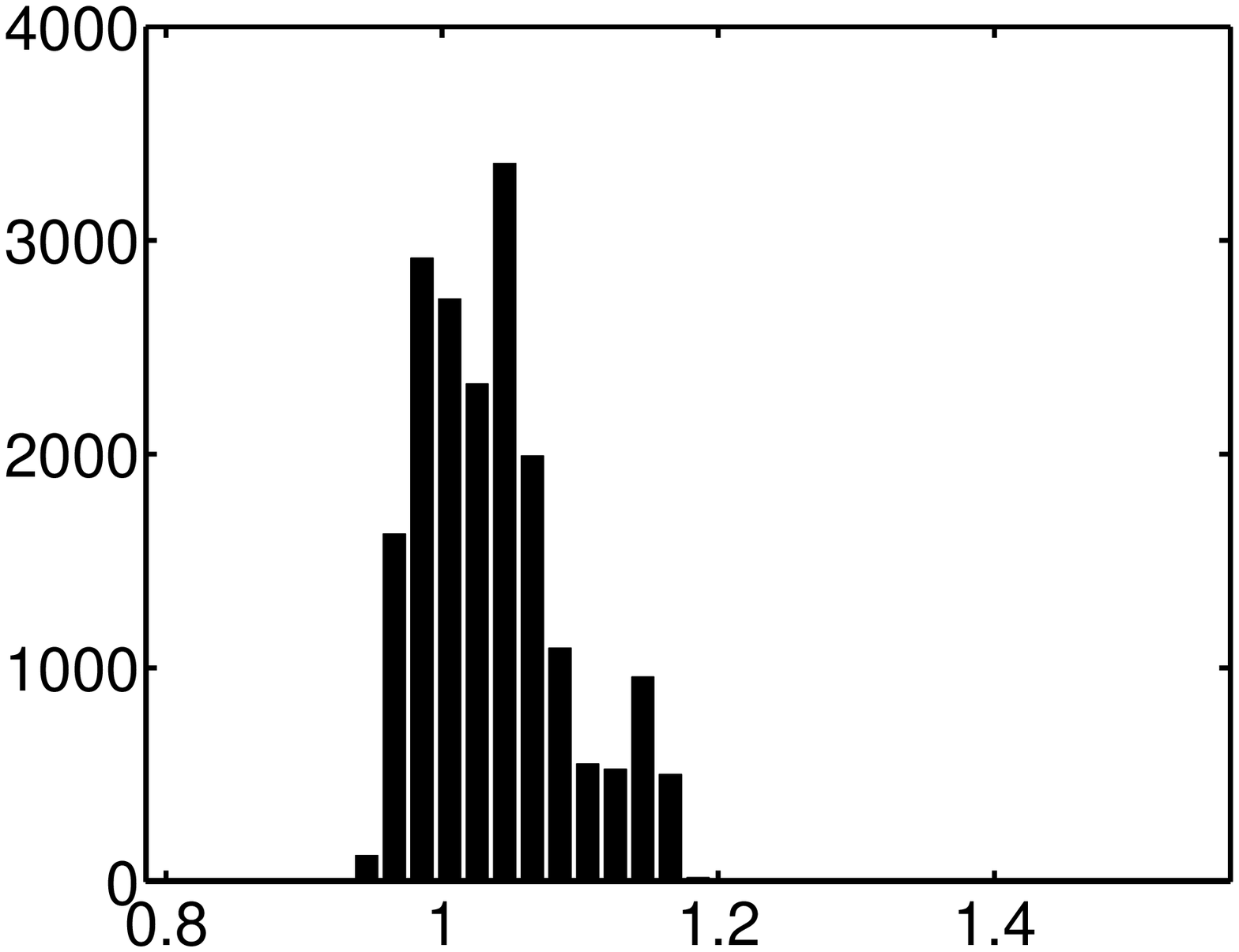} \\
          \includegraphics[width=\figWidth cm,height=3.5cm]{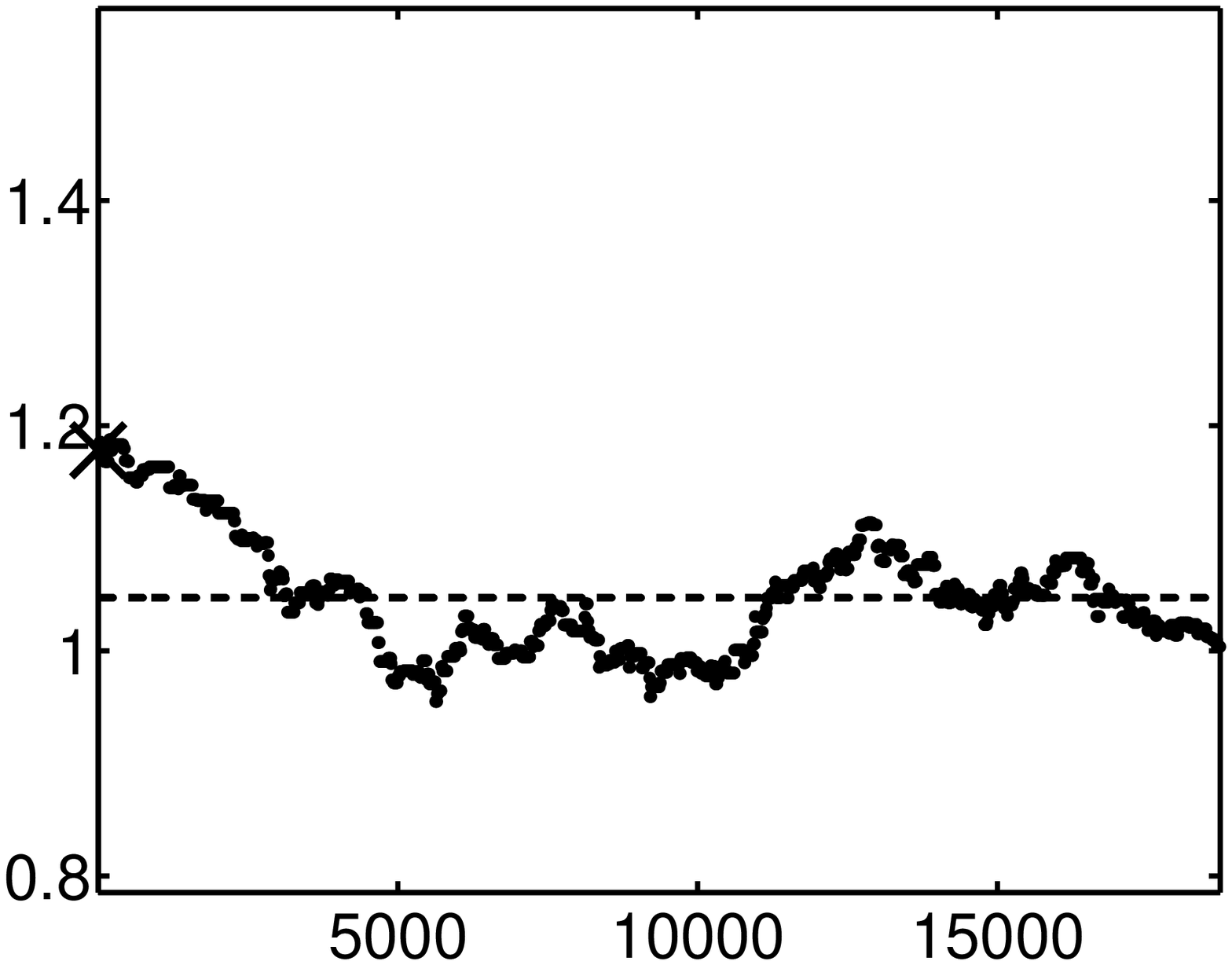}        
      \end{tabular}
      \label{fig:Phi}
    }
    \caption{Histogram and chain for the PSF parameters $\wA$ in
      Fig.~\ref{fig:Wa}, $\wB$ in Fig.~\ref{fig:Wb} and $\varphi$ in
      Fig.~\ref{fig:Phi}. The symbol $\times$ localizes the initial
      value and the dashed line corresponds to the true value. The
      x-axis for the histograms and the y-axis of the chain are limits
      of \aprio law. }
    \label{fig:intrumentParam}
\end{figure}

\newpage{}

\begin{figure}[h]
    \centering

    \subfigure[$(\G1,\wA)$]{\includegraphics[width=\textwidth/3]{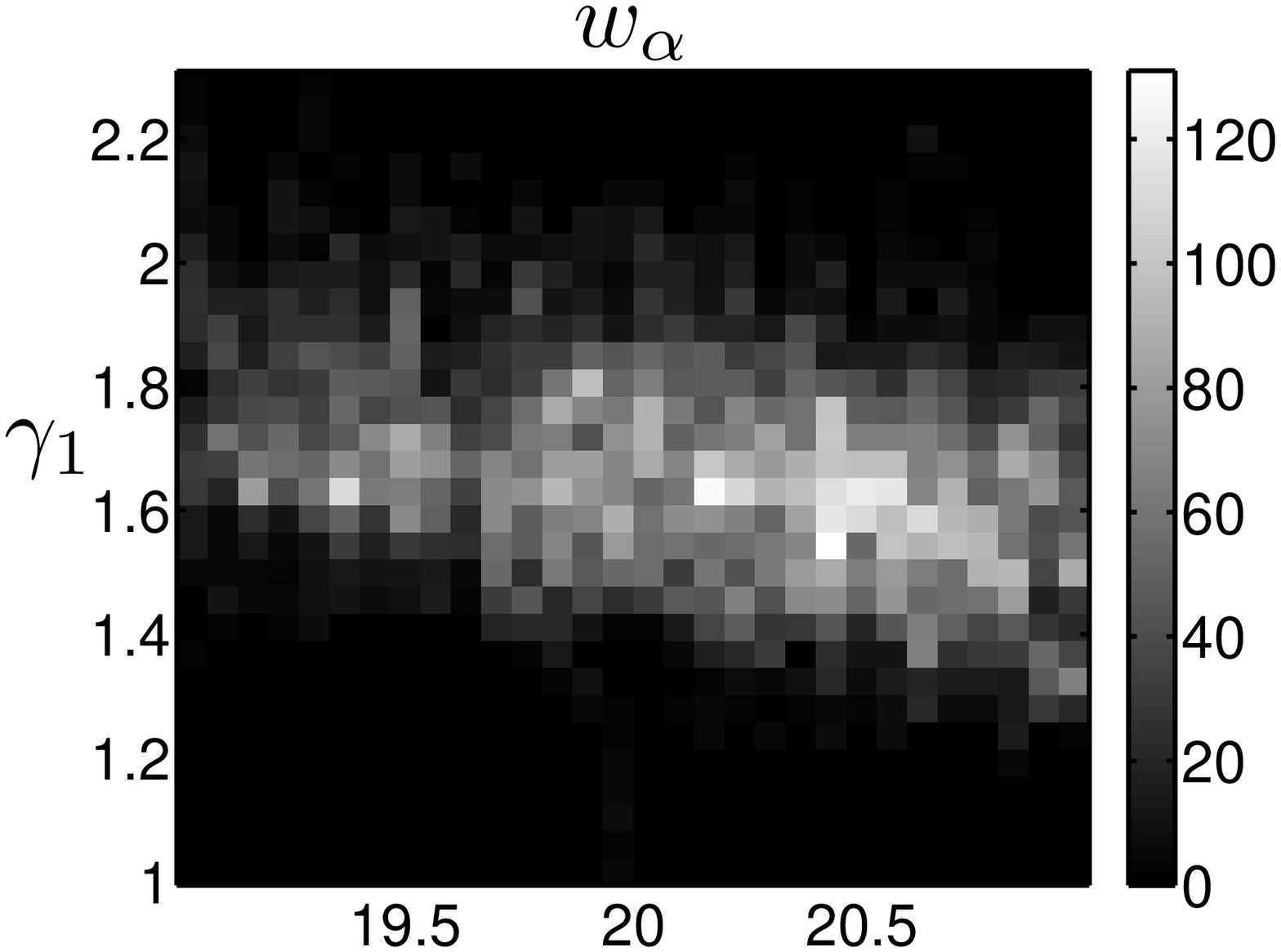}
      \label{fig:JointG1Wa}}
    \subfigure[$(\G1,\wB)$]{\includegraphics[width=\textwidth/3]{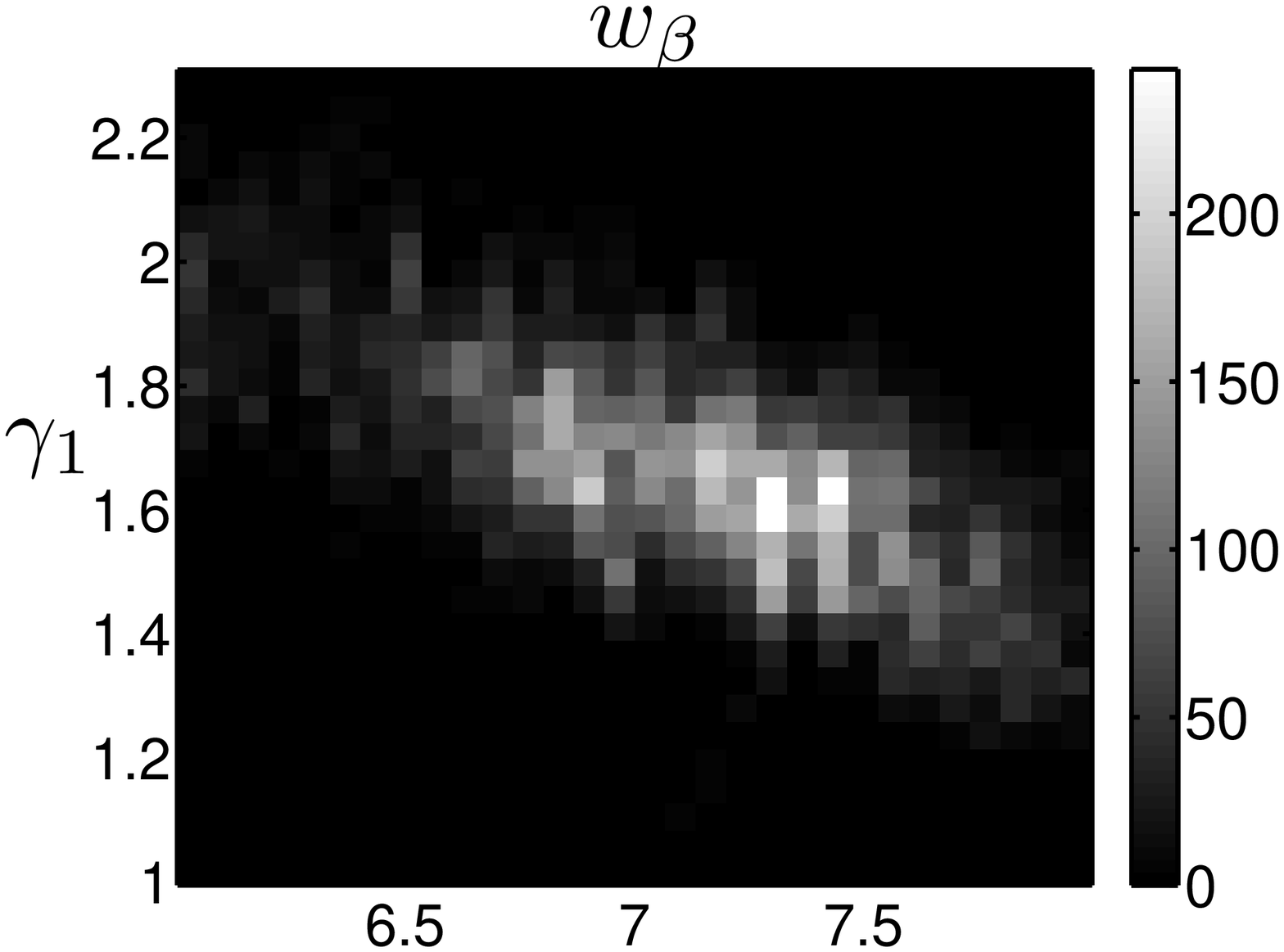}
      \label{fig:JointG1Wb}}

    \caption{Joint histograms for the couple $(\G1,\wA)$ and
      $(\G1,\wB)$ in Figs.~\ref{fig:JointG1Wa} and~\ref{fig:JointG1Wb}
      respectively.  The x-axis and y-axis are the parameter value.}
    \label{fig:JointHist}
\end{figure}

\begin{figure}[H]
    \centering

    \subfigure[Data]{
      \begin{tabular}{c}
          \includegraphics[width=\figWidthMed cm]{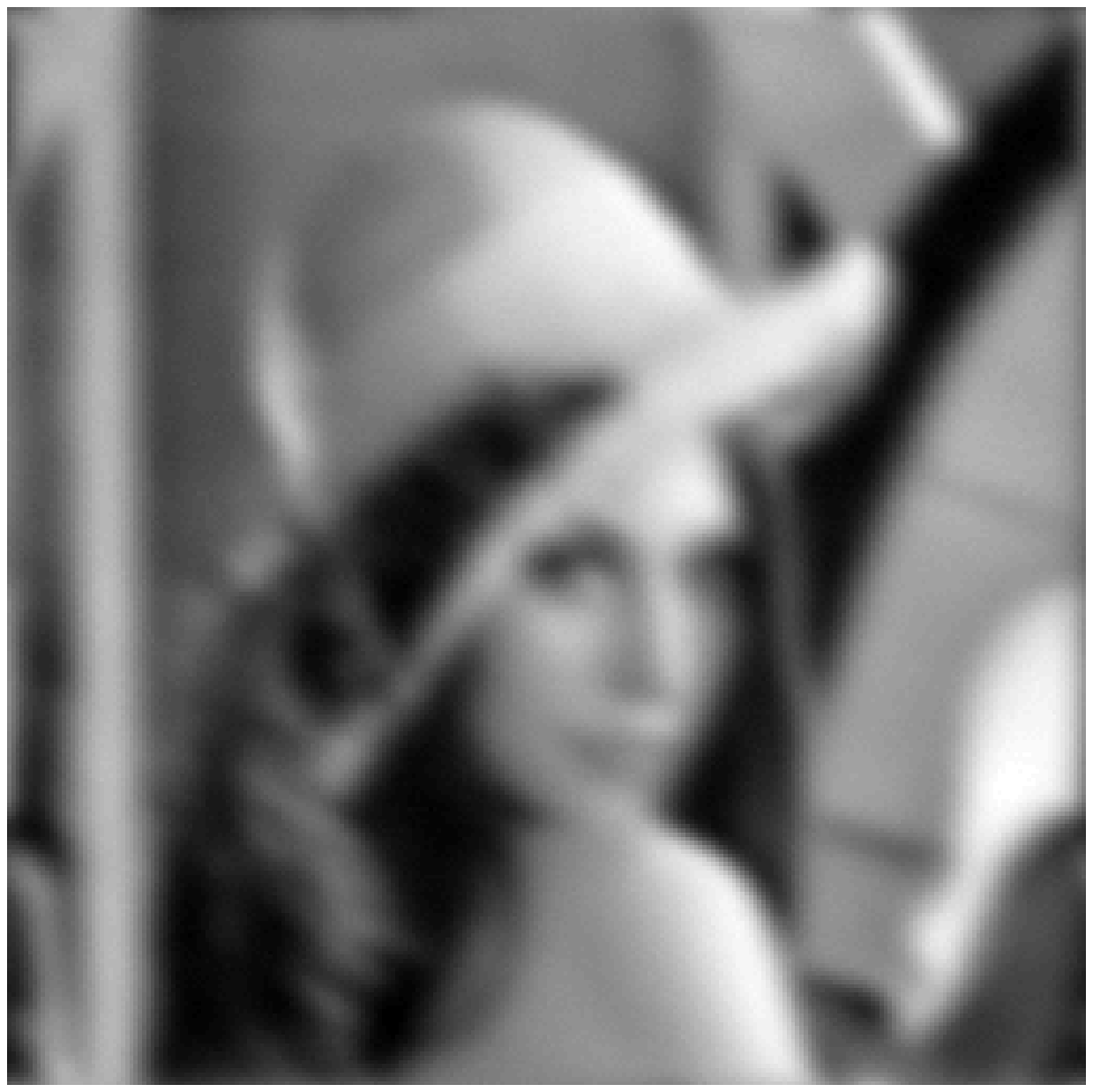}\\
          \includegraphics[width=\figWidthMed cm]{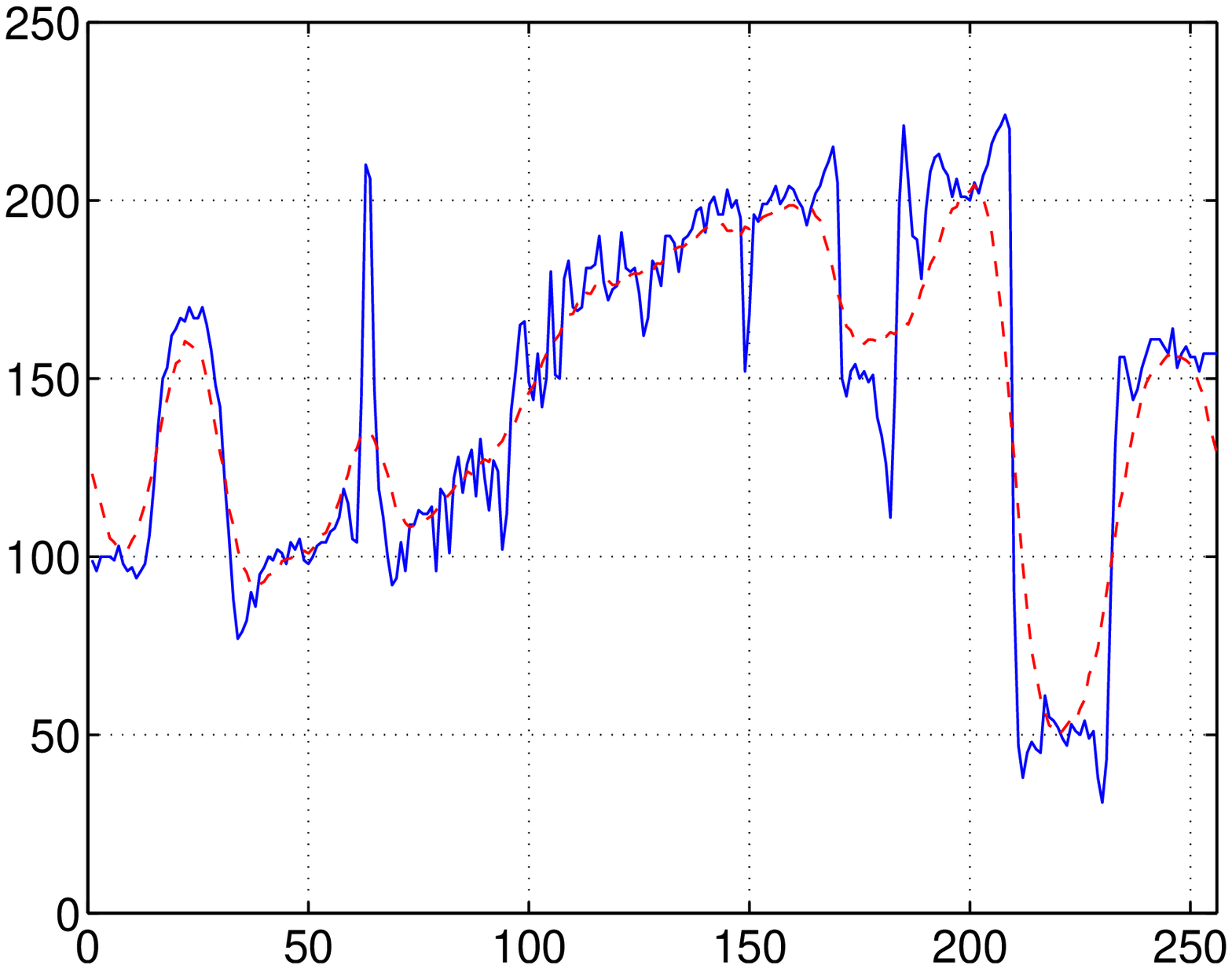}
      \end{tabular}
      \label{fig:trueLena}
    } \hspace{2cm}
    \subfigure[Estimated image]{
      \begin{tabular}{c}
          \includegraphics[width=\figWidthMed cm]{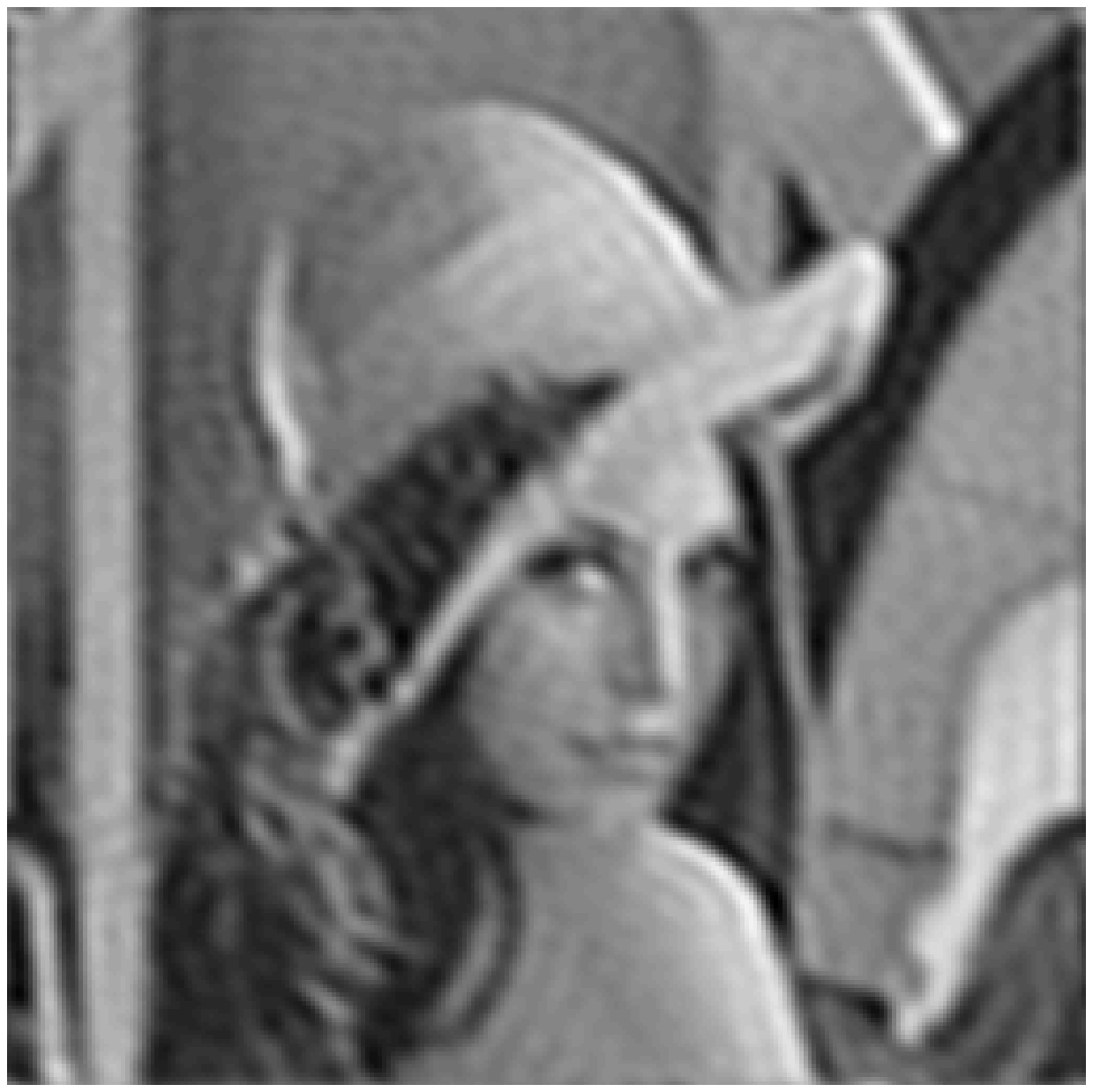} \\
          \includegraphics[width=\figWidthMed cm]{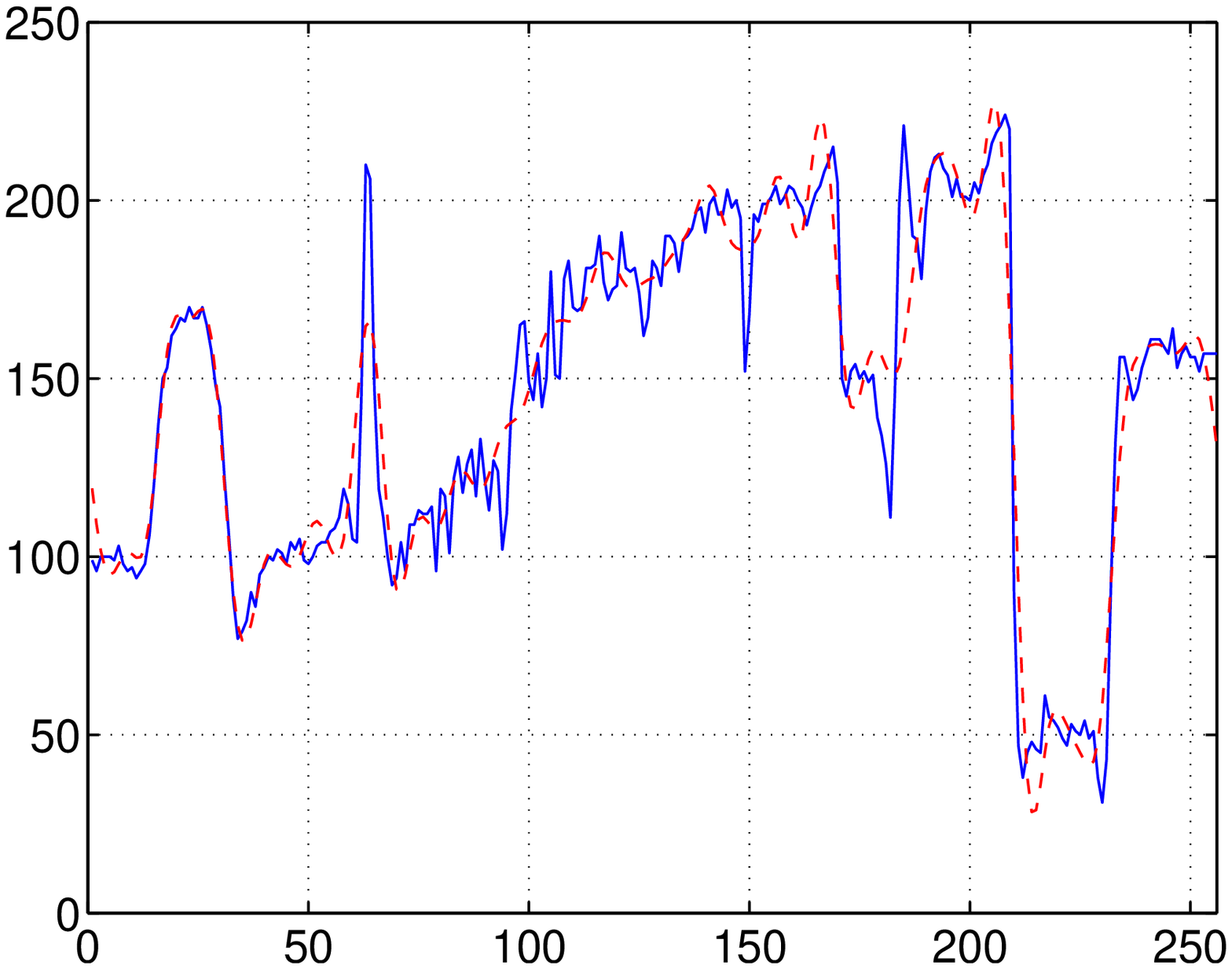}
      \end{tabular}
      \label{fig:eapLena}
    }
    \caption{Observed image Fig. \ref{fig:trueLena} and restored image
      Fig. \ref{fig:eapLena}. Profiles correspond to the 68-th
      line. The solid line is the true profile. Dashed line correspond
      to data in Fig. \ref{fig:trueLena} and estimated profiles in
      Fig. \ref{fig:eapLena}.}
    \label{fig:lena}
\end{figure}

\clearpage{}

\newpage{}

\begin{figure}[H]
    \centering
    \begin{algorithmic}[1]
        \State Initialisation of $\left[ \Im^{(0)}, \gammab^{(0)},
            \wb^{(0)}, k = 0 \right]$

        \Repeat

        \texttt{\% Sample of $\Im$}

        \State $\Sigmab \gets \GN^{(k)} |\LH|^2 + \GO^{(k)} |\LO|^2 +
        \G1^{(k)} |\LD|^2 $

        \State $\mub \gets \GN^{(k)} \Sigmab^{-1} \LH^*~ \Data$

        \State $\Im^{(k)} \gets \mub + \Sigmab^{-1/2}.*\mathtt{randn}$

        \texttt{\% Sample of} $\gammab$

        \State $\GN^{(k)} \gets \mathtt{gamrnd}(\alpha_\epsilon, \beta_\epsilon)$

        \State $\G1^{(k)} \gets \mathtt{gamrnd}(\alpha_1, \beta_1)$

        \State $\GO^{(k)} \gets \mathtt{gamrnd}(\alpha_0, \beta_0)$

        \texttt{\% Sample of} $\wb$

        \State $\wb_{\pD} \gets \mathtt{rand}*(\ab - \bb) + \ab$

        \State $J \gets \GN \left( \| \Data - \Lambdab_{\Hb}\,\Im \|^2
            - \|\Data - \Lambdab_{\Hb, \wb_{\pD}}\,\Im\|^2 \right)/2$

        \State \textbf{if} $\log (\mathtt{rand}) < \min \{J,0\}$
        \textbf{then} 

        \State \indent $\wb^{(k)} \gets \wb_{\pD}$

        \State \indent $ \LH \gets \Lambdab_{\Hb, \wb_{\pD}}$

        \State \textbf{else}

        \State \indent $\wb ^{(k)} \gets \wb^{(k-1)}$

        \State \textbf{end if}

        \texttt{\% Empirical mean}

        \State $k \gets k+1$

        \State $\rond{\bar \xb}^{(k)} \gets \sum_i \Im^{(i)}/k$

        \Until{$|\bar\xb^{(k)} - \bar \xb^{(k-1)}|/|\bar\xb^{(k)}| \leq \text{criterion}$}

    \end{algorithmic}
    \caption{Pseudo-code algorithm. $\mathtt{gamrnd, rand}$ and
      $\mathtt{randn}$ draw samples of gamma variable, uniform
      variable, and zero-mean unit-variance white complex Gaussian
      vector respectively.}
    \label{fig:algo}
\end{figure}

\end{document}